%% file: cmame17_paper.tex
\documentclass[final,3p,times,authoryear]{elsarticle}

\input{aux/preamble.tex}

\input{aux/macros.tex}

\usepackage{amssymb}

\journal{Computer Methods in Applied Mechanics and Engineering}
\begin{document}
\begin{frontmatter}

\title{Coupling Brain-Tumor Biophysical Models and Diffeomorphic Image Registration}
\author[ustutt]{Klaudius Scheufele}
\ead{klaudius.scheufele@ipvs.uni-stuttgart.de}
\author[uhouston]{Andreas Mang}
\ead{andreas@math.uh.edu}
\author[berkeley]{Amir Gholami}
\ead{amirgh@berkeley.edu}
\author[upenn]{Christos Davatzikos}
\ead{christos.davatzikos@uphs.upenn.edu}
\author[utexas]{\\George Biros}
\ead{biros@ices.utexas.edu}
\author[ustutt]{Miriam Mehl\footnote{Declarations of interest: none}}
\ead{miriam.mehl@ipvs.uni-stuttgart.de}

\address[ustutt]{University of Stuttgart, IPVS, Universit\"atstra{\ss}e 38, 70569 Stuttgart, Germany}
\address[uhouston]{University of Houston, Department of Mathematics, 3551 Cullen Blvd., Houston, Texas 77204-3008, USA}
\address[utexas]{University of Texas, ICES, 201 East 24th St, Austin, Texas 78712-1229, USA}
\address[berkeley]{University of California Berkeley, EECS, Berkeley, CA 94720-1776, USA}
\address[upenn]{Department of Radiology, University of Pennsylvania School of Medicine, 3700 Hamilton Walk, Philadelphia, PA 19104, USA}

\input{00-abstract.tex}

\begin{keyword}
biophysically constrained diffeomorphic image registration \sep tumor growth \sep atlas registration \sep adjoint-based methods \sep parallel algorithms
\MSC[2010] 49K20 
\sep 49M15 
\sep 35K57 
\sep 65K10 
\sep 68W10 
\end{keyword}
\end{frontmatter}

\section{Introduction}
\label{s:intro}
\input{01-00-intro.tex}

\section{Notation}
\label{s:notation}
\input{02-notation.tex}

\section{Formulation}
\label{s:formulation}
\input{03-formulation.tex}

\section{Numerical Methods}
\label{s:methods}
\input{04-methods.tex}

\section{Numerical Experiments}
\label{s:results}
\input{05-00-results.tex}

\section{Conclusion}
\label{s:conclusion}
\input{06-conclusion.tex}

\section{Acknowledgements}
\label{s:acknowledgements}
This material is based upon work supported by AFOSR grants
FA9550-17-1-0190; by NSF grant CCF-1337393; by the
U.S. Department of Energy, Office of Science, Office of Advanced
Scientific Computing Research, Applied Mathematics program under Award
Numbers DE-SC0010518 and DE-SC0009286; by NIH grant 10042242; by DARPA
grant W911NF-115-2-0121; and by the Technische Universit\"{a}t
M\"{u}nchen---Institute for Advanced Study, funded by the German
Excellence Initiative (and the European Union Seventh Framework
Programme under grant agreement 291763).  Any opinions, findings, and
conclusions or recommendations expressed herein are those of the
authors and do not necessarily reflect the views of the AFOSR,
DOE, NIH, DARPA, and NSF. Computing time on the High-Performance Computing
Centers (HLRS) Hazel Hen system (Stuttgart, Germany) was provided by an
allocation of the federal project application ACID-44104.
Computing time on the Texas Advanced Computing Centers Stampede system was
provided by an allocation from TACC and the NSF.

\bibliographystyle{elsarticle-harv}
\bibliography{mia_library}

\FloatBarrier
\cleardoublepage
\appendix
\section{Supplementary Material} \label{s:appendix}
\input{07-appendix.tex}

\end{document}

%% file: aux/preamble.tex
\usepackage{amsmath,amssymb}
\usepackage[sc]{mathpazo}
\usepackage{algorithm,algorithmic}
\usepackage[sort,compress]{cite}
\usepackage{graphicx}
\usepackage{multirow}
\usepackage{makecell}
\usepackage{textcomp}
\usepackage{rotating}
\usepackage{siunitx}
\usepackage{paralist}
\usepackage{booktabs}
\usepackage{nicefrac}
\usepackage{placeins}
\usepackage{lscape}
\usepackage[table]{xcolor}
\usepackage{arydshln}
\usepackage{geometry}
\usepackage{mathtools}
\usepackage{enumerate}
\usepackage[hyperpageref]{backref}
\usepackage[labelfont=bf,textfont=it,belowskip=0pt,aboveskip=5pt,tableposition=top]{caption}
\usepackage[colorlinks=true,
citecolor=blue,
filecolor=black,
linkcolor=blue,
urlcolor=blue]{hyperref}

\graphicspath{{figures/}{tables/}}
\newcounter{runidnum}

\newcolumntype{R}{>{\columncolor{gray!40}}r}
\newcolumntype{L}{>{\columncolor{gray!40}}l}
\newcolumntype{C}{>{\columncolor{gray!40}}c}
\newcommand{\highlightrow}{\rowcolor{gray!20}}

\DeclareMathOperator*{\argmin}{arg\,min}

\newcommand{\Tfor}{\mathcal{T}^{\mbox{\scriptsize{fwd}}}}
\newcommand{\Tinv}{\mathcal{T}^{\mbox{\scriptsize{inv}}}}
\newcommand{\Rfor}{\mathcal{R}^{\mbox{\scriptsize{fwd}}}}
\newcommand{\Rinv}{\mathcal{R}^{\mbox{\scriptsize{inv}}}}
\newcommand{\geom}{\mathcal{G}}

%% file: aux/macros.tex
\newcommand{\figref}[1]{Fig.~\ref{#1}}
\newcommand{\tabref}[1]{Tab.~\ref{#1}}
\newcommand{\secref}[1]{\S\ref{#1}}

\newcommand{\nf}[1]{\nicefrac{#1}}

\newcommand{\diceB}{\ensuremath{\text{DICE}_B}} 
\newcommand{\diceT}{\ensuremath{\text{DICE}_T}} 
\newcommand{\elltwoB}{\ensuremath{\mu_{B,L^2}}} 
\newcommand{\elltwoT}{\ensuremath{\mu_{T,L^2}}} 
\newcommand{\elltwoV}{\ensuremath{e_{\vect{v},L^2}}} 
\newcommand{\relG}{\ensuremath{\|\vect{g}\|_{\text{rel}}}} 
\newcommand{\elltwoINIT}{\ensuremath{e_{c0,L^2}}} 
\newcommand{\rtIT}{T^{it}} 
\newcommand{\rtTOT}{T^{total}} 
\newcommand{\rtTUM}{T^{tu}_{inv}} 
\newcommand{\rtREG}{T^{reg}_{inv}} 

\newcolumntype{R}{>{\columncolor{gray!20}}r}
\newcolumntype{L}{>{\columncolor{gray!20}}l}

\newcommand\pernum[1]{\num[round-precision=1]{#1}}

\newcommand{\D}[1]{\ensuremath{\mathcal{#1}}}    

\newcommand{\ns}[1]{\ensuremath{\mathbb{#1}}}
\newcommand{\fs}[1]{\ensuremath{\mathcal{#1}}}

\newcommand{\idiv}{\ensuremath{\mbox{div}\,}}
\newcommand{\igrad}{\ensuremath{\nabla}}
\newcommand{\ilap}{\rotatebox[origin=c]{180}{$\nabla$}}

\newcommand{\half}[1]{\frac{#1}{2}}

\renewcommand{\d}[1]{\mathop{}\!\mathrm{d}#1}
\newcommand{\dt}{\d{t}}

\newcommand{\p} {\partial}

\newcommand{\vect}[1]{\boldsymbol{#1}} 
\newcommand{\mat}[1]{\boldsymbol{#1}}  

\newcommand{\bipa}{\begin{inparaenum}[(\itshape i\upshape)]}
\newcommand{\eipa}{\end{inparaenum}}

\newcommand{\bipasub}{\begin{inparaenum}[(\itshape a\upshape)]}
\newcommand{\eipasub}{\end{inparaenum}}

\newcommand{\ipoint}[1]{\textit{\textbf{#1}}}

\newcommand{\defeq}{\ensuremath{\mathrel{\mathop:}=}}

\definecolor{sred}{cmyk}{0.01,0.98,0,0.2} 
\reversemarginpar
\newcommand{\mmargin}[1]{{\marginpar{\em\tiny #1}}}\renewcommand{\mmargin}[1]{}

%% file: 00-abstract.tex
\begin{abstract}
We present the SIBIA (Scalable Integrated Biophysics-based Image Analysis) framework for joint image registration and biophysical inversion  and  we  apply it to analyse MR images of glioblastomas (primary brain tumors). 
Given the segmentation of a normal brain MRI and the segmentation of a cancer patient MRI, we wish to determine tumor growth parameters and a registration map so that if we ``grow a tumor'' (using our tumor model) in the normal segmented image and then register it to the segmented patient image, then the registration mismatch is as small as possible. We call this  ``\emph{the coupled problem}'' because it two-way couples the biophysical inversion and registration problems.  In the image registration step we solve a large-deformation diffeomorphic registration problem parameterized by an Eulerian velocity field.  In the biophysical inversion step we estimate parameters in a reaction-diffusion tumor growth model that is formulated as a  partial differential equation  (PDE).  In SIBIA, we couple these two steps in an iterative manner.
We first presented the components of SIBIA in  \emph{ ``Gholami et al, Framework for Scalable Biophysics-based Image Analysis,  IEEE/ACM Proceedings of the SC2017''}, in which we derived parallel distributed memory algorithms and software modules for the \emph{decoupled} registration and biophysical inverse problems.
In this paper, our contributions are  the introduction of a PDE-constrained optimization formulation of the coupled problem, the derivation of the optimality conditions, and the derivation of a Picard iterative scheme for the solution of the coupled problem. In addition, we perform several tests to experimentally assess the performance of our method on synthetic and clinical datasets. We demonstrate the convergence of the SIBIA optimization solver in different usage scenarios. We demonstrate that using SIBIA, we can accurately solve the coupled problem in three dimensions ($256^3$ resolution) in a few minutes using 11 dual-x86 nodes.
\end{abstract}

%% file: 01-00-intro.tex
\begin{figure*}[t!]
\centering
\includegraphics[scale=1]{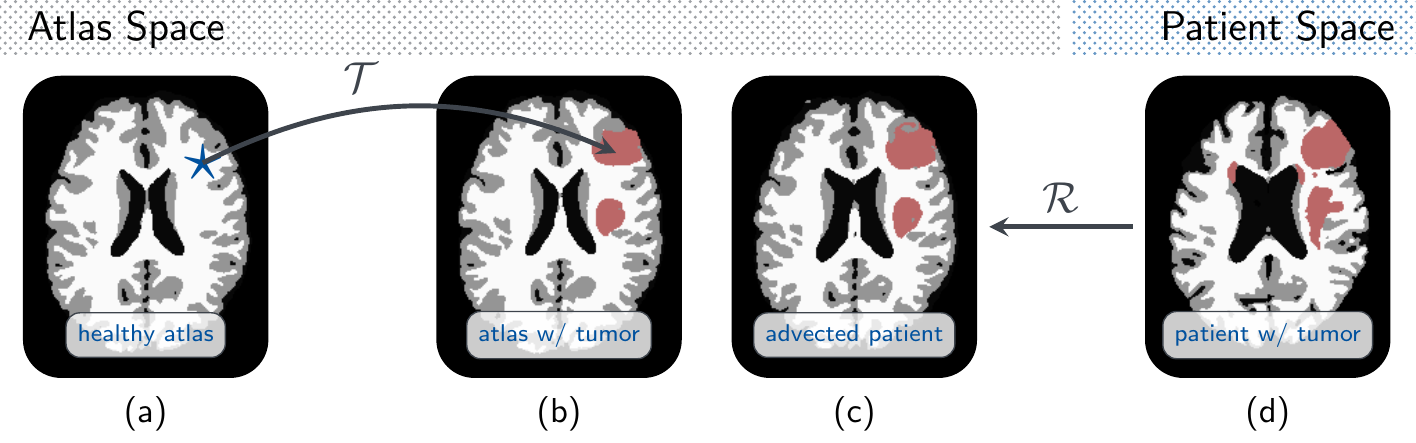}
\caption{Here we summarize the joint registration and biophysical inversion in SIBIA. \textbf{(a)} shows the segmented healthy brain (the atlas), \textbf{(b}) tumor-bearing atlas brain generated by biophysical simulation, \textbf{(c}) patient tumor registered to atlas (in atlas space), \textbf{(d)} tumor-bearing patient brain. The inputs are images (a) and (d). The outputs are the tumor growth parameters, the registration parameters, and images (b) and (c). In SIBIA, we compute tumor-growth and registration parameters so that images (b) and (c) are as similar as possible. In this example, the biophysical tumor growth parameters are the initial conditions for a reaction-diffusion equation. The registration is parameterized using an Eulerian framework and a velocity field that is used to advect image (d) to image (c).}
\label{f:brain-pipeline}
\end{figure*}

We present the SIBIA framework that comprises a mathematical formulation, algorithms, and software for joint image registration and biophysical inversion. Given a volumetric segmentation of a magnetic resonance imaging (MRI) dataset of a glioma (primary brain tumor) patient, SIBIA simultaneously registers this image to a segmented MRI dataset of a normal brain (an atlas) and fits a biophysical tumor growth model.
%
Such a joint registration-biophysical inversion approach is motivated by two practical problems:
\begin{inparaenum}[(1)]
\item Often, we need to calibrate complex macroscopic biophysical PDE models using a single-time snapshot dataset. Since the biophysical model is typically a dynamical system, it is unclear how to fit model parameters using just a single-time dataset since we don't have the initial state. Replacing the missing second patient snapshot by a healthy atlas
and coupling registration between patient and atlas with the tumor growth model attempts to resolve this problem.
\item We want to register normal MR images to MR images with abnormalities. This normal-to-abnormal registration finds applications in surgical planning (e.g., mapping structural and functional information to a patient image), longitudinal studies, followups, or population studies.
\end{inparaenum}

We focus on applying SIBIA to the registration of segmented normal images to segmented MR images of glioma patients. 
Many image analysis workflows for gliomas involve fitting biophysical models for tumor characterization and prognosis~\citep{Swanson:2008a,Rahman:2017a}, for image registration~\citep{Mohamed:2006a,Kwon:2014a,Zacharaki:2009a}, and for image segmentation~\citep{Bakas:2015a,Gooya:2012a,Prastawa:2009a}. In many methods for these image analysis and biophysical modeling problems, an essential step is the solution of an inverse tumor-growth problem, i.e., for instance the calculation of the initial conditions for tumor growth; this process can be combined with registration and segmentation as in~\citep{Gooya:2012a}.

\subsection{Statement of the problem and summary of the approach}
The specific scenario we are interested in is summarized in~\figref{f:brain-pipeline}. We are given a single snapshot
of the multi-modal MRI of a glioma patient. Let us assume that we have obtained a segmentation of the MRI in healthy tissue and abnormal tissue, i.e., we have labels for white matter, gray matter, cerebrospinal fluid (CSF), cerebellum, ventricles, and tumor
\footnote{In reality, the tumor has subclasses like edema, necrotic tumor, enhancing tumor, and others. Since we're using a very simple tumor model, we don't consider these subclases here.}.
Since the tumor-growth problem is typically a time dependent problem, we need at least two time snapshots to invert for parameters, ideally we would like to have a pre-cancer image of the healthy brain of the patient, the white, gray, and CSF structure, which we don't have, though. The basic idea to tackle this issue is to use a pre-segmented normal healthy brain (atlas) as a means to create a second snapshot. This is the step that requires registration. The best way to describe this is to consider the \emph{``forward problem''}: Assume we are given the (haelthy) atlas, a tumor-growth model, and a deformation map (represented as advection with a given velocity field). Then, we first apply the tumor growth model to the healthy image and produce a new segmentation that has \emph{both} healthy tissue and tumor. In a second step, we warp this image using the given deformation map. This image is the tumor-plus-deformation warped atlas. In the \emph{``inverse problem''}, we seek to compute the tumor-growth parameters and deformation parameters so that the tumor-plus-deformation warped atlas image matches the patient image.


Thus, SIBIA combines the following steps:
\begin{inparaenum}[(i)]
\item a forward tumor solver growing a virtual (or synthetic) tumor in the segmented healthy atlas brain,
\item the inverse tumor solver determining the initial condition for tumor growth given an observation at a later time,
\item a forward linear advection solver that implicitly computes the warped image given the velocity field, and
\item a large-deformation diffeomorphic image registration solver.
\end{inparaenum}
Both inversion and registration are optimization problems.

\subsection{Contributions}
%
\begin{enumerate}[(i)]
\item We formulate a coupled tumor growth and registration optimization problem in \secref{s:formulation} and derive its gradient.
\item We derive the first order optimality conditions for the joint registration and biophysical inversion for brain tumor growth.
\item Instead of using a gradient descent method, we propose a Picard iteration scheme for solving the system based on modular tumor and image registration components, a generic idea that can be extended to other problems.
\item We conduct numerical experiments on synthetic and real data that demonstrate the validity and efficiency of our approach, in particular
we show that the Picard iteration reduces the gradient and converges to a local minimum.
\item We examine the sensitivity of our solver on the choice of the tumor model variant. We consider three variants of the tumor model: a reaction model, a reaction-diffusion model, and a simple radial basis approximation (no time evolution; the tumor is simply approximated at the observation time). All these variants can be used depending on the goal of the analysis.
\end{enumerate}

\subsection{Limitations and Open Issues}
\label{s:limitations}
Several limitations and unresolved issues remain. The main limitation
is that we don't have a theoretical proof that the Picard scheme
converges%
\footnote{The
optimization problem we're solving is highly non-convex, so multiple
solutions may exist. Our solver convergences to a  local
minimum.}
, we only provide empirical evidence. A second limitation is that our tumor
model does not include mass-effect, we use a simple phenomenological reaction-diffusion
model~\citep{Swanson:2000a,Swanson:2002a,Murray:1989a}. This model, although
it has been proven to be quantitatively useful for image analysis and
tumor characterization~\citep{Mang:2012b,Swanson:2008a,Jackson:2015a,Lima:2016a}, has limited predictive capabilities.
Its predictive capabilities are very limited. We are currently working on integrating a mass-effect, which is considered critical for characterizing tumor aggressiveness~\citep{Kyriacou99a,yankeelov-miga13,Mohamed:2006a} as well as richer reaction-diffusion models.  SIBIA can naturally be extended to these more complex
models although of course more tests will be necessary to demonstrate
convergence of the Picard scheme.

\subsection{Literature review}

\input{01-01-literature.tex}

\subsection{Outline}

\input{01-02-outline.tex}

%% file: 01-01-literature.tex
We presented fast algorithms for the individual components of SIBIA in~\citep{Gholami:2017a} were we introduced the key computational kernels (reaction-diffusion, advection, and inverse solution based on FFT and particle-in-mesh methods)  and demonstrated their scalability on very large images and distributed-memory architectures. Our problem formulation results in a mixed-type PDE-constrained optimization problem that poses significant numerical challenges. We refer to~\citep{Biegler:2003a,Borzi:2012a,Herzog:2010a,Hinze:2009a} for a general introduction into PDE-constrained optimization and to~\citep{Angelini:2007a,Bauer:2013a,Mang:2017c} for reviews on its application to medical image analysis. 
We will limit ourselves to the work most closely related to ours in the following.\\

The first  component of SIBIA is image registration. We refer to~\citep{Modersitzki:2004a,Sotiras:2013a} for a general overview of medical image registration. Registering the atlas to the patient requires finding correspondences between two topologically different images---one with tumor and one without tumor. The key issue here is the ill-defined correspondence arising from the presence of a tumor in only one of the images to be registered. A simple strategy to deal with this issue is to consider the tumor area as non-informative and mask it from the optimization~\citep{Henn:2004a,Stefanescu:2004a,Brett:2001a} or to relax the registration in the area affected by the tumor~\citep{Parisot:2014a}. This, however, yields poor registration quality for tumors with severe tissue deformation due to mass effect. Another strategy is to simultaneously invert for the deformation map and a drift in intensity representing the imaging abnormality associated with the tumor~\citep{Li:2010a,Li:2012a}, which can also be applied to other registration problems with topological differences (for instance, pre- and postoperative image registration~\citep{Kwon:2014a}). While it may produce acceptable results for the purpose of atlas-based segmentation and registration, it can not be used in the context of model prediction---which is our ultimate goal. 
Our prior work on diffeomorphic image registration~\citep{Mang:2015a,Mang:2016a,Mang:2016c,Mang:2017a,Mang:2017b} forms the basis for the proposed methodology based on the pioneering work in~\citep{Christensen:1996a,Trouve:1998a,Beg:2005a} (see~\citep{Mang:2015a} for additional references).

The second component of SIBIA is the calibration of brain tumor models to medical imaging data~\citep{Gholami:2016a,gholami_thesis,Gholami:2017a,yankeelov-miga13,Miga98a}. The associated PDE operator is a parabolic, non-linear, reaction-diffusion equation~\citep{Murray:1989a,Swanson:2002a}. Despite its phenomenological character, this model is capable of generating simulations that are in good agreement with observations of abnormalities in standard MR imaging data and has been used by other groups besides ours~\citep{Clatz:2005a,Hogea:2007a,Harpold:2007a,Konukoglu:2010a,Konukoglu:2010b,Le:2016a,Lima:2016a,Mosayebi:2012a,Menze:2011a,Mang:2012b,Swanson:2008a}. Our implementation features inversion operators for the initial tumor concentration, for diffusivity parameters, or for the growth rate~\citep{Gholami:2016a}. Related optimal control formulations can, e.g., be found in~\citep{Colin:2014a,Hogea:2008b,Knopoff:2013a,Knopoff:2017a,Liu:2014a,Mang:2012b,Quiroga:2015a,Quiroga:2016a,Wong:2015a}. Unlike most existing approaches (with the exception of~\citep{Kwon:2014b}), our parameterization of the problem~\citep{Gholami:2016a,gholami_thesis} allows us to invert for multifocal tumors. Numerical methods for solving the tumor calibration problem are based either on point estimate or  Bayesian inference methods~\citep{Lima:2016a,lima-oden-e17}. Bayesian inference  characterizes the uncertainty in the model parameters. In this paper, we just focus on methods for  point estimates using an adjoint formulation. Our approach can be extended to the Bayesian setting~\citep{flath-ghattas-e11}. 

The integration of biophysical brain tumor simulations with deformable image registration is not new~\citep{Gooya:2012a,Hogea:2008a,Mohamed:2006a,Zacharaki:2008a,Zacharaki:2008b,Zacharaki:2009a}. In~\citep{Mohamed:2006a,Zacharaki:2008a,Zacharaki:2008b,Zacharaki:2009a}, a purely mechanical model for tumor progression was used. The two key limitations of the work in~\citep{Mohamed:2006a,Zacharaki:2008a,Zacharaki:2008b,Zacharaki:2009a} are \bipa\item that the model is oversimplified; it did not provide the capabilities to generate tumors with complex shapes \item that these models do not provide information  about the progression and infiltration of cancerous cells into surrounding healthy tissue.\eipa The proposed formulation does not share these limitations. The work that is closest to ours is~\citep{Bakas:2015a,Kwon:2014a,Hogea:2008a,Gooya:2012a}. Here, the authors present a framework for joint segmentation, registration, and tumor modelling. Likewise to ours, their approach is based on a PDE constrained optimization problem, where the constraint is a non-linear, mixed-type reaction-diffusion-advection equation for the tumor cell density.

What sets our work apart are \bipa\item the solver (we do not iterate on both control variables simultaneously; we perform a block elimination and iterate resulting in an interleaved optimization on the controls using globalized Newton--Krylov solvers as opposed to derivative free-optimization~\citep{Gooya:2012a,Hogea:2008a,Mohamed:2006a,Zacharaki:2008a,Zacharaki:2008b,Zacharaki:2009a,Wong:2015a});  \item an efficient parallel implementation~\citep{Mang:2016c,Gholami:2017a}; \item the derivation of the optimality systems for the fully coupled problem; \item the parameterization for the initial condition of the tumor, which not only allows us to represent multifocal tumors but also to significantly simplify the PDE constraint without loosing segmentation accuracy; \item and finally the integration with a state-of-the-art algorithm for constrained large deformation diffeomorphic image registration~\citep{Mang:2015a,Mang:2016a,Mang:2016c,Mang:2017b}\eipa.

%% file: 01-02-outline.tex
We present the proposed formulation for biophysically constrained diffeomorphic image registraiton in \secref{s:formulation}. We show that the optimality systems of our problem are complex, space-time, multi-physics operators that pose significant numerical challenges. We further motivate and explain our choices for the forward operators for the tumor model and the registration in \secref{sec:tumor} and \secref{sec:registration} and present the associated inversion operators. In \secref{s:methods}, we discuss the algorithmic details. We summarize the setup for the experiments in \secref{s:setup} and present numerical results on real and synthetic data in \secref{s:results} and conclude with \secref{s:conclusion}. Additional material is provided in \secref{s:appendix}.

%% file: 02-notation.tex
Before presenting the models used for tumor growth and image registration, we summarize the notation used throughout the manuscript.

\paragraph{Segmentation labels and probability maps} For each healthy brain tissue type, i.e., gray matter (GM), white matter (WM), and cerebrospinal fluid (CSF; which includes the ventricles), we use a
separate probability map. We represent these probability maps as a space-time vector field
\begin{equation}
\vect{m}(\vect{x},t) =
\left( m_i(\vect{x},t) \right)_{i=1,\ldots,3} \in\ns{R}^3 \; \mbox{ with } \;
m_1 = m_{GM}, m_2 = m_{WM}, m_3 = m_{CSF}
\label{eq:vec_m}
\end{equation}

\noindent with $m_i(\vect{x},t) \in [0,1]$ defined on the space-time interval $\Omega\times[0,1]^3$ with boundary $\p\Omega$. The domain occupied by brain tissue (healthy or unhealthy) is denoted by $\Omega_B\subset\Omega$. The fourth probability map is $c(\vect{x},t) \in [0;1]$, also defined in $\Omega_B$; $c$ is the output of the tumor forward simulation and represents the probability to encounter cancerous cells at a location $\vect{x}$ at time $t$.
%
For convenience, we also define the space-time domains $U = \Omega \times (0,1]$ and $\bar{U} = \Omega \times [0,1)$.

Labels, i.e., characteristic functions calculated based on given threshold values
for tumor and healthy tissue types are used only in the evaluation of our results to compute
Dice coefficients.

\paragraph{Patient and atlas data} Data of the actual patient image are marked with a subscript $T$ (template, $\vect{m}_T, c_T$), data of patient images advected to the atlas domain with a subscript $P$ ($\vect{m}_P, c_P$), and those of the atlas brain with a subscript $A$ ($\vect{m}_A, c_A$).

\paragraph{Dependency on time and space}
In most formulations, we do not explicitly include the dependency on the spatial position $\vect{x}$, but only the time dependency. For instance, $c_A(0)$ denotes the initial tumor probability map defined in the atlas image, whereas $c_A(1)$ is the tumor at time $t=1$ (solution of the tumor forward problem). This is the point in time associated with the patient image.
%
Probability maps in the atlas image evolve in the non-dimensional (tumor growth) time interval $[0,1]$.
For diffeomorphic image registration, we introduce a \emph{pseudo-time variable} $t\in[0,1]$ and invert for a stationary velocity field $\vect{v}(\vect{x})$; $t$ does not have a physical meaning; we associate $t=0$ with the undeformed patient image (image to be registered; template image) and $t=1$ with its deformed representation. We make more explicit definitions below.

\paragraph{Vector notation}
Given a vector field $\vect{m} \in \ns{R}^3$, we compute $\igrad \vect{m} = (\partial_j \vect{m}_i)_{i,j=1,2,3}, \in\,  \ns{R}^{3,3}$.
%
That is, given a velocity field $\vect{v} \in \ns{R}^3$, $\igrad \vect{m}\ \vect{v}\ \in \ns{R}^3$ indicates a matrix-vector multiplication. The standard scalar product in $\mathbb{R}^3$ is denoted by ``$\cdot$'' and the outer product between two vector fields will be denoted by ``$\otimes$''. In addition, we define the following inner products:
\begin{align}
&\langle\vect{m},\tilde{\vect{m}}\rangle_{L^2(\Omega)^3}
\coloneqq \sum_{i=1}^3\langle m^i, \tilde{m}^i \rangle_{L^2(\Omega)},
&\|\vect{m}\|^2_{L^2(\Omega)^3}
\coloneqq\sum_{i=1}^3\|m^i\|^2_{L^2(\Omega)}.
\label{eq:scalar_vect_norm_vect}
\end{align}

%% file: 03-formulation.tex
\medskip

We propose a new coupled formulation based on diffeomorphic image registration and inverse tumor growth simulation. As an input, we take
the tumor-free atlas brain geometry $\vect{m}_A(\vect{x},0)$ (reference image, healthy atlas) and the tumor-bearing patient geometry $\vect{m}_T(\vect{x})$ (template image, patient with tumor) plus the patient's tumor probability map $c_T(\vect{x})$ (template tumor).
Our formulation is based on optimal control theory and results in a coupled PDE-constrained optimization
problem. We invert for a stationary velocity field $\vect{v}(\vect{x}) \defeq (v^1(\vect{x}),v^2(\vect{x}),v^3(\vect{x}))^\mathsf{T}$ (establishes the spatial correspondence between the patient and atlas image), a parameter vector $\vect{p}$ (parametrizes the simulated tumor in the tumor-free atlas image), and a mass-source map $w(\vect{x})$ (controls the computed deformation pattern) as follows:
\begin{subequations}
\label{eq:global_opt}
\begin{eqnarray}
\min_{\vect{v},\vect{p},w} \fs{J}(\vect{v},p,w) & \mbox{ with } & \fs{J}(\vect{v},p,w) \defeq \mathcal{D}_c(c_{A},c_{P}) + \mathcal{D}_{\vect{m}}(\vect{m}_A,\vect{m}_P) + \beta_{\vect{p}} \D{S}_{\vect{p}} + \beta_{\vect{v}} \D{S}_{\vect{v}} + \beta_w \D{S}_w  \;  \\ & \mbox{ subject to } \notag
\label{e:global_objective}
\end{eqnarray}
\begin{align}
\p_t c_A - \idiv \mat{k} \igrad c_A - f(c_A) &=  0 &&\mbox{in}~U, 
\label{eq:glob:tumor:fwd} \\
c_A(0) &= \Phi p &&\mbox{in}~\Omega,
\label{eq:glob:tumor:initcond-1} \\
\partial_t \vect{m}_{P} + \nabla \vect{m}_{P}\ \vect{v} &= \vect{0} && \mbox{in}~U, 
\label{eq:glob:reg:state:geom}\\
\vect{m}_P(0) &= \vect{m}_T &&\mbox{in}~\Omega,
\label{eq:glob:reg:state-initcond-1} \\
\p_t c_P + \igrad c_P \cdot\vect{v} &= 0 && \mbox{in}~U, 
\label{eq:glob:reg:state:c} \\
c_P(0) &= c_T &&\mbox{in}~\Omega,
\label{eq:glob:reg:state-c-initcond-2} \\
\idiv \vect{v} &= w && \mbox{in}~U 
\label{eq:glob:reg:divfree} \\
%
\vect{m}_A(1) & = \vect{m}_A(0)(1-c_A(1)) &&
\mbox{in}~\Omega\label{eq:glob:coupling-eq:geom}
\end{align}
\end{subequations}

\noindent with periodic boundary conditions on $\partial\Omega$. The objective functional $\mathcal{J}$ in~\eqref{e:global_objective} consists of the following building blocks:

\begin{enumerate}[(i)] \setlength\itemsep{-.5pt}
\item the two driving $L^2$-distance measures
\begin{align*}
&\mathcal{D}_c(c_{A},c_{P}) \defeq \half{1} \|c_A(1) - c_P(1)\|^2_{L^2(\Omega)}, \; \mbox{ and } \;
\mathcal{D}_{\vect{m}}(\vect{m}_A,\vect{m}_P) \defeq \half{1} \|\vect{m}_A(1) - \vect{m}_P(1)\|^2_{L^2(\Omega)^3}
\end{align*}
that measure the discrepancy between the simulated tumor in atlas space $c_A(1)$ (solution of the tumor forward problem~\eqref{eq:glob:tumor:fwd} with initial condition~\eqref{eq:glob:tumor:initcond-1} parametrized by $\vect{p}$) and the transported probability map $c_P(1)$ of cancerous cells for the patient data (solution of the registration forward problem~\eqref{eq:glob:reg:state:c} with initial condition~\eqref{eq:glob:reg:state-c-initcond-2}) and
the discrepancy between the healthy tissue probability maps in atlas space with tumor (calculated
according to~\eqref{eq:glob:coupling-eq:geom}) and the transported probability maps of healthy tissue
for the patient data (solution of the registration forward problem~\eqref{eq:glob:reg:state:geom}
with initial condition~\eqref{eq:glob:reg:state-initcond-1});

\item three regularization operators balanced against the discrepancy measures $\D{D}_c$ and $\D{D}_{\vect{m}}$ based on regularization weights $\beta_j > 0$, $j\in\{v,w,p\}$ involving
\begin{enumerate}[(a)] \setlength\itemsep{-.5pt} 
\item a regularization operator $\D{S}_{\vect{p}}$ for $\Phi\vect{p}$ (an $L^2$-norm; defined in \eqref{e:tumor_solo} in \secref{sec:tumor}),
\item a regularization operator $\D{S}_{\vect{v}}$ for $\vect{v}$ (an $H^1$ Sobolev norm; defined in \eqref{e:reg_regularization_v} in \secref{sec:registration}), and
\item a regularization operator $\D{S}_w$ for $w$ (an $H^1$ Sobolev norm; defined in \eqref{e:reg_regularization_v} in \secref{sec:registration}).
\end{enumerate}
\end{enumerate}

The remaining parameters are defined as follows: $\mat{k}:\Omega_B \rightarrow\ns{R}^{3\times 3}$ is a diffusion tensor field parametri\-zed by scalar weights that specify the diffusivity in WM and GM (see \citep{Gholami:2016a} for details) and $\rho>0$ is the growth parameter for the logistic growth function $f(c_A) \defeq \rho c_A(1-c_A)$ in \eqref{eq:glob:tumor:fwd}. Both $\mat{k}$ and $\rho$ have different values in gray matter and white matter and vanish everywhere else. Although our implementation is general, in all our experiments $\mat{k}$ is just an isotropic diagonal tensor (a scalar function times the identity tensor). The image registration velocity $\vect{v}$ is used to transport both the brain geometry $\vect{m}_T$ and the tumor concentration $c_T$ from patient to atlas space. \eqref{eq:glob:coupling-eq:geom} defines the model for the effect of tumor growth on the brain geometry $\vect{m}_A(0)$ in the atlas space. 
We solve this coupled inverse problem with gradient-based optimization. We use the method of Lagrange multipliers to transform the constrained  problem~\eqref{eq:global_opt} into an unconstrained one. The Lagrangian of~\eqref{eq:global_opt} reads
\begin{equation}
\begin{split}
&\mathcal{L}_G\left( c_A, c_P, \vect{m}_P, \vect{m}_A(1), \alpha, \lambda_c, \vect{\lambda}_{\vect{m}}, \nu, \vect{\xi}, \vect{p}, \vect{v}, w \right)
  =  \fs{J}(\vect{p},\vect{v},w)  \\[1mm] 
  &+ \left\langle \alpha(0), (c_A(0) - \Phi \vect{p})  \right\rangle_{L^2(\Omega)}
   + \int_0^1\left\langle \alpha, \partial_t c_A - \nabla \cdot \left(\mat{k} \nabla c_A\right) - f(c_A) \right\rangle_{L^2(\Omega)} \dt\\[1mm]
  &+ \int_0^1 \left\langle \vect{\lambda}_{\vect{m}}, \partial_t \vect{m}_{P} + \nabla \vect{m}_{P}\ \vect{v} \right\rangle_{L^2(\Omega)^3} \dt
   + \left\langle \vect{\lambda}_{\vect{m}}(0), \vect{m}_P(0)-\vect{m}_T \right\rangle_{L^2(\Omega)^3} \\[1mm]
  &+ \int_0^1 \left\langle \lambda_c, \partial_t c_P + \nabla c_P \cdot \vect{v} \right\rangle_{L^2(\Omega)} \dt
   + \left\langle \lambda_c(0), c_P(0)-c_T \right\rangle_{L^2(\Omega)}\\[1mm]
  &+ \int_0^1 \left\langle \nu, \idiv \vect{v} - w\right\rangle_{L^2(\Omega)} \dt
   + \langle \vect{\xi}, \vect{m}_A(1) - (1 - c_A(1))\vect{m}_A(0) \rangle_{L^2(\Omega)^3},
\end{split}
\label{e:lagrangian}
\end{equation}
with the state fields $c_A, c_P, \vect{m}_P,$ and $\vect{m}_A(1)$, the adjoint fields $\alpha, \lambda_c, \vect{\lambda}_{\vect{m}}, \nu,$ and $\vect{\xi}$, and the inversion fields $\vect{p}, \vect{v},$ and $w$. 
The strong form of the \ipoint{first-order optimality conditions} for Equation~\ref{e:lagrangian} is given by the following equations:

\begin{subequations}\label{e:first_order_opt_global}
\begin{align}
& \mbox{\textit{tumor forward} } & \delta_{\alpha} \mathcal{L}_G &= 0:    &\; \mbox{ given by~\eqref{eq:glob:tumor:fwd} and \eqref{eq:glob:tumor:initcond-1}} & \\
& \mbox{\textit{tumor adjoint} } & 
\delta_{c_A} \mathcal{L}_G &= 0:    & -\partial_t \alpha - \nabla \cdot \mat{k}\nabla  \alpha -\alpha \rho + 2\alpha \rho c_A &= 0
& &\mbox{in}~\bar{U}, \label{adj_T}\\
&& \delta_{c_A(1)} \mathcal{L}_T &= 0: & c_P(1) - c_A(1) - \vect{\xi}\vect{m}_A(0) - \alpha(1) &= 0  & &\mbox{in}~\Omega.
\label{adj_T_init} \\
& \mbox{\textit{registration forward} } & \delta_{\vect{\lambda}} \mathcal{L}_G &= 0:    &\;
\mbox{ given by~\eqref{eq:glob:reg:state:geom} -- \eqref{eq:glob:reg:divfree}} & \\
& \mbox{\textit{registration adjoint} } &
\delta_{\vect{m}_P} \mathcal{L}_G &= 0:    & -\partial_t \vect{\lambda}_{\vect{m}}
-\idiv (\vect{\lambda}_{\vect{m}} \otimes \vect{v})  &= \vect{0}
& &\mbox{in}~\bar{U}, \label{adj_R_m} \\
&&\delta_{\vect{m}_P(1)} \mathcal{L}_G &= 0:    & \vect{m}_{A}(1) - \vect{m}_{P}(1) - \vect{\lambda}_{\vect{m}}(1) &= \vect{0}
& &\mbox{in}~\Omega, \label{adj_R_m_init}\\
&&\delta_{c_P} \mathcal{L}_G &= 0:    & -\partial_t \lambda_{c} - \idiv (\lambda_{c} \vect{v})
&= 0 & &\mbox{in}~\bar{U}, \label{adj_R_c} \\
&&\delta_{c_P(1)} \mathcal{L}_G &= 0:    & c_A(1) - c_P(1) - \lambda_c(1) &= 0 & &\mbox{in}~\Omega,
\label{adj_R_c_init} \\
%
& \mbox{\textit{adjoint coupling} } &
\delta_{\vect{m}_A(1)} \mathcal{L}_G &= 0:    & \vect{m}_{P}(1) - \vect{m}_{A}(1) - \vect{\xi} &= \vect{0}
& &\mbox{in}~\Omega. \label{e:xi} \\
& \mbox{\textit{tumor inversion} } &
\delta_{\vect{p}} \mathcal{L}_G &= 0:    & \igrad_{\vect{p}} \mathcal{S}_{\vect{p}}(\vect{p})
- \Phi^T \alpha(0) &= \vect{0} & &\mbox{in}~\Omega.
\label{opt_grad_T} \\
& \mbox{\textit{registration inversion} } &
\delta_{\vect{v}} \mathcal{L}_G &= 0:    & \igrad_{\vect{v}} \mathcal{S}_{\vect{v}}(\vect{v})
+ \D{K}[\int_0^1 (\igrad \vect{m}_P)^T \vect{\lambda}_{\vect{m}}
+ \nabla c_P \lambda_{c}\, \dt] &= \vect{0}. & &
\label{opt_grad_R}
\end{align}
\end{subequations}

\noindent Note, that $(\igrad \vect{m})^T \vect{\lambda} = \sum_i^3 \lambda_i(\partial_1 m_i, \partial_2m_i, \partial_3m_i)^T$ and that we haven't given an equation for $w$. The operator $\D{K}$ in~\eqref{opt_grad_R} is a pseudo-differential operator that is derived by eliminating  $w$ in~\eqref{eq:glob:reg:divfree}. We discuss it further in \secref{sec:registration}.
In summary, the first-order optimality conditions
~\eqref{e:first_order_opt_global} comprise a system of non-linear partial differential equations, which is quite formidable since it has 11 fields 
in addition to the tumor initial condition parameters $\vect{p}$. Given $\vect{m}_T$, $c_T$, $\vect{m}_A(0)$, and $\mat{k}$ and $\rho$ (in that atlas space at $t=0$), we need to solve the first-order optimality PDEs for the state, adjoint, and inversion variables.\\
We use an elimination method (or also called a \emph{``reduced-space method''}~\citep{Nocedal:2006a,Borzi:2012a}) to solve these equations.  That is, we assume that the \emph{state} and \emph{adjoint equations} are fulfilled exactly and require vanishing variations of the Lagrangian with respect to the inversion variables $\vect{v}$, $w$, and $\vect{p}$ for an admissible solution of our problem. In our algorithm, we  iterate only on  $\vect{p}$ and $\vect{v}$ as we eliminate $w$. 
Given $\vect{p}$ and $\vect{v}$, we wish to compute the gradient of $\mathcal{J}$,  i.e., $\vect{g}_p$ and
$\vect{g}_v$ used to update $\vect{p}$ and $\vect{v}$, respectively. The \ipoint{ gradient computation} involves
the following steps:
\begin{enumerate} \setlength\itemsep{-.5pt}
  \item Solve the forward tumor growth and registration equations~\eqref{eq:glob:tumor:initcond-1}--\eqref{eq:glob:coupling-eq:geom} for  the state variables $c_A(t)$, $c_P(t)$, $\vect{m}_P(t)$, $\vect{m}_A(1)$.
  \item Compute the coupling adjoint variable $\xi$ from equation~\eqref{e:xi}.
  \item Solve the adjoint tumor equations~\eqref{adj_T} and \eqref{adj_T_init} for $\alpha(t)$.
  \item Solve the adjoint registration equations~\eqref{adj_R_m}--\eqref{adj_R_c_init} for $\vect{\lambda}_{\vect{m}}(t)$ and $\lambda_c(t)$.
  \item 
  Evaluate the gradients using the inversion equations~\eqref{opt_grad_T}  and \eqref{opt_grad_R} at $\vect{v}$ and $\vect{p}$:
  \begin{equation} \label{e:reduced_gradient}
   \vect{g}_{\vect{v}} = \beta_{\vect{v}}\igrad_{\vect{v}} \mathcal{S}_{\vect{v}}(\vect{v}) + \D{K}[\int_0^1 (\igrad \vect{m}_P)^T \vect{\lambda}_{\vect{m}} + \nabla c_P\ \lambda_{c}\, \dt],
   ~~~~~~~ \vect{g}_{\vect{p}} = \beta_{\vect{p}}\igrad_{\vect{p}} \mathcal{S}_{\vect{p}}(\vect{p}) - \Phi^T \alpha(0).
 \end{equation}
\end{enumerate}
At a stationary point $\vect{g} = (\vect{g}_{\vect{v}}$, $\vect{g}_{\vect{p}})^T$ vanishes. In~\secref{sec:tumor} and \secref{sec:registration}, we present more details for the tumor forward problem~\eqref{eq:glob:tumor:fwd},~\eqref{eq:glob:tumor:initcond-1} and the registration forward problems~\eqref{eq:glob:reg:state:geom},~\eqref{eq:glob:reg:state:c},~\eqref{eq:glob:reg:state-initcond-1},~\eqref{eq:glob:reg:state-c-initcond-2}, respectively. In addition, we formulate separate inverse problems for tumor growth and image registration that are used as high-level components in our coupling algorithm presented in~\ref{sec:picard}. The latter is based on a Picard iteration using tumor and registration as modular components instead of a gradient descent or Newton method with line search for the coupled
problem~\eqref{eq:global_opt}.

\subsection{The Tumor Model}
\label{sec:tumor}

  \ipoint{Forward Tumor Problem.} We model the tumor growth based on the population density $c_A(\vect{x},t)$. Two main phenomena are included: \ipoint{proliferation} of cancerous cells and the \ipoint{net migration} of cancerous cells into surrounding healthy tissue~\citep{Murray:1989a}\footnote{Displacement of healthy tissue as a result of tumor growth (mass effect) is currently neglected in our model.}. The proliferation model is a logistic growth function $f(c_A) \defeq \rho\, c_A(1-c_A)$ with reaction coefficient $\rho(\vect{x}) \defeq \rho_f\, \rho_0(\vect{x})$. $\rho_f$ denotes the scaling of the spatially variable characteristic growth rate parameters defined by white and gray matter, i.e., $\rho_0(\vect{x}) \defeq \rho_w\, m_{WM}(\vect{x}) + \rho_g\, m_{GM}(\vect{x})$. The migration model is based on an inhomogeneous (potentially anisotropic) diffusion process with diffusion coefficient $\mat{k}(\vect{x}) \defeq k_f\, k_0(\vect{x})\mat{I} + k_a\, \mat{T}(\vect{x})$. Here, $k_f$ and $k_a$ are the scaling factors for the isotropic and anisotropic parts of the diffusion tensor, $\mat{T}(\vect{x})$ is a weighted diffusion tensor modelling anisotropy. In our test cases in \secref{s:results}, we always use isotropic diffusion, i.e., $k_a=0$. The isotropic part is $k_0(\vect{x}) \defeq k_{w}m_{WM}(\vect{x}) + k_gm_{GM}(\vect{x})$. This yields a \emph{non-linear} parabolic PDE with non-constant
coefficients for the tumor concentration $c$ given by
\begin{subequations}
\label{e:tumor_solo}
\begin{align}
\p_t\, c_A - \idiv \mat{k} \igrad c_A - f(c_A)  &=  0 && \mbox{in}~\Omega_B \times (0,1], \label{eq:solo:tumor:fwd} \\
\partial_n c_A & =  0  && \mbox{on}~\partial \Omega_B \times (0,1], \label{eq:tumor_boundary_solo} \\
c_A(0) &= \Phi \vect{p} && \mbox{in}~\Omega_B. \label{eq:solo:tumor:initcond-1}
\end{align}
\end{subequations}

\noindent We use a parametrization $\Phi\vect{p}$ for the tumor initial condition $c_A(0)$ as originally in \eqref{eq:solo:tumor:initcond-1} in an $n_{\vect{p}}$-dimensional space spanned by a Gaussian basis functions, i.e., $\vect{p} \in\ns{R}^{n_{\vect{p}}}$,
$\Phi \vect{p} := \sum_{i=1}^{n_{\vect{p}}} \Phi_i p_i$ with Gaussian basis functions $\Phi_i: \Omega_B \rightarrow \ns{R}$.
We set the Gaussians in CSF to zero to prevent a spurious diffusion of cancerous cells into the area associated with CSF.

%

For notational convenience, we represent the process of solving~\eqref{e:tumor_solo} with the operator
\begin{equation}
\Tfor(\vect{p}) \defeq c_A(1)
\label{eq:tumor_for_operator}
\end{equation}

\noindent mapping the parametrization $\vect{p}$ of the initial conditions in~\eqref{eq:solo:tumor:initcond-1} to the tumor density $c_A(1)$ at time $t=1$. This simple model is by no means predictive on its own, but is the de-facto standard approach when it comes to modeling tumor progression as seen in medical imaging~\citep{Swanson:2000a,Swanson:2008a,Clatz:2005a,Harpold:2007a,Konukoglu:2010a,Hogea:2008a}. Some results available in the literature have suggested that this model can offer (to some extent) predictive capabilities when integrated with medical imaging information~\citep{Swanson:2000a,Tomer:2014a}. Its usefulness is in segmentation and registration algorithms that use normal atlas information and in producing features (e.g., tumor parameters) to augment image-based features for tumor staging and prognosis. Note that unlike~\secref{s:formulation}, we have stated the tumor problem in $\Omega_B$ with Neumann boundary conditions~\eqref{eq:tumor_boundary_solo}, which is  the actual biophysical problem. In our numerical implementation, however, we extend $\mat{k}$ by a small parameter, set $\rho = 0$ in $\Omega \setminus \Omega_B$,  discretize in $\Omega$ using periodic boundary conditions and use a penalty approach to approximate the boundary conditions at $\partial \Omega_B$. One can show that this `'fictitious domain method`' approximates~\eqref{e:tumor_solo} and as we refine the discretization it converges to the correct solution (compare~\citep{Hogea:2008a,Gholami:2016a}). \\

\ipoint{Inverse Tumor Problem.} In the inverse tumor problem, we seek an initial condition for the forward tumor problem that recovers a given tumor concentration $c_P(1)$ at time $t=1$ as good as possible, i.e., we solve the minimization problem
\begin{subequations}
\label{e:tumor_inv_solo}
\begin{eqnarray}
 \min_{\vect{p}} \fs{J}_T(\vect{p}) & \mbox{ with } &
\fs{J}_T(\vect{p}) \defeq\mathcal{D}_c(c_A,c_P)
+ \half{\beta_{\vect{p}}}\|\Phi \vect{p}\|^2_2
\label{e:varopt:objective_T}
\end{eqnarray}

\noindent subject to the constraints \eqref{eq:solo:tumor:fwd}, \eqref{eq:tumor_boundary_solo}, and \eqref{eq:solo:tumor:initcond-1}. This defines the inverse tumor operator
\begin{equation}
\Tinv(c_P(1)) \defeq \vect{p} = \argmin_{\vect{q}} \fs{J}_T(\vect{q}).
\label{e:inv_tumor}
\end{equation}
\end{subequations}

\noindent Note that the inverse tumor problem does --- in contrast to~\eqref{e:global_objective} --- only minimize the mismatch between the observed and the predicted tumor cell population density; the healthy
tissue mismatch is not used as the tumor solver only 'knows' the altas.
In addition,~\eqref{e:tumor_inv_solo} defines a concrete instance for the tumor regularization operator $\mathcal{S}_{\vect{p}}$ in~\eqref{eq:global_opt}. Neglecting 
the boundary conditions, the Lagrangian of~\eqref{e:tumor_inv_solo} reads
\begin{flalign*}
\mathcal{L}_T\left(c_A, \vect{p}, \alpha \right) = \fs{J}_T(\vect{p})
+\left\langle \alpha(0),c_A(0) - \Phi \vect{p}\right\rangle_{L^2(\Omega_B)}
 +\int_0^1\left\langle \alpha, \partial_t c_A
- \nabla \cdot \mat{k} \nabla c_A
- \rho c_A\left(1-c_A\right) \right\rangle_{L^2(\Omega_B)} \dt
\end{flalign*}

\noindent We require vanishing variations of the Lagrangian $\mathcal{L}_T$ with respect to the inversion variable $\vect{p}$ for an admissible solution to~\eqref{e:tumor_inv_solo}. This yields the first order
optimality condition of the tumor problem:
\begin{subequations}
\begin{align}
& \mbox{\textit{tumor forward} } & \delta_{\alpha} \mathcal{L}_T &= 0:    &\; \mbox{ given by~\eqref{e:tumor_solo}} & \\
& \mbox{\textit{tumor adjoint} } & \delta_{c_A} \mathcal{L}_T &= 0: & -\partial_t \alpha - \nabla \cdot \mat{k}\nabla  \alpha
-\alpha \rho + 2\alpha \rho c_A &= 0 & &\mbox{in}~\Omega_B \times [0,1)\label{adj_T_solo} \\
&& \delta_{c_A(1)} \mathcal{L}_T &= 0: & c_P(1) - c_A(1) - \alpha(1) &= 0 & &\mbox{in}~\Omega_B \label{adj_T_init_solo}, \\
& \mbox{\textit{tumor inversion} } & \delta_{\vect{p}} \mathcal{L}_T &= 0: & \beta_{\vect{p}}\Phi^\mathsf{T}\Phi \vect{p} - \Phi^{\mathsf{T}} \alpha(0) &= \vect{0} & &\mbox{in}~\Omega_B,\label{opt_grad_T_solo}.
\end{align}
\end{subequations}
\noindent 
Similarly to the forward problem, the adjoint equation are discretized by extending $\mat{k}$ and setting $\rho=0$ in $\Omega \setminus \Omega_B$ and using periodic boundary conditions in $\p\Omega$.

\subsection{The Registration Problem}
\label{sec:registration}

The input for our formulation are not image intensities~\citep{Mang:2015a,Mang:2016a,Mang:2017a,Mang:2017b} but the probability maps for tissue classes (see \eqref{eq:vec_m}; WM, GM, CSF, and tumor). The formulation we propose here is suited for general problems that involve the registration of vector fields. The \emph{template image} (image to be registered) $\vect{m}_T:\bar{\Omega}\rightarrow\ns{R}^3$ is given by the probability maps of the patient's healthy anatomy in all areas except the part hidden by the tumor. We treat the probability map for the tumor, $c_T(\vect{x})$, as an individual entity to make the coupled formulation in~\eqref{eq:global_opt} more accessible.
We register the three probability maps for healthy tissue and the tumor concentrations.\\

\ipoint{Advective Image Transformation (Forward Problem).} Given some template image $\vect{m}_T(\vect{x})$, some tumor concentration $c_T(\vect{x})$, and a \emph{stationary velocity field} $\vect{v}(\vect{x})$, the forward problem describes the advective transformation of $\vect{m}_T$ and $c_T$ in a pseudo-time interval $[0,1]$:
\begin{subequations}
\label{eq:reg_solo}
\begin{align}
\partial_t \vect{m}_P + \nabla \vect{m}_P\ \vect{v}
&= \vect{0} && \mbox{in}~\Omega \times (0,1],
\label{eq:solo:reg:state:geom}\\
\vect{m}_P(0) & = \vect{m}_T && \mbox{in}~\Omega,
\label{eq:solo:reg:state-initcond-1} \\
\p_t c_P + \igrad c_P \cdot\vect{v} &= 0
&& \mbox{in}~\Omega \times (0,1],
\label{eq:solo:reg:state:c} \\
c_P(0) &= c_T && \mbox{in}~\Omega
\label{eq:solo:reg:state-c-initcond-2}
\shortintertext{with periodic boundary conditions. We augment this formulation by an additional constraint on the divergence of $\vect{v}$ to better control the Jacobian of the computed deformation map (see \citep{Mang:2016a} for additional details):}
\idiv \vect{v} &= w && \mbox{in}~\Omega,
\label{eq:solo:reg:divfree}
\end{align}
\end{subequations}
Solving~\eqref{eq:reg_solo} defines the implicitly given operator
\begin{equation}
\Rfor \left( \vect{v}, \vect{m}_T, c_T \right)
\defeq \left( \vect{m}_P(1), c_P(1) \right)
\label{eq:registration_for_operator}
\end{equation}

\noindent that maps the template images $\vect{m}_T$ and $c_T$ to images $\vect{m}_P$ and $c_P$ defined at pseudo-time $t=1$.
For simplicity, we will later slightly abuse our notation and use the operator $\Rfor$ also for the advection of only $\vect{m}$, $c$ or one of the components of $\vect{m}$, i.e., $m_{WM}$, $m_{GM}$, or $m_{CSF}$.\\

\ipoint{Image Registration (Inverse Problem).} In the inverse problem, i.e., the actual image registration, we look for a velocity $\vect{v}$ that advects the given template (patient) images to images that are as close as possible to the corresponding images in the atlas brain (denoted with a subscript $A$). That is, we solve the following minimization problem:
\begin{subequations}
\label{e:reg_inv_solo}
\begin{eqnarray}
\min_{\vect{v}} \fs{J}_R(\vect{v})  & ~\text{with}~ & 
\fs{J}_R(\vect{v}) \defeq
        \mathcal{D}_{\vect{m}}(\vect{m}_{A},\vect{m}_{P}) + \mathcal{D}_c(c_{A},c_{P})
      + \beta_{\vect{v}}\D{S}_{\vect{v}}(\vect{v}) + \beta_w \D{S}_w(w) 
\label{e:varopt:objective_R}
\end{eqnarray}

\noindent subject to \eqref{eq:reg_solo}.
The regularization in \eqref{e:varopt:objective_R} is a \emph{smoother} for $\vect{v}$ and $w$ and given by an $H^1$-seminorm for $\vect{v}$, and an $H^1$-norm for $w$ (i.e., for $\idiv\vect{v}$ according to~\eqref{eq:solo:reg:divfree}), respectively:
\begin{align}
\D{S}_{\vect{v}}(\vect{v}) = \half{1}\int_{\Omega}
\sum_{i=1}^3| \nabla v^i(\vect{x})|^2\d\Omega, && \D{S}_w(w) = \half{1}\int_{\Omega}
|\nabla w(\vect{x})|^2 + |w|^2\d\Omega. \label{e:reg_regularization_v}
\end{align}


\noindent Overall, this defines the (inverse) image registration operator
\begin{equation}
\Rinv \left( \vect{m}_A(1), c_A(1), \vect{m}_T, c_T  \right) \defeq \vect{v} = \argmin_{\vect{u}} \fs{J}_R(\vect{u}). \label{e:reg_op}
\end{equation}
\end{subequations}

\noindent The \emph{Lagrangian} for our image registration problem reads
\begin{flalign*}
\mathcal{L}_R\left(c_P, \vect{m}_P, \vect{v}, \vect{\lambda}_{\vect{m}}, \lambda_c\right) &= \fs{J}_R(\vect{v}) \\
&+ \int_0^1 \left\langle\vect{\lambda}_{\vect{m}}, \partial_t \vect{m}_{P} + \igrad \vect{m}_P\ \vect{v} \right\rangle_{L^2(\Omega)^3} \dt
 + \left\langle \vect{\lambda}_{\vect{m}}(0), \vect{m}_P(0)-\vect{m}_T \right\rangle_{L^2(\Omega)^3}\\
&+ \int_0^1 \left\langle \lambda_c, \partial_t c_P + \igrad c_P \cdot \vect{v} \right\rangle_{L^2(\Omega)} \dt
 + \left\langle \lambda_c(0), c_P(0)-c_T\right\rangle_{L^2(\Omega)} \\
&+ \int_0^1 \left\langle \nu, \idiv\vect{v} - w \right\rangle_{L^2(\Omega)}\dt.
\end{flalign*}
Taking variations, we obtain the first-order optimality conditions~\citep{Mang:2016a} for the registration problem:
\begin{subequations}
\begin{align}
& \mbox{\textit{registration forward} } & \delta_{\vect{\lambda}_{\vect{m}}, \lambda_c} \mathcal{L}_R &= 0:    &\; \mbox{ given by~\eqref{eq:reg_solo}} & \\
& \mbox{\textit{registration adjoint} } & \delta_{\vect{m}_P} \mathcal{L}_R &= 0: &\; -\partial_t \vect{\lambda}_{\vect{m}} - \idiv (\vect{\lambda}_m \otimes \vect{v}) &= \vect{0} & &\mbox{in}~U,\label{adj_R_m_solo} \\
&& \delta_{\vect{m}_P(1)} \mathcal{L}_R &= 0: &\; \vect{m}_A(1) - \vect{m}_P(1) - \vect{\lambda}_{\vect{m}}(1)  &= \vect{0}
& &\mbox{in}~\Omega,\label{adj_R_m_init_solo} \\
&& \delta_{c_P} \mathcal{L}_R &= 0: &\;
-\partial_t \lambda_{c} - \idiv (\lambda_{c} \vect{v}) &= 0 & &\mbox{in}~U, \label{adj_R_c_solo} \\
&& \delta_{c_P(1)} \mathcal{L}_R &= 0: &\;  c_A(1) - c_P(1) - \lambda_c(1)  &= 0 & &\mbox{in}~\Omega \label{adj_R_c_init_solo}, \\
& \mbox{\textit{registration inversion}} & \delta_{\vect{v}} \mathcal{L}_R &= 0: & \beta_{\vect{v}} \ilap \vect{v} +\D{K}[\int_0^1(\igrad\vect{m}_P)^T\vect{\lambda}_{\vect{m}} + \igrad c_P\ \lambda_c \dt] &= \vect{0}
& &\mbox{in~} \Omega.  \label{opt_grad_R_solo}
\end{align}
\end{subequations}
%
As mentioned before, the operator $\D{K}$ is a pseudo-differential operator and comes from an elimination step for the inversion variable $w$ in~\eqref{eq:glob:reg:divfree}. For the case of exact incompressibility ($w=0$), it is the Leray projection $\D{K}(\vect{u}) \defeq \vect{u} + \igrad \ilap^{-1}\idiv \vect{u}$; 
for a non-zero $w$, the projection operator becomes slightly more involved; we refer to~\citep{Mang:2015a,Mang:2016a} for additional details.

%% file: 04-methods.tex
The main contribution of this paper is a Picard iteration scheme for~\eqref{eq:global_opt}: Instead of
solving~\eqref{eq:global_opt} using a gradient descent
or Newton scheme, we use a modular approach, in which we combine tumor growth inversion and diffeomorphic registration models in an (interleaved) Picard iteration scheme. This scheme iteratively improves both the tumor initial condition's parametrization $\vect{p}$ and the registration velocity $\vect{v}$. This allows us to establish a coupling of both components in an easy, stable, modular, and efficient way; the submodules can be exchanged as required. For instance, our simple tumor solver can be replaced by more sophisticated approaches. Our results presented in~\secref{s:results} demonstrate that the Picard iteration is a powerful approach 
that reduces the gradient as given in~\eqref{e:reduced_gradient}.

The algorithms for the individual subblocks have been published in~\citep{Mang:2015a,Mang:2016a,Gholami:2017a,Gholami:2016a,Gholami:2016b,Mang:2016c,Mang:2017b}. The main ingredients can be described as follows:
\begin{inparaenum}[(i)]
  \item All PDEs are spatially discretized in $\Omega = [0,2\pi]^3$.
  \item All spatial derivatives ($\igrad, \idiv$, and higher derivatives) are computed using 3D Fourier transforms.
  \item Although in the formulation we present the derivatives in the strong form, in our implementation we use a discretize-then-optimize approach for the tumor equations and an optimize-then-discretize approach for the registration.
  \item The solution of pure advection equations is done using a semi-Lagrangian time-stepping scheme to avoid stability issues and small time-steps.
  \item We use Krylov and matrix-free Newton methods for linear and nonlinear solvers.
  \item We use a Picard iteration scheme for the coupled optimization problem, without line search\footnote{The Newton-type iterations for the sub-problems registration and tumor inversion are enhanced with a globalizing line search method.}.
\end{inparaenum}

In the following subsections, we give more details on the proposed Picard iteration scheme, and then give a short overview of the numerical methods used to solve the tumor and image registration forward and inverse problems.

\subsection{The Picard Iteration Algorithm}
\label{sec:picard}

To initialize our Picard iteration, we start with an initial guess for $\vect{p}\in\ns{R}^{n_{\vect{p}}}$ and $\vect{v}$ (both zero), compute the center of mass of the given patient tumor $c_T$  and place the Gaussian basis functions for the tumor initial
condition parametrization as a regular grid around the center of mass. In our current algorithm, the grid of Gaussian basis
functions is fixed throughout the Picard iterations, i.e., we do not re-compute the center of mass of the tumor or advect
the Gaussian basis functions with the velocity $\vect{v}$ (this is ongoing work). The algorithm per Picard iteration
executes the following steps:
\begin{enumerate} \setlength\itemsep{1pt}
\item Given some $\vect{p}$, grow a tumor in the atlas space. This seeds the atlas brain with a tumor; update the probability
maps at $t=1$ in the atlas space using~\eqref{eq:glob:coupling-eq:geom}. To define the Picard iteration, it is useful to introduce the operator $\geom$
\begin{equation}
\geom (c_A(1)) := \left( \vect{m}_A(0)(1-c_A(1)), c_A(1) \right).
\label{eq:geometry_coupling}
\end{equation}
\item Register the probability maps (for the anatomical regions and the tumor) given for the patient's image with the updated probability maps in the atlas space.
\item Use the computed velocity $\vect{v}$ to transport the patient data (probability maps for the anatomical regions and the tumor) to the reference atlas space. This gives us a new $\vect{m}_P(1)$ and $c_P(1)$.
\item Use the transported (i.e., updated) $c_P(1)$ within the tumor inversion to find a better $\vect{p}$.
\item Check for convergence. If the convergence criteria are fulfilled, stop. If not, go back to step 1 (i.e., continue iterating).
\end{enumerate}

Some important facts to note are:
\begin{inparaenum}[(i)]
\item We use the solution $\vect{v}$ from the former iteration as an initial guess for the next iteration to reduce the runtime (warm start).
\item We perform a continuation in the regularization parameter $\beta_{\vect{v}}$ (we start with a large value and successively reduce it). We explain this below.
\item We do not have a proof for the convergence of the proposed Picard scheme to a minimizer of the fully-coupled optimization problem in~\eqref{eq:global_opt}. Such a proof is beyond the scope of the present paper and remains for future work. We provide numerical evidence that shows that our scheme reduces the gradient of the fully-coupled problem. We discuss this below.
\end{inparaenum}

Using the operators defined in~\secref{s:formulation}, the described algorithm corresponds to the Picard iteration
\begin{equation*}
\vect{p}^{k+1}\!=\!\Tinv\circ\Rfor\!\left(\Rinv\!\left(\geom\circ\Tfor(\vect{p}^k),\vect{m}_T,c_T\right)\!,c_T\right)
\end{equation*}
\noindent for the parametrization $\vect{p}$ of the tumor initial conditions $c_A(0)$ in the atlas brain with probability maps $\vect{m}_T$ and $c_T$ for an individual patient as input (we use $\Rfor$ here to only advect the patient's tumor $c_T$).

We have implemented the gradient of the coupled problem in \secref{s:formulation} in order to verify that our Picard iteration actually reduces the gradient of the global optimization problem. An implementation of a scheme that iterates simultaneously on both control variables (i.e., solves the global problem), requires more work and will be addressed in a follow-up paper.  We show experimentally in various settings that our algorithm is effective and generates registration results that are in excellent agreement with the patient data. Next, we give additional details on the numerical methods used in SIBIA.\\

\ipoint{Parameter continuation.}
We use a combination of Tikhonov-type regularization and parameter continuation
(additional details can be found in~\citep{Mang:2015a}) not only to stabilize the registration
problem, but also to identify an adequate regularization parameter $\beta_{\vect{v}}$ for the $H^1$-Sobolev norm for $\vect{v}$
in every Picard iteration as follows. We specify a lower admissible bound on the determinant of the deformation gradient and
start the Picard iteration with $\beta_{\vect{v}}^0 = 1$. In each iteration $k$, the candidate $\beta_{\vect{v}}^k$ is reduced by
one order of magnitude until the specified lower bound for the determinant of the deformation gradient is breached for a
candidate $\beta_{\vect{v}}^k$. The registration solver disregards the associated solution, sets the candidate value for
$\beta_{\vect{v}}^k$ to $\beta_{\vect{v}}^k \leftarrow \beta_{\vect{v}}^{k-1} - (\beta_{\vect{v}}^{k-1} - \beta_{\vect{v}}^k)/2$ and
restarts the inversion with the velocity of the former Picard iteration as initial guess. This process is repeated until the lower bound for the determinant of the deformation gradient is no longer violated. If no violation of the determinant of the deformation gradient was detected during the registration solve, we finalize the current Picard iteration and proceed with the next iteration. 
\\

\ipoint{Stopping conditions.} We finalize our Picard iteration either if $\beta_{\vect{v}}$ reaches the prescribed minimum\footnote{This lower bound is chosen based on numerical experience and is a safeguard against numerical instabilities that can occur if $\vect{v}$ becomes highly irregular (see~\citep{Mang:2017b} for a discussion).} or if the requested $\beta_{\vect{v}}$ results in a violation of a user defined lower bound on the determinant of the deformation gradient (see above for further explanation). In both cases, we execute two additional iterations with the final value for $\beta_{\vect{v}}$.

\subsection{Numerics for the tumor inversion and registration sub-blocks}

The optimality conditions of both~\eqref{e:varopt:objective_T} and~\eqref{e:varopt:objective_R} are complex, multi-component, non-linear operators for the state, adjoint, and control fields.  We employ an inexact, globalized, preconditioned Gauss--Newton--Krylov method for both problems.
In reduced space methods, the second order optimality conditions have a very similar structure as the first order optimality conditions; we can employ the same numerical strategies we use for the solution of the PDE operators in the first order optimality conditions (see below).
We refer to~\citep{Gholami:2016a,Mang:2015a,Mang:2017b,Mang:2017c} for details on the reduced space Hessian system and its
discretization. \\


\ipoint{Numerical solution and discretization of the PDE operators.} To discretize forward and adjoint tumor and registration
problems in space and time, we use regular grids consisting of $N_0\times N_1\times N_2$, $N_i\in\ns{N}$ grid points. For all spatial differential operators, we use a spectral projection scheme as described in~\citep{Gholami:2017a,Mang:2016c}. The mapping between the spatial and spectral coefficients is done using FFTs (the implementation of the parallel FFT library is presented in~\citep{Gholami:2016b}). Corresponding to the spectral collocation scheme, we assume that the functions in our formulation (including images) are periodic and continuously differentiable and apply filtering operations and periodically extensions of the discrete data to meet these requirements.

%
To solve the forward and adjoint tumor problem, we use an unconditionally stable, second-order Strang-splitting method, where we split the right-hand side of~\eqref{eq:solo:tumor:fwd} and \eqref{adj_T_solo} into diffusion and reaction (see~\citep{Hogea:2008a,Gholami:2016a} for details). The diffusion sub-steps are solved using an implicit Crank--Nicolson method. The solver is a preconditioned conjugate gradient method  with a fixed tolerance of $\num{1E-6}$. 
The reaction sub-steps are solved analytically.
We enforce a positivity constraint on the initial tumor concentration before we apply the forward operator to make sure that $c(\vect{x},t)$ is in $[0,1]$ for all $\vect{x}\in\Omega$ and $t\in[0,1]$. This thresholding operation is necessary since the parametrization allows for negative concentrations for the initial condition $c_A(0)$ (the coefficient vector $\vect{p}$ can have negative entries) and the logistic growth function would amplify these negative values in time.

In the image registration module, we solve the hyperbolic transport equations in~\eqref{eq:reg_solo} based on a semi-Lagrangian scheme~\citep{Mang:2016c,Mang:2017b}. Semi-Lagrangian schemes are a hybrid between Lagrangian and Eulerian schemes. They are unconditionally stable and, like Lagrangian schemes (see~\citep{Mang:2017a} for an example in the context of diffeomorphic registration), require the evaluation of the space-time fields that appear in our optimality conditions at off-grid locations defined by the characteristic associated with $\vect{v}$. We compute the values of these space-time fields at the off-grid locations with a cubic Lagrange polynomial interpolation model. The time integration for computing the characteristic and for the solution of ordinary differential equations along it (if necessary) is based on a second order accurate Runge--Kutta scheme. We refer to~\citep{Mang:2016c,Mang:2017b} for a precise definition of these computations. \\

\ipoint{Newton--Krylov solver.} We initialize our solvers for both reduced gradient systems~\eqref{opt_grad_T_solo}
and~\eqref{opt_grad_R_solo} with a zero initial guess and use an inexact, globalized, preconditioned Gauss--Newton--Krylov
method. We terminate the tumor inversion if the relative change of the norm of the reduced gradient in~\eqref{opt_grad_T_solo} is below a user defined threshold $\text{opttol}_T>0$. The reference gradient is the gradient obtained for the zero initial guess for $\vect{p}$ in the first Picard iteration. For the registration problem, we use a combination of the relative change of \bipa\item the norm of the gradient in~\eqref{opt_grad_R_solo}, \item the objective in~\eqref{e:varopt:objective_R} and \item the control variable $\vect{v}$, all controlled by a single parameter $\text{opttol}_R>0$, as a stopping criterion\eipa. We also specify a maximal number of Newton iterations ($\text{maxit}_{N,T}$ and $\text{maxit}_{N,R}$) and a lower bound of $\num{1E-6}$
for the absolute norm of the gradient as a safeguard against a prohibitively high number of iterations. Details for the stopping conditions can be found in~\citep{Mang:2015a,Modersitzki:2009a}; see~\citep[305\,ff.]{Gill:1981a} for a discussion.

We invert the inner linear KKT systems using a matrix-free PCG method. We terminate the PCG method when we either reach a predefined tolerance for the relative residual norm of the PCG method or exceed the maximum number of iterations ($\text{maxit}_{K,T}$ and $\text{maxit}_{K,R}$). We perform inexact solves~\citep{Dembo:1983a, Eisenstat:1996a} with a tolerance that is proportional to the norm of the reduced gradient of our problem. The key idea here is to not invert the Hessian accurately if we are far
from an optimum (large gradient norm). Details about this approach can be found in~\citep[p.~165ff.]{Nocedal:2006a}.\\

\subsection{Parallel Algorithms and Computational Kernels}

Our parallel implementation is described in detail in~\citep{Mang:2016c,Gholami:2017a}. We use a 2D pencil decomposition for 3D FFTs~\citep{Grama:2003a,Czechowski:2012a} to distribute the data among processors. We denote the number of MPI tasks by $P = P_0P_1$. Each MPI task gets a total of $N_0/P_0 \times N_1/P_1 \times N_2$ values. We use the open source library AccFFT~\citep{Gholami:2016b}, which supports parallel FFT on CPU and GPU for both single and double precision computations. For parallel linear algebra operations, we use PETSC as well as it's optimization interface TAO \citep{Balay:2016a,Munson:2015a}. The other computational kernel besides FFTs is the cubic Lagrange polynomial interpolation model used for the semi-Lagrangian time integration of hyperbolic transport equations. We refer to~\citep{Mang:2016a,Gholami:2017a} for additional details on the parallel implementation of this interpolation kernel. We provide an algorithmic complexity analysis for the registration in~\citep{Mang:2016c} and for the FFT in \citep{Gholami:2016a}.

%% file: 05-00-results.tex

We test the performance of our method on synthetically generated  and clinical datasets.
Both are based on real MR neuroimaging data for the brain tissue probability maps.

\ipoint{ATAV.} This first class of synthetic test cases is used as a proof of concept to assess the convergence of our Picard scheme, its robustness with respect to the tumor model, and the sensitivity of tumor reconstruction quality in terms of the tumor model and its parameters. It uses analytic tumor and analytic velocity (fully synthetic; ground truth for tumor parameters and velocity is known); see \secref{s:ATAV_results} for details. It comes in three different flavours:
\bipa \item \ipoint{ATAV-REAC} with reaction only and diffusion disabled,
      \item \ipoint{ATAV-DIF}  with reaction-duffusion model, and
      \item \ipoint{ATAV-LD}   with reaction only and diffusion disabled and a small number $n_{\vect{p}}$ of Gaussian basis functions for the initial condition.\footnote{LD stands for low-dimensional. That is, we use only very few active Gaussian basis functions with a limited support in a small neighbourhood surrounding the center of mass of the patient's tumor. For the other test cases, we allow the grid for the Gaussian to cover the entire area identified as tumor in the patient data.}
\eipa 

\ipoint{RTRV.} This second second class of test cases uses real clinical images used in the study described in~\citep{Gooya:2012a}. It uses real patient data, diffusion disabled and enabled; see \secref{s:RTRV_results} for details.

\subsection{General Setup}
\label{s:setup}
\ipoint{Common Parameters.}
%
We list all model and numerical parmeters that are common to all test cases in
\tabref{tab:tc-parameters_eq} and those that are specific for the test cases in
\tabref{tab:tc-parameters}. In the following, we shortly motivate some of the more involved
choices:

$\beta_{\vect{p}}$ has been determined experimentally for a purely synthetic test case with
image resolution $N_i=128$ and $n_{\vect{p}}=125$ Gaussian basis function. For variations of $N_i$ and $n_{\vect{p}}$ in our test cases, we observed that the inversion is not sensitive with respect to $\beta_{\vect{p}}$. Accordingly, we fixed it for all test cases.  Smaller values did not further reduce the tumor mismatch in our Picard iteration scheme. To initialize the Gaussian basis, the sub-domain for the grid of basis functions is chosen a priori for the synthetic test cases and determined automatically for the real patient data (cases RTRV) to cover the actual tumor volume in an optimal way: We set the center of the grid for the parameterization to the center of mass of the tumor. The support of the domain covered by the basis functions is the $\ell^\infty$-ball that covers the entire patient's tumor in $\ns{R}^3$. We adjust the standard deviation $\sigma$ of the Gaussian functions and the distance of their centers automatically. The values are chosen such that their quotient remains constant, and equal to the value used in the synthetic cases. This allows us to ensure a similar conditioning for $\Phi$ and to use a fixed $\beta_{\vect{p}}$.


The choice for $\vect{n}_p$ depends on the appearance of the patient's tumor (shape and size). We need to be able to parameterize sufficiently complex initial conditions to recover tumors with a complex appearance. Accordingly, we invert for a varying number of parameters $n_{\vect{p}}$ on a case-by-case basis ($n_{\vect{p}}\in \{8,125,343\}$ (see \tabref{tab:tc-parameters}).
For the class of synthetic test cases (ATAV), we use an image resolution of $128^3$ to be
able to perform more experiments in limited time, for the real patient data, we use $256^3$
to ensure high accuracy.


\begin{table}[hb!]
\caption{\label{tab:tc-parameters_eq}Common parameters used in all test cases: $\text{opttol}_R$, $\text{opttol}_T$ are the
convergence tolerances for registratioin and tumor inversion (\eqref{opt_grad_R_solo} and \eqref{opt_grad_T_solo}); $\beta_{\vect{p}}$ is the regularization parameter for the tumor inversion
(see~\eqref{e:varopt:objective_T}); $d$ is
the spacing between the Gaussian basis function, $\sigma$ their standard deviation;
$\beta_{\vect{v}}^0$ and $\beta_{\vect{v}}^{\text{min}}$ are the initial and final values for
the $\beta$-continuation scheme as described in \ref{sec:picard} applied in image registration,
determined based on which values have been shown to yield best results in numerical tests for the RTRV test problem.}
\begin{small}
\begin{center}
\begin{tabular}{rrrrrr}
\toprule
$\text{opttol}_R$ & $\text{opttol}_T$ & $\beta_{\vect{p}}$ & $d$ & $\beta_{\vect{v}}^0$ &  $\beta_{\vect{v}}^{\text{min}}$ \\\midrule
 & & & \\
$\num{1E-3}$ & $\num{1E-3}$ & $\num{2.5E-4}$ & $1.5 \sigma$ & $1$ & $\num{1E-03}$ \\
\bottomrule
\end{tabular}
\end{center}
\end{small}
\end{table}

\begin{table*}
\caption{\label{tab:tc-parameters}Summary of the parameters for the generation of the synthetic test cases and the inversion. We report values for the following parameters: $N_i$, $i=1,2,3$ denotes the grid size used for the discretization of the problem; $\rho_w$ and $\rho_g$ are the characteristic reaction factors for white and gray matter, $\rho_f$ is the overall reaction scaling factor (see~\secref{sec:tumor});
$k_w$ and $k_g$ are the characteristic diffusion parameters for white and gray matter, $k_f$ the overall scaling parameter for the isotropic part of the inhomogenious diffusion coefficient for net migration of cancerous cells into surrounding tissue;
$n_{\vect{p}}$ is the number of Gaussian for the parameterization of the tumor initial condition, $\sigma$ is the standard deviation of the associated Gaussian basis functions, and $\text{maxit}_i = (\text{maxit}_{i,N}, \text{maxit}_{i,K})$ denotes the maximum number of Newton iterations and Krylov iterations (for the KKT system) for the tumor inversion ($i=T$) and registration ($i=R$), respectively,}
\begin{small}
\begin{center}
\begin{tabular}{rrrrrrr}
\toprule
                  & ATAV-REAC      & ATAV-DIF       & ATAV-LD                            & RTRV                \\\midrule
$N_i$             & 128            & 128            & 128                                & 256                 \\
$\rho_w$ & $1$ & $1$ & $1$ & $1$ \\
$\rho_g$ & $0$ & $0$ & $0$ & $0.2$ \\
$\rho_{f}$        & 15             & $\{5,10,15\}$  & $\{ 5,10,15 \}$                    & $\{ 0, 5, 10, 15\}$ \\
$k_w$ & $1$ & $1$  & $1$  & $1$ \\
$k_g$ & $0$ & $0$  & $0$  & $0.1$ \\
$k_f$             & 0              & $\{0,\num{1E-2}\}$   & $\{\num{1E-2},\num{1E-1}\}$  & $\{0,\num{1E-2}\}$  \\
$n_{\vect{p}}$    & 125            & 125            & 8                                  & $\{125, 343\}$      \\
$\sigma$          & $\nf{\pi}{10}$ & $\nf{\pi}{10}$ & $\nf{\pi}{15}$                     & auto\\
$\text{maxit}_T$  & $(50, 100)$    & $(30,60)$      & $(30,60)$                          & $(30, 30)$          \\
$\text{maxit}_R$  & $(50, 80)$     & $(50, 80)$     & $(50, 80)$                         & $(10, 20)$ \\
\bottomrule
\end{tabular}
\end{center}
\end{small}
\end{table*}

\ipoint{Data.} For the generation of the synthetic cases, we use normal brain imaging data obtained at the University of Pennsylvania. For the test case on real imaging data, we use the data available after the first iteration of GLISTR~\citep{gooya-davatzikos-biros11a,Gooya:2012a,Bakas:2015a}. The data for these results are the patient data used in the study presented in~\citep{Gooya:2012a,Bakas:2015a}. We consider six datasets from this repository (patient IDs: AAMH, AAAN, AAAC, AAMP, AAQD and AAWI). The original datasets have more labels than we use in our Picard iterations. In particular, they contain background (BG), white matter (WM), gray matter (GM), cerebellum (CB), cerebrospinal fluid (CSF), ventricles (VE), edema (ED), enhancing tumor (ENH), and necrotic tumor (NEC). We construct the labels (WM), (GM), (CSF), (BG) and (TU) by integrating \bipa\item (CB) into (BG), \item (VE) into (CSF), and \item (ENH), (NEC) and (ED) into (TU)\eipa.

For all brains, we use the labels white matter (WM), gray matter (GM), cerebrospinal fluid (CSF), and background (BG). The motivation for introducing an additional label BG is technical. We
have to ensure the partition of unity across all probability maps for each $\vect{x}$ in $\Omega$, i.e., all labels have to sum up to one. For example, for the atlas data at $t=1$ we have
\[
\forall\vect{x}\in\Omega : c_A(\vect{x},1) + \sum_{i=1}^4 m_{A,i}(\vect{x},1) \overset{!}{=} 1
\]
\noindent Note that $m_{BG}$ is not used as a label in the image registration formulation \eqref{e:varopt:objective_R}. Glial matter is integrated in BG.
\\

\ipoint{Performance Measures.} We perform the registration from the patient space to the atlas space. However, we report all performance measures in the patient space, since the patient space is the relevant space from an applications point of view (for atlas based segmentation).\footnote{If we perform cohort studies, this is different (see \citep{Gooya:2012a} for an example).} An important point to note here is that velocity based image registration offers an immediate access to the inverse of the deformation applied to an image; we can essentially solve the forward problem with a negative velocity to obtain the action of the inverse deformation map. We use the following measures to assess the performance of our approach:

\noindent The relative mismatch/residual $\elltwoB$ between patient anatomy and atlas anatomy after registration ($i=1$: GM; $i=2$: WM; $i=3$: CSF), and the relative mismatch/residual $\elltwoT$ between patient tumor and atlas tumor after registration:
\begin{align*}
\elltwoB \defeq \frac{\sum_{i=1}^3\|\Rfor(-\vect{v},m_{A,i}(1)) - m_{T,i}\|_2}{\sum_{i=1}^3 \| m_{A,i}(0) - m_{T,i} \|_2} &&
\elltwoT \defeq \frac{\|\Rfor(-\vect{v},c_A(1))-c_T\|_2}{\|c_T(1)\|_2}
\end{align*}
\noindent The Dice coefficient $\text{DICE}_{l,B}$ for the individual label maps (generated by thresholding; see below) associated with the probability maps for $l=WM$, $l=GM$, and $l=CSF$, for the patient and atlas anatomy, as well as the average Dice coefficient $\diceB$ across all labels:
\begin{align*}
\text{DICE}_{l,B}c \defeq 2\frac{|H(\Rfor(-\vect{v}, m_{A,l}(1)))\cap H(m_{T,l})|} {|H(\Rfor(-\vect{v},m_{A,l}(1)))| + |H(m_{T,l})|}, &&
\diceB = \sum_{l=1}^3\text{DICE}_{l,B} / 3
\end{align*}
where $|\,\cdot\,|$ is the cardinality of the set and $H$ is a characteristic function of a label with threshold $0.5$, i.e.,
\[
H(u(\vect{x})) \defeq
\left\{
\begin{array}{lr}
1 & \text{for } u(\vect{x}) \geq 0.5,\\
0 & \text{else},
\end{array}\right.
\]
for all $\vect{x}\in\Omega$.
We also report values for the Dice coefficient computed for the probability maps of the tumor, denoted by $\text{DICE}_T$.
Further, we monitor the relative change of the gradient for the coupled problem (see \secref{s:formulation}) for the final iteration $k$:
\[
\relG
\defeq \|\vect{g}^k\|_2 / \|\vect{g}^0\|_2.
\]
where $\vect{g}^k$ is the gradient of the coupled optimization problem~\eqref{eq:global_opt} after the $k$th Picard iteration and $\vect{g}^0$ the gradient for the initial guess.
Finally, the relative $\ell^2$-error for the computed velocity and the initial condition (under the assumption we know the true velocity $\vect{v}^\star$ and true tumor parameters $\vect{p}^\star$; synthetic test problem) at the final ($k$th Picard) iteration:
\begin{align*}
\elltwoV \defeq \|\vect{v}^\star - \vect{v}^k\|_2 / \|\vect{v}^\star\|_2 &&
\elltwoINIT \defeq \|c_A^\star(0) - c_A^k(0)\|_2/\| c_A^\star(0)\|_2.
\end{align*}

\ipoint{Hardware.}
The runs for all test cases were executed on the Tier-1 supercomputer HazelHen at the High Performance Computing Center HLRS in Stuttgart (\url{www.hlrs.de}), a Cray XC40 system with a peak performance of $7.42$ Petaflops comprising $7,712$ nodes with Xeon E5-2680 v3 processors and $24$ cores on two sockets per node. The nodes are connected via an Aries interconnect.
For data sizes of $N_i = 128$, $i=0,1,2$, we use $3$ nodes with $64$ MPI tasks and for $N_i = 256$, $i=0,1,2$, we use $11$ nodes and $256$ MPI tasks. 

\subsection{Test Case \ipoint{ATAV}}
\label{s:ATAV_results}
\input{05-01-results_ATAV_all.tex}






\subsection{Test Case \ipoint{RTRV}}
\label{s:RTRV_results}
\input{05-06-results_RTRV.tex}

%% file: 05-01-results_ATAV_all.tex

\sisetup{round-mode=places,round-precision=2}

\ipoint{Purpose.} This experiment is a proof of concept. We test the numerical accuracy of our scheme, identify the inversion accuracies we can ideally expect (i.e., the errors we get if we use the forward operators to generate the observations for our coupled inversion), and study the convergence of our solver.\footnote{In general, considering real data, our forward model is---at best---a crude approximation of the actual observations. That is, first of all, we can not expect our registration algorithm to recover a bio-physically meaningful deformation between the brain anatomies of different subjects (such a model does not even exist). Second, our forward tumor model is far from being realistic (see \secref{sec:tumor} for a discussion).}. With ATAV-REAC
and ATAV-DIF, we in addition test the sensitivity of our approach with respect to pertubations
in the model and model parameters. In ATAV-LD, we restruct the admissible initial condition parametrization for tumor growth to a very low dimensionality in order to exclude fully grown tumors as initial conditions. We examine, whether this increases the sensitivity of the reconstruction quality for the tumor with respect to correct model parameters.\\

\ipoint{Setup.} ATAV is based on real brain geometries, but uses a tumor grown with our forward solver in a synthetically generated patient image. We use a resolution of $N_i = 128$. We choose $\vect{p}=\vect{p}^\star$, which defines $c_A(0)=c_A^\star(0)$, grow a tumor in the tumor-free atlas, which gives $c_A^\star(1)$, choose $\vect{v}=\vect{v}^\star$, deform the atlas with the grown tumor by advecting the probability maps with the negative velocity $\vect{v}^\star$. This gives us $\vect{m}_T$ and $c_T$ (patient image with tumor). The velocity $\vect{v}^\star$ is generated by registering two tumor-free images of two different individuals ($\beta_{\vect{v}}^\star = \num{1E-4}$). The center of mass for the synthetic tumor is set to $(x^1,x^2,x^3) = 2\pi\cdot(0.285, 0.36, 0.5)$. As initial condition for the artificial tumor generation, we enable two of the Gaussians at the center of the grid of Gaussians. See~\secref{s:setup} for further details on the parameters.
As a baseline, we also report results for the sole registration of the healthy anatomy (i.e., neglecting the tumor forward solve to generate the data). In addition to that, we quantify the numerical error of our scheme for solving the transport equations (forward registration). This is done by solving the forward problem twice, once with the original and once with the reverted (negative) velocity. The associated error is given by
\begin{equation}
\frac{\left\|c_A(1)^\star -\Rfor\left(-\vect{v}^\star,\Rfor\left(\vect{v}^\star,c_A(1)^\star\right)\right)\right\|_2}{\|c_A(1)^\star\|_2} = \num{9.372E-02}
\label{e:advection_error}
\end{equation}

Whereas we disable diffusion ($k_f=0$) in ATAV-REAC, we use the full tumor model including
diffusion ($k_f \neq 0$) for ATAV-DIF. In ATAV-REAC, the same growth rates $\rho_w, \rho_g$ with scaling $\rho_f$ and $k_w, k_g$ with scaling $k_f$ are used for growing the tumor and for the inversion to reconstruct the initial condition. In ATAV-DIF, we use values for the reaction and diffusion coefficients for the inversion in the Picard iterations, that are either the same or differ from those used for the generation of the tumor. In ATAV-LD, we use the full reaction-diffusion tumor model and enforce the initial tumor to be small, i.e., only invert for $n_{\vect{p}} = 8$ parameters, which results in a grid of $2 \times 2 \times 2$ Gaussians that cover the true initial condition of the artificially grown tumor. We expect this to increase the sensitivity of our inversion with respect to the tumor parameters (we can not fully represent the whole patient tumor purely based on a linear combination of the basis functions). For the inversion, we again consider a variety of models and model parameter combinations, which includes the use of the `'correct`' (ground truth) tumor parameters. We also consider the case in which we completely neglect the tumor model (i.e., $\rho_f = k_f=0$) and just invert for the basis.\\

\begin{figure*}[h!]
\includegraphics[scale=.85]{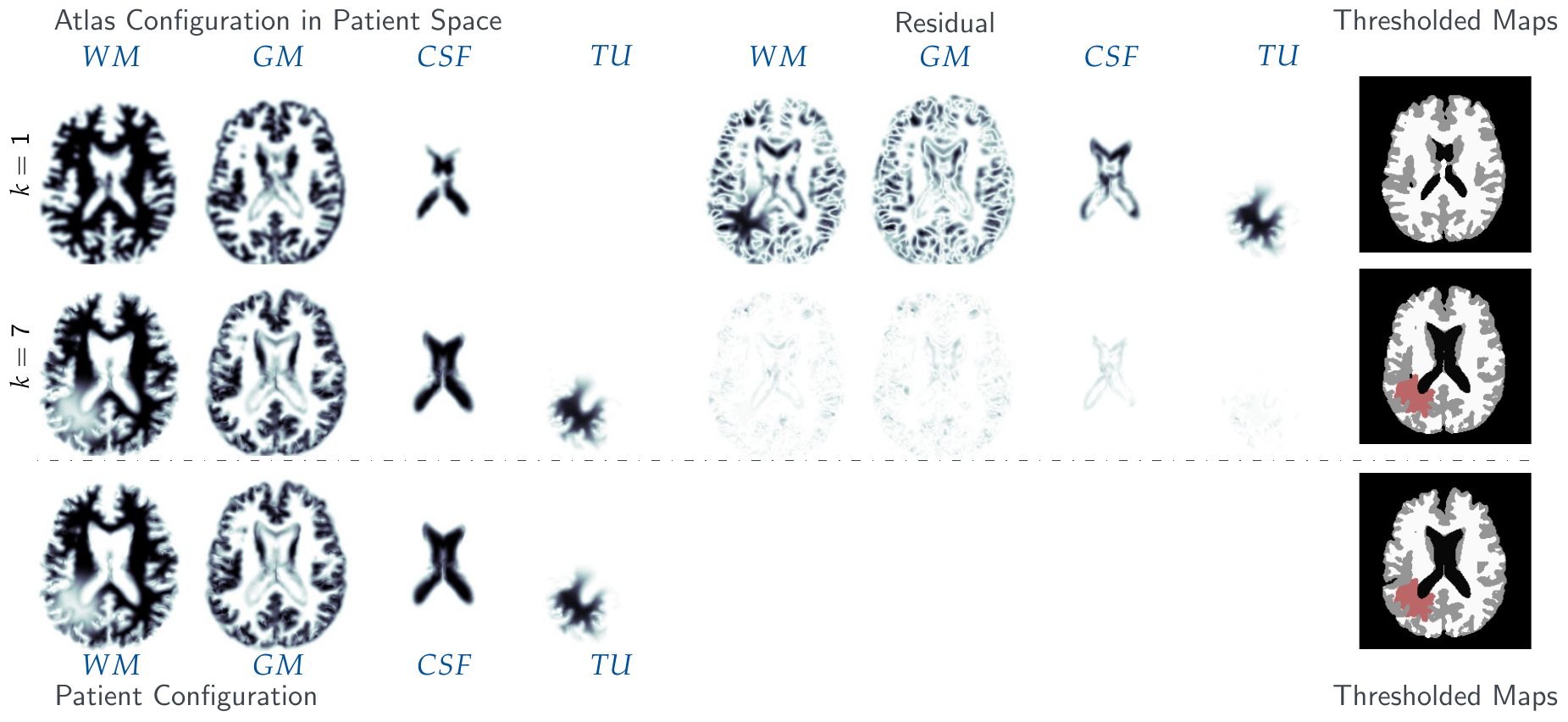} \\
\centering
\begin{small}
\begin{tabular}{rrLlLlllll}
\toprule
iteration & $\beta_{\vect{v}}$ & $\elltwoB$         & $\diceB$           & $\elltwoT$         & $\diceT$            & $\elltwoV$         & $\delta\vect{v}$    & $\elltwoINIT$   & $\relG$           \\
\midrule
initial   &  --                & \num{1.000000e+00} & \num{7.098000e-01} & \num{1.000000e+00} & \num{0.000000e+00}  & \num{1.000000e+00} & --                  & \num{1.000e+00} & \num{1.000000e+00}\\
$1$       &  \num{1e+00}       & \num{9.371000e-01} & \num{7.270000e-01} & \num{1.000000e+00} & \num{0.000000e+00}  & \num{9.595000e-01} & --                  & \num{2.543e-01} & \num{9.976000e-01}\\
$2$       &  \num{1e-01}       & \num{6.303000e-01} & \num{8.012000e-01} & \num{2.262000e-01} & \num{8.625000e-01}  & \num{8.495000e-01} & \num{9.770000e-01}  & \num{2.047e-01} & \num{5.913000e-02}\\
$3$       &  \num{1e-02}       & \num{3.668000e-01} & \num{8.878000e-01} & \num{1.466000e-01} & \num{9.192000e-01}  & \num{6.311000e-01} & \num{9.332000e-01}  & \num{1.648e-01} & \num{3.070000e-02}\\
$4$       &  \num{1e-03}       & \num{1.706000e-01} & \num{9.396000e-01} & \num{7.910000e-02} & \num{9.661000e-01}  & \num{3.351000e-01} & \num{6.624000e-01}  & \num{1.398e-01} & \num{1.953000e-02}\\
$5$       &  \num{1e-04}       & \num{1.570000e-01} & \num{9.470000e-01} & \num{6.552000e-02} & \num{9.738000e-01}  & \num{3.594000e-01} & \num{3.649000e-01}  & \num{1.358e-01} & \num{1.322000e-02}\\
$6$       &  \num{1e-04}       & \num{1.567000e-01} & \num{9.470000e-01} & \num{6.389000e-02} & \num{9.731000e-01}  & \num{3.594000e-01} & \num{1.445000e-05}  & \num{1.330e-01} & \num{1.211000e-02}\\
$7$       &  \num{1e-04}       & \num{1.566000e-01} & \num{9.470000e-01} & \num{6.291000e-02} & \num{9.729000e-01}  & \num{3.594000e-01} & \num{1.446000e-05}  & --              & \num{1.213000e-02}\\
\bottomrule
\end{tabular}
\end{small}\\[2mm]
\caption{Results for the \textbf{analytic tumor\,/\,analytic velocity reaction-only (ATAV-REAC)} test case; ground truth: ($\rho_f=15$, $\rho_w=1$, $\rho_g=0$, $k_f=0$, $\vect{p}=\vect{p}^\star$, $\vect{v}=-\vect{v}^\star$). The figure shows probability maps for the labels of the healthy atlas brain and the patient brain with tumor generated from a tumor grown in the atlas and known atlas to patient advection velocity (see text for details; axial-slice $64$). We show the initial configuration for the problem (top row; iteration $k=1$), the final configuration after joint registration and tumor inversion (middle row; iteration $k = 7$; the atlas image probability maps are transported to the patient space), and the target patient data (reference image; bottom row). Each row contains (from left to right) the probability maps for WM, GM, CSF, and TU, the residual differences (if available) between the probability maps, and a hard segmentation based on the given probabilities for the individual tissue classes. The table on the bottom provides quantitative results for the inversion. We report the average mismatch for the probability maps for the brain tissue labels $\elltwoB$ and the tumor $\elltwoT$, the mean DICE coefficient for brain tissue $\diceB$ and tumor $\diceT$, respectively. The reconstruction quality is given in terms of convergence of $\vect{v}^k$ and $c_A^k(0)$ towards the ground truth $\vect{v}^\star$ and $c_A^\star(0)$, respectively ($\elltwoV$ and $\elltwoINIT$). We can not expect this error to go to zero for several reasons. First, we loose information when we construct the test case (zero gradients in the intensity of the image), second our numerical solver introduces errors (in particular, the solver for the transport equations).
We in addition to that report the change in update in the velocity $\vect{v}$ across successive iterations $\delta\vect{v} = \|\vect{v}^k - \vect{v}^{k-1} \| / \|\vect{v}^{k-1}\|$. Finally, we also list the relative norm of the gradient for the coupled problem in~\eqref{eq:global_opt} ($\relG$).}
\label{fig:mismatch_syn-ATAV-nx-128-H1SN-jacbound1E-3_betaIt-1_vdiff_c0diff}
\end{figure*}

\ipoint{Results.} We report results for the registration of anatomy (without tumor) in \tabref{tab:reg-error-syn-tc-mov-cov}. Results for the inversion in ATAV-REAC are presented in \figref{fig:mismatch_syn-ATAV-nx-128-H1SN-jacbound1E-3_betaIt-1_vdiff_c0diff} and assess the reconstruction quality (Dice and residuals) and the reduction of the reduced gradient with respect to the iteration index. We also report the error between the ground truth $c_A(0)^{\star}=\Phi\vect{p}^\star$ and $\vect{v}^\star$ and the estimated iterates as well as the relative norm of the gradient of the fully coupled problem in~\eqref{eq:global_opt}.

For ATAV-DIF, quantitative results for the inversion are shown in \tabref{tab:ATAV-nonzero-diffusion}, qualitative results can be found in the supplementary material in \figref{fig:ATAV-nonzero-diffusion}. In addition to reconstruction quality and gradient reduction, we list the runtime for the Picard scheme per iteration and the percentage spent in each individual solver (tumor and registration), respectively. Note that tumor and registration runtimes do not add up to 100\% as further parts of the code such as the calculation of the reduced gradient and the steering of the Picard iteration are not included in the measurements.

\begin{figure}[h!]
\centering
\includegraphics[width=0.48\textwidth]{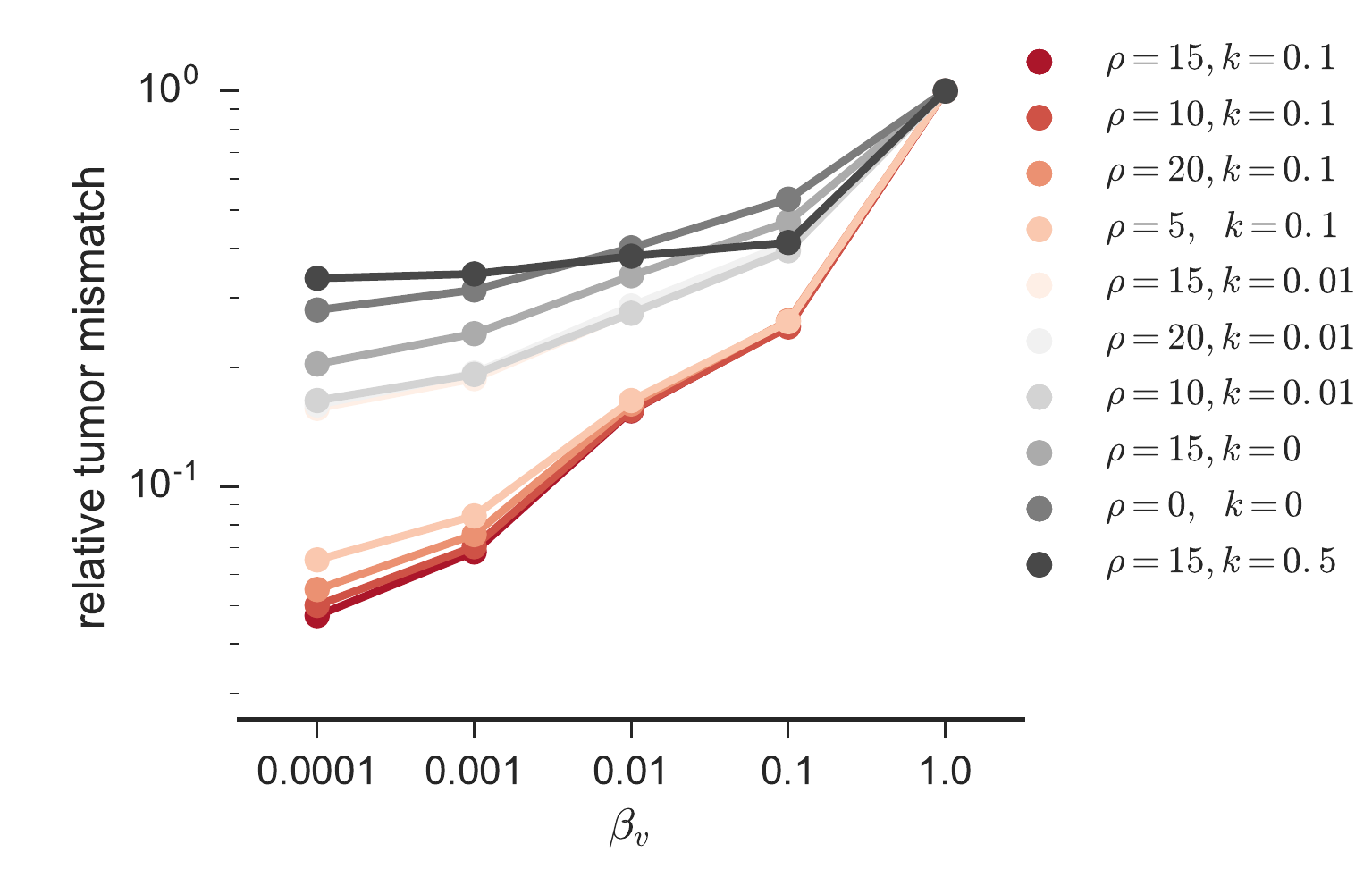}
\caption{\label{fig:ATAV-SPARSE-MM-curves}Mismatch reduction as a function of the regularization parameter $\beta_{\vect{v}}$ for the \textbf{analytic tumor with analytic velocity, diffusion and low-dimensional initial condition (ATAV-LD)} test case; ground truth: ($\rho_f=15$, $\rho_w=1$, $\rho_g=0$, $k_f=\num{1.0E-1}$, $k_w=1$, $k_g=0$, $\vect{p}=\vect{p}^\star$ (in patient domain), $\vect{v}$ N/A). We use an initial condition parameterized with only $n_{\vect{p}}=8$ Gaussians ($\sigma=\nicefrac{\pi}{15}$) and an analytic tumor with non-zero diffusion. The plot shows the relative mismatch for the tumor probability map $\elltwoT$ in patient space. Note that we reduce $\beta_{\vect{v}}$ by a factor of ten in each Picard iteration; the plot shows the mismatch reduction over the Picard iterations if read from right to left.}
\end{figure}

\begin{figure}[ht!]
\centering
\includegraphics[width=0.85\textwidth]{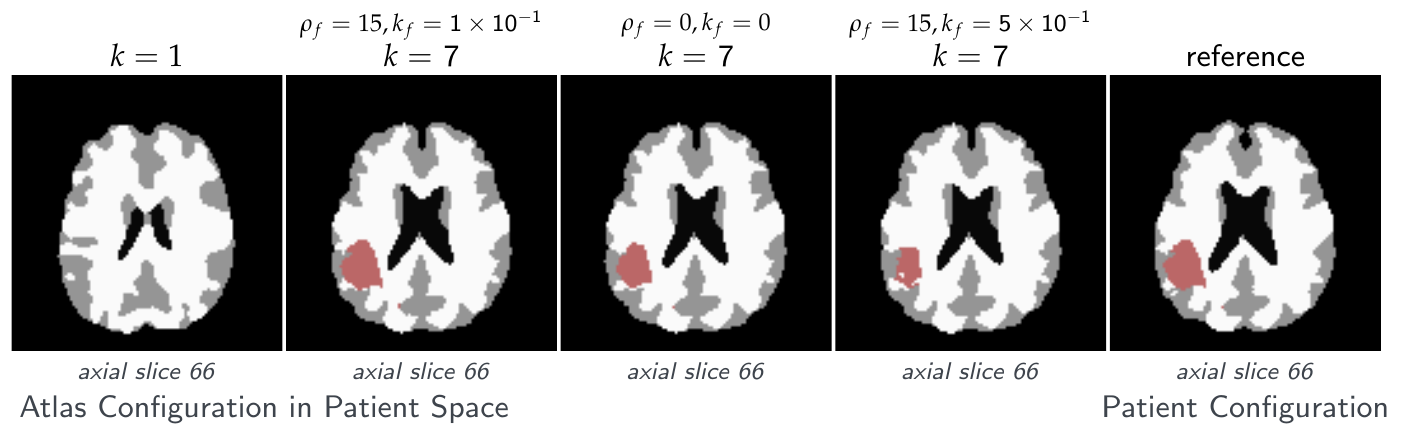}
\caption{Tumor and brain labels for the \textbf{analythic tumor with analythic velocity with non-zero diffusion and low-dimensional initial condition (ATAV-LD)}
test case; ground truth: ($\rho_f=15$, $\rho_w=1$, $\rho_g=0$, $k_f=\num{1.0E-1}$, $k_w=1$, $k_g=0$, $\vect{p}=\vect{p}^\star$ (in patient domain), $\vect{v}$ N/A); for the 'correct' tumor parameters $\rho_f=15$, $k_f=\num{1.0E-1}$, and the two different settings $\rho_f=15$, $k_f=\num{5.0E-1}$ and $\rho_f=k_f=0$.}
\label{fig:ATAV-nonzero-diffusion-sparse-ax-only}
\end{figure}

For ATAV-LD, we show simulation results in~\figref{fig:ATAV-nonzero-diffusion-sparse-ax-only} (for sagittal and coronal slices, \figref{fig:ATAV-nonzero-diffusion-sparse}) and report quantitative results in \tabref{tab:mismatch_ATAV-sparsity-of-initial-condition_non-zero-diffusion}. We plot the trend of the relative tumor mismatch with respect to the regularization parameter $\beta_{\vect{v}}$ for the Sobolev norm for the registration velocity (and by that the Picard iteration index) in \figref{fig:ATAV-SPARSE-MM-curves}.\\

\begin{table}[h!]
\caption{Reference results for geometry registration only between healthy atlas and healthy patient. The table shows values for the relative mismatch for the geometry ($\elltwoB$) and the  associated Dice coefficient $\diceB$ as well as the relative $\ell^2$-error for the reconstruction of the velocity field $\elltwoV$ with respect to the ground truth $\vect{v}^\star$.\label{tab:reg-error-syn-tc-mov-cov}}
\centering
\begin{small}
\begin{tabular}{llll}\toprule
$\text{maxit}_R$ & \elltwoB        & $\diceB$        &  $\elltwoV$      \\\midrule
$(50,80)$        & \num{1.784e-01} & \num{9.367E-01} &  \num{3.588e-01} \\
$(10,20)$        & \num{1.675e-01} & \num{9.364E-01} &  \num{3.141e-01} \\
\bottomrule
\end{tabular}
\end{small}
\end{table}

\begin{table*}[h!]
\caption{Results for the \textbf{analytic tumor\,/\,analytic velocity with non-zero diffusion (ATAV-DIF)} test case; ground truth: ($\rho_f=10$, $\rho_w=1$, $\rho_g=0$, $k_f=\num{1.0E-2}$, $k_w=1$, $k_g=0$, $\vect{p}=\vect{p}^\star$, $\vect{v}=-\vect{v}^\star$). We report values for the (summed) norm of the residual between the respective probability maps for the different brain tissue classes $\elltwoB$ and tumor $\elltwoT$ in patient space, the mean Dice coefficient for brain tissue $\diceB$ and tumor $\diceT$, respectively, as well as the relative norm of the gradient $\relG$ for the global coupled problem~\eqref{eq:global_opt}. We report results for different values of $\rho_f \in \{5,10,15\}$ used in the inversion ($\rho_f=10$ is the ground truth). We report the time spent per iteration (in seconds; top run) or in total (in seconds; bottom runs) for the entire Picard inversion, and the amount of that time spent in the tumor inversion and image registration (in percent (top run); in seconds (bottom runs)), respectively.  Note that the latter sums up to less than $100\%$ as we do not explicitly measure time spent in additional coupling functionality and forward solvers. These runs are performed using 64 MPI tasks on three nodes of \emph{HazelHen} (see \secref{s:hardware} for details). The top block shows the course of the inversion with respect to the Picard iteration index for the correct parameters (ground truth) for $\rho_f$ and $k_f$. The four rows on the bottom show the final result for our Picard scheme for different parameter and model combinations.}
\label{tab:ATAV-nonzero-diffusion}
\centering
\begin{small}
\begin{tabular}{crLlLlllll}
\toprule
iterations & $\beta_{\vect{v}}$ & $\elltwoB$         & $\diceB$           &  $\elltwoT$        & $\diceT$           & $\relG$            & $\rtIT$\,[s]       & $\rtTUM$\,[\%]      & $\rtREG$ [\%] \\
\midrule
\multicolumn{10}{l}{\emph{non-zero diffusion with ground truth parameters $\rho_f = 10, k_f=\num{1E-2}$}}\\
\midrule
ref  &  --           & \num{1.000000e+00} & \num{7.135000e-01} & \num{1.000000e+00} & \num{0.000000e+00} & \num{1.000000e+00} &  \\
$1$  &  \num{1e+00}  & \num{9.320000e-01} & \num{7.316000e-01} & \num{1.000000e+00} & \num{0.000000e+00} & \num{9.943000e-01} & \num{2.112000e+03} & \pernum{99.668561}  & \pernum{0.164962}  \\
$2$  &  \num{1e-01}  & \num{6.515000e-01} & \num{8.005000e-01} & \num{2.082000e-01} & \num{8.558000e-01} & \num{7.454000e-02} & \num{2.884000e+02} & \pernum{96.012483}  & \pernum{2.144938}  \\
$3$  &  \num{1e-02}  & \num{3.794000e-01} & \num{8.875000e-01} & \num{1.473000e-01} & \num{9.116000e-01} & \num{3.869000e-02} & \num{3.943000e+02} & \pernum{91.808268}  & \pernum{6.913518}  \\
$4$  &  \num{1e-03}  & \num{1.745000e-01} & \num{9.394000e-01} & \num{8.080000e-02} & \num{9.629000e-01} & \num{2.454000e-02} & \num{3.980000e+02} & \pernum{72.663317}  & \pernum{26.080402} \\
$5$  &  \num{1e-04}  & \num{1.601000e-01} & \num{9.475000e-01} & \num{5.929000e-02} & \num{9.781000e-01} & \num{1.770000e-02} & \num{2.003000e+02} & \pernum{35.411882}  & \pernum{62.106840} \\
$6$  &  \num{1e-04}  & \num{1.600000e-01} & \num{9.475000e-01} & \num{5.772000e-02} & \num{9.769000e-01} & \num{1.720000e-02} & \num{1.335000e+02} & \pernum{46.561798}  & \pernum{49.580524} \\
$7$  &  \num{1e-04}  & \num{1.599000e-01} & \num{9.474000e-01} & \num{5.627000e-02} & \num{9.758000e-01} & \num{1.733000e-02} & \num{7.099000e+01} & \pernum{0.0000000}  & \pernum{92.999014} \\
\midrule
\multicolumn{7}{l}{\emph{varying $\rho_f \in \{5,\boldsymbol{10},15\}, k_f=\num{1E-2}$, ground truth $\rho_f = 10, k_f=\num{1E-2}$}} & $\rtTOT$\,[s]       & $\rtTUM$\,[s]      & $\rtREG$ [s] \\
\midrule
$\rho_f = 5\phantom{0}$ &  \num{1e-04}  & \num{1.610000e-01}  & \num{9.462000e-01}  & \num{6.484000e-02}  & \num{9.674000e-01}  & \num{2.207000e-02}  & \num{4.374850e+03}  & \num{3.947620e+03}  & \num{3.967840e+02} \\
\highlightrow
$\rho_f = 10$ &  \num{1e-04}  & \num[math-rm=\mathbf]{1.599000e-01}  & \num[math-rm=\mathbf]{9.474000e-01}  & \num[math-rm=\mathbf]{5.627000e-02}  & \num[math-rm=\mathbf]{9.758000e-01}  & \num{1.733000e-02}  & \num{3.597490e+03}  & \num{3.166190e+03}  & \num{4.008240e+02} \\
$\rho_f = 15$ &  \num{1e-04}  & \num{1.613000e-01}  & \num{9.477000e-01}  & \num{6.393000e-02}  & \num{9.723000e-01}  & \num{1.411000e-02}  & \num{3.957320e+03}  & \num{3.527350e+03}  & \num{4.004410e+02} \\
\midrule
\multicolumn{7}{l}{\emph{$\rho_f 10, k_f=0$, ground truth $\rho_f = 10, k_f=\num{1E-2}$}} & $\rtTOT$\,[s]       & $\rtTUM$\,[s]      & $\rtREG$ [s] \\
\midrule
$\rho_f = 10$ &  \num{1e-04}  & \num{1.602000e-01}  & \num{9.477000e-01}  & \num{5.951000e-02}  & \num{9.711000e-01}  & \num{1.742000e-02}  & \num{4.252200e+02}  & \num{2.253460e+01}  & \num{4.009880e+02} \\
\bottomrule
\end{tabular}
\end{small}
\end{table*}

\begin{table*}
\caption{Results for the \textbf{analytic tumor with analytic velocity with diffusion and low-dimensional initial condition (ATAV-LD)} test case;  ground truth: ($\rho_f=15$, $\rho_w=1$, $\rho_g=0$, $k_f=\num{1.0E-1}$, $k_w=1$, $k_g=0$, $\vect{p}=\vect{p}^\star$ (in patient domain), $\vect{v}$ N/A). We use a low-dimensional parameterization for the initial condition with $n_{\vect{p}}=8$ Gaussians ($\sigma=\nicefrac{\pi}{15}$). We report the (summed) mismatch for the probability maps for the brain tissue $\elltwoB$ and tumor $\elltwoT$ in patient space, the mean Dice coefficient for hard segmentations of brain tissue ($\diceB$) and tumor ($\diceT$), respectively, and the relative norm of the gradient ($\relG$) for the coupled problem \eqref{eq:global_opt}. We assess the final state of the reconstruction using different values for $\rho_f \in \{5,10,15, 20\}$ and $k_f \in \{0, \num{1E-1}\}$. Absolute timings are given for the tumor inversion and image registration, respectively using 64 MPI tasks on three nodes of HazelHen.  Note that the latter sums up to less than the reported total run time as we do not explicitly measure time spent in additional coupling functionality and forward solvers. We always reach the target value of $\beta_{\vect{v}} = \num{1E-4}$.}
\label{tab:mismatch_ATAV-sparsity-of-initial-condition_non-zero-diffusion}
\centering
\begin{small}
\begin{tabular}{llrrLlLlllll}
\toprule
$\rho$ & $k$  & It      & $\elltwoB$ & $\diceB$ & $  $\elltwoT$ $ & $\diceT$ & $\relG$  & $\rtTOT$\,[s] & $\rtTUM$\,[s] & $\rtREG$\,[s] \\
\midrule
     &            & initial & \num{1.000000e+00} & \num{7.176000e-01} & \num{1.000000e+00} & \num{0.000000e+00} & \num{1.000000e+00} & -- & -- & -- \\
\midrule
 $0$ & $0$          & final & \num{1.961000e-01} & \num{9.448000e-01} & \num{2.795000e-01} & \num{8.477000e-01} & \num{4.157000e-02} & \num{3.776320e+02} & \num{3.330000e+00} & \num{3.768990e+02} \\
$15$ & $0$          &       & \num{1.865000e-01} & \num{9.461000e-01} & \num{2.036000e-01} & \num{9.157000e-01} & \num{1.702000e-02} & \num{3.776650e+02} & \num{2.189700e+00} & \num{3.780200e+02} \\
\highlightrow
$15$ & $\num{1E-1}$ &       & \num[math-rm=\mathbf]{1.693000e-01} & \num[math-rm=\mathbf]{9.477000e-01} & \num[math-rm=\mathbf]{4.675000e-02} & \num[math-rm=\mathbf]{9.649000e-01} & \num{1.731000e-02} & \num{4.724960e+03} & \num{4.171100e+03} & \num{4.593420e+02} \\
$15$ & $\num{5E-1}$ &       & \num{1.901000e-01} & \num{9.412000e-01} & \num{3.365000e-01} & \num{4.714000e-02} & \num{1.782000e-02} & \num{6.670500e+03} & \num{5.538300e+03} & \num{4.289570e+02} \\
$15$ & $\num{1E-2}$ &       & \num{1.789000e-01} & \num{9.467000e-01} & \num{1.572000e-01} & \num{9.313000e-01} & \num{1.507000e-02} & \num{2.233390e+03} & \num{1.504600e+03} & \num{4.309730e+02} \\
 $5$ & $\num{1E-1}$ &       & \num{1.706000e-01} & \num{9.462000e-01} & \num{6.502000e-02} & \num{9.567000e-01} & \num{2.451000e-02} & \num{1.959840e+03} & \num{1.504300e+03} & \num{3.870510e+02} \\
$10$ & $\num{1E-1}$ &       & \num{1.695000e-01} & \num{9.467000e-01} & \num{4.976000e-02} & \num{9.620000e-01} & \num{2.006000e-02} & \num{3.900050e+03} & \num{2.902200e+03} & \num{3.725620e+02} \\
$10$ & $\num{1E-2}$ &       & \num{1.794000e-01} & \num{9.465000e-01} & \num{1.647000e-01} & \num{9.257000e-01} & \num{1.910000e-02} & \num{2.174100e+03} & \num{1.625260e+03} & \num{3.655810e+02} \\
$20$ & $\num{1E-1}$ &       & \num{1.695000e-01} & \num{9.477000e-01} & \num{5.445000e-02} & \num{9.559000e-01} & \num{2.018000e-02} & \num{3.073610e+03} & \num{2.617200e+03} & \num{3.881380e+02} \\
$20$ & $\num{1E-2}$ &       & \num{1.797000e-01} & \num{9.467000e-01} & \num{1.603000e-01} & \num{9.184000e-01} & \num{1.197000e-02} & \num{2.032620e+03} & \num{4.514000e+02} & \num{3.781680e+02} \\
\bottomrule
\end{tabular}
\end{small}
\end{table*}

\ipoint{Observations.} The most important observation are \bipa\item that the reconstructed data (tumor and registered anatomy) is in excellent agreement with the patient data, and \item we are able to reduce the reduced gradient~\eqref{e:reduced_gradient} by two orders of magnitude in less than iterations of our Picard scheme for ATAV-REAC and significantly more than a factor of $50$ for all experiments in ATAV-DIF and ATAV-LD\eipa.

The numerical error for the advection in~\eqref{e:advection_error} is \num{9.373e-02}.
The relative mismatch for the anatomy obtained for our iterative Picard scheme is in the order of the advection error for the forward image registration problem. We can also see that the reconstruction of $c_A(0)^\star$ and $c_A(1)^\star$ seems to be bounded by this error. In fact, due to the advection error that leads to a mismatch in this order in the atlas domain, this is the best we can expect without over-fitting the data. Similar observations can be made if we compare
the inversion results
with the results obtained for the registration for healthy brains (neglecting the tumor simulations) reported in \tabref{tab:reg-error-syn-tc-mov-cov}. Hence, the quality of tumor reconstruction is comparable to the quality of pure image registration between the healthy geometries. This is an excellent result that clearly demonstrates the potential of our approach. The obtained Dice coefficient for the brain anatomy is in the order of what we see for the sole registration of healthy anatomies.\footnote{In fact, it is even slightly better for ATAV-REAC. This slight increase might be a consequence of numerical inaccuracies in our scheme, discrepancies in the number of Newton-steps and Krylov iterations taken.}
Note that the mismatch between the true velocity $\vect{v}^\star$ and the recovered velocity $\vect{v}^k$ reported for ATAV-REAC is due to the fact that image registration is an inherently ill-posed problem: the velocity can only be reconstructed exactly in image areas with non-zero gradients and if there are only non-zero intensity differences between the images to be registered in areas that do correspond to one another\footnote{As in any formulation based on an $L^2$-distance functional, it is the mismatch between the reference and template image and the gradient of the deformed template image that drive the optimization (see \eqref{opt_grad_R_solo}).}. In addition, we ask for the reconstruction of a vector field from scalar data.

The Dice coefficient for the brain anatomy increases from \num{7.098000e-01} to \num{9.470000e-01} in ATAV-REAC, where we obtain a final Dice coefficient of \num{9.729000e-01} for the tumor. The results for ATAV-DIF and
ATAV-LD show that the used model and the dimensionality of the initial conditions do not have
a significant impact on the quality of the inversion (Dice and mismatch).

We can furthermore see that we can significantly reduce the norm of the reduced gradient~\eqref{e:reduced_gradient} to  \num{1.213000e-02}. We can also see that once we have reached the target regularization parameter $\beta_{\vect{v}} = \num{1E-4}$ we do not make any more progress. The update for the velocity tends to zero, the changes in the reduced gradient are small and the error measures (residual and Dice) do no longer change significantly.

For ATAV-DIF, we see that we obtain a slightly better mismatch (\num{5.627000e-02}) and Dice coefficient (\num{9.758000e-01}) for the tumor probability map if we use the correct $\rho_f$. These results suggest that we can identify the correct $\rho_f$ if we run multiple inversions for the initial condition. However, we note that these differences are small (the mismatch is between \num{5.951000e-02} and \num{5.627000e-02} and the Dice is between \num{9.711000e-01} and \num{9.758000e-01}, respectively, for the tumor). Overall, we obtain an excellent agreement between patient data and atlas data irrespective of the model choice. These observations are confirmed by careful visual inspection of \figref{fig:ATAV-nonzero-diffusion} (supplementary material; shows only results for the `'correct`' tumor parameters).
%
This can be explained by the parameterization of the initial condition. We can reconstruct complex tumor shapes even with a simple model. Overall, this indicates that using a reaction-only model might be sufficient for pure diffeomorphic image registration\footnote{This is clearly not the case when targeting the design of computational framework to aid clinical  decision making by generating model based prediction of a patient's future tumor state. In this case, we quite certainly have to use more complex, high-fidelity models. We discuss this in more detail in \secref{s:conclusion}.}, where a comparison of the total run time for the last row in \tabref{tab:ATAV-nonzero-diffusion} to the total runtimes attained when enabling diffusion shows that we can save a factor of 10 in runtime by disabling diffusion.
Another interesting behavior within our scheme is that, during the first few iterations, most time is spent in the tumor inversion, whereas, as we reduce $\beta_{\vect{v}}$, the registration does most of the work. This is to be expected, since the runtime of our scheme (more precisely, the condition number of the Hessian) for diffeomorphic registration, is not independent of $\beta_{\vect{v}}$~\citep{Mang:2015a,Mang:2017a}.

The most important observation for ATAV-LD is that we can identify the correct pair of diffusion and reaction rate that has been used to generate the synthetic test based on slightly more pronounced differences in the overall mismatch and Dice scores than in ATAV-DIF.
%
The sensitivity with respect to changes in the diffusion parameter $k_f$ is larger than with respect to changes in the growth rate $\rho_f$. We expect this much more pronounced dependence on the diffusion model if we parameterize the initial condition on a grid with a smaller support, since we need the cancerous cells to diffuse to areas more distant to the tumor center in order to be able to reconstruct the whole tumor. The trend of the relative mismatch for the tumor in \figref{fig:ATAV-SPARSE-MM-curves} highlights this effect. The curves almost cluster in terms of the different choices for the diffusion coefficient; we overall achieve significantly better mismatch if we choose the correct diffusion coefficient. We can also see that the curves plateau much earlier in cases where we use the wrong parameter combinations. The tumor mismatch takes values between \num{3.365000e-01} and \num{4.675000e-02} with a Dice score of \num{4.714000e-02} to \num{9.649000e-01}.\\


\ipoint{Conclusion:} We conclude that our Picard scheme is efficient and converges to a valid (local) minimum in the search space (we reduce the relative global gradient, the distance measure, and significantly increase the Dice coefficient). We attain an excellent agreement between the data (patient tumor and geometry) and the predicted state (transported atlas geometry and predicted tumor) with a final Dice coefficient of \num{9.470000e-01} and \num{9.729000e-01} for the labels of the anatomy and tumor In ATAV-REAC and only slightly worse values in ATAV-DIF and
ATAV-LD.

The integration of a diffusion model into our inversion is very costly, at least for our current implementation. Designing a more efficient forward solver for the diffusion operator requires more work. We could demonstrate that our parameterization of the initial condition allows us to generate high-fidelity registration results, especially for the healthy anatomy, irrespective of the model that has been used to generate the data. These are clearly preliminary results, but they provide some evidence that reaction-only models might be sufficient for pure image registration (something that is quite certainly not the case if we target parameter identification and tumor growth prediction)~\citep{Hogea:2008a}.

We can also observe that there are subtle differences in the reconstruction quality of the tumor if we use the `'correct`' growth rate for the inversion for ATAV-DIF, that are more
pronounced in ATAV-LD. Overall, we conclude that we \bipa\item can neglect the diffusion model in the context of diffeomorphic registration and compensate the resulting loss in accuracy
by a higher-dimensional Gaussian basis, \item might be able to identify appropriate growth rates if we run multiple inversions for different parameters with a low-dimensional Gaussian basis
\eipa.


\FloatBarrier

%% file: 05-06-results_RTRV.tex
\ipoint{Purpose.} We test our approach on real data of patients diagnosed with glioma tumors and study the registration quality for a variety of parameter choices for the tumor growth model.\\

\ipoint{Setup.} This test scenario consists of real patient brains with real tumors for which we do not know any parameters. The patient datasets are the first proposal for a patient segmentation produced in the first iteration of GLISTR~\citep{Gooya:2012a}. We provide additional details about this databasis in \secref{s:setup}.
We have to identify the support of the domain spanned by the Gaussian basis functions for the tumor initial condition parametrization as well as their spacing $d$, and the standard deviation $\sigma$ for any unseen patient. This is done automatically. As some of the real tumors are multifocal, we use $n_{\vect{p}}=343$ Gaussians in our first set of experiments on all six brains for the initial condition parametrization.
In contrast to the synthetic test cases ATAV-REAC, ATAV-DIF and ATAV-LD, we allow the tumor to grow also in gray matter instead of in white matter only, but with a reaction parameter that is five times smaller than in white matter (see \tabref{tab:tc-parameters} for details). We use a variety of models and parameter settings in our Picard scheme for the two patients AAMH and AAAN to not only assess the performance of our method but also study its sensitivity towards parameter changes and model complexity.
We vary the reaction parameter $\rho_f$ between $0$ and $15$ and choose the diffusion coefficient $k_f$ either as $0$ or $\num{1.0E-2}$. See \secref{s:setup} for additional details on the setup of the test case and the parameters.\\

\begin{figure*}
\centering
\begin{minipage}{0.49\textwidth}
\textit{AAMH}\\
\includegraphics[scale=.85]{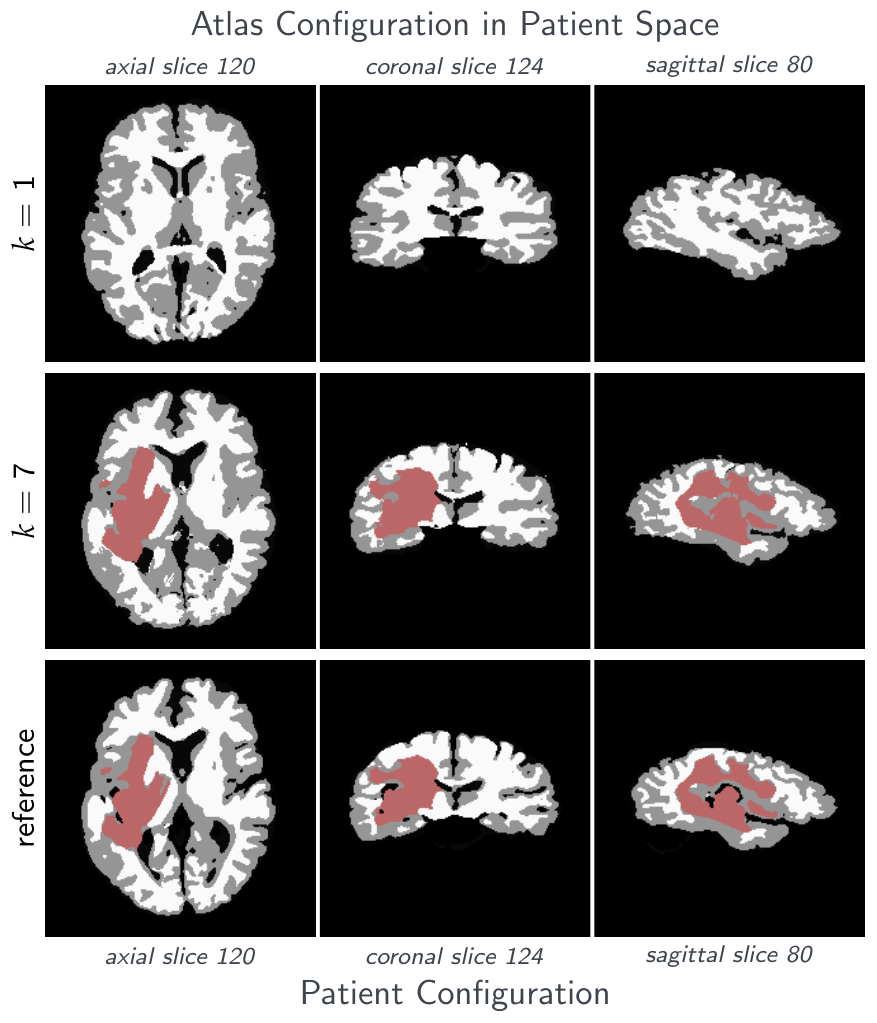}
\end{minipage}
\begin{minipage}{0.49\textwidth}
\textit{AAAN}\\
\includegraphics[scale=.85]{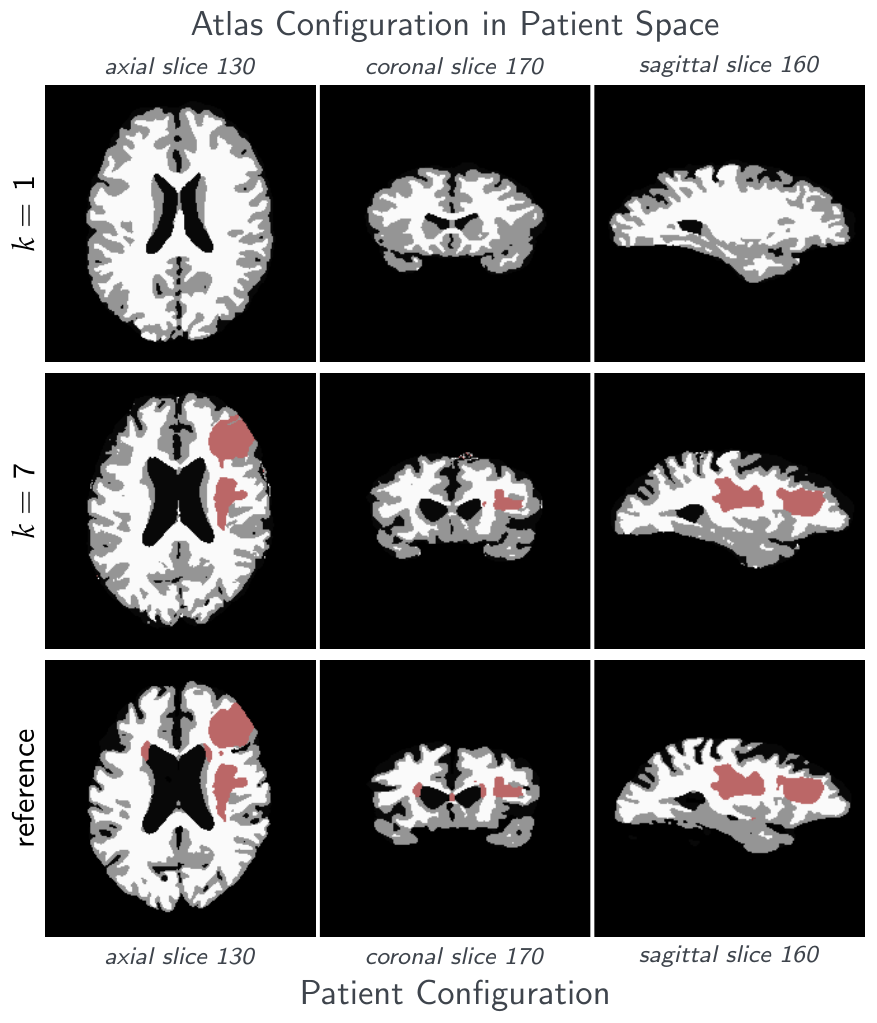}
\end{minipage}\\
\caption{Tumor and brain labels for the \textbf{real tumor with real velocity (RTRV)} test case;  ground truth ($\rho$ N/A, $\mat{k}$ N/A, $\vect{p}$ N/A, $\vect{v}$ N/A); patients {\bf AAMH, AAAN}.
        We set the parameters for the tumor solver to $\rho_f=15$, $k_f=0$ (reaction-only). We use $n_{\vect{p}}=343$ Gaussians for the inversion. The top row shows the original atlas image. The bottom
        row shows the patient image. The row in the middle shows the solution for our coupled scheme.}
\label{fig:RTRV-6Patients_ax_cor_sag_its_1}
\end{figure*}

\begin{figure*}
\centering
\begin{minipage}{0.49\textwidth}
\textit{AAAC}\\
\includegraphics[scale=.85]{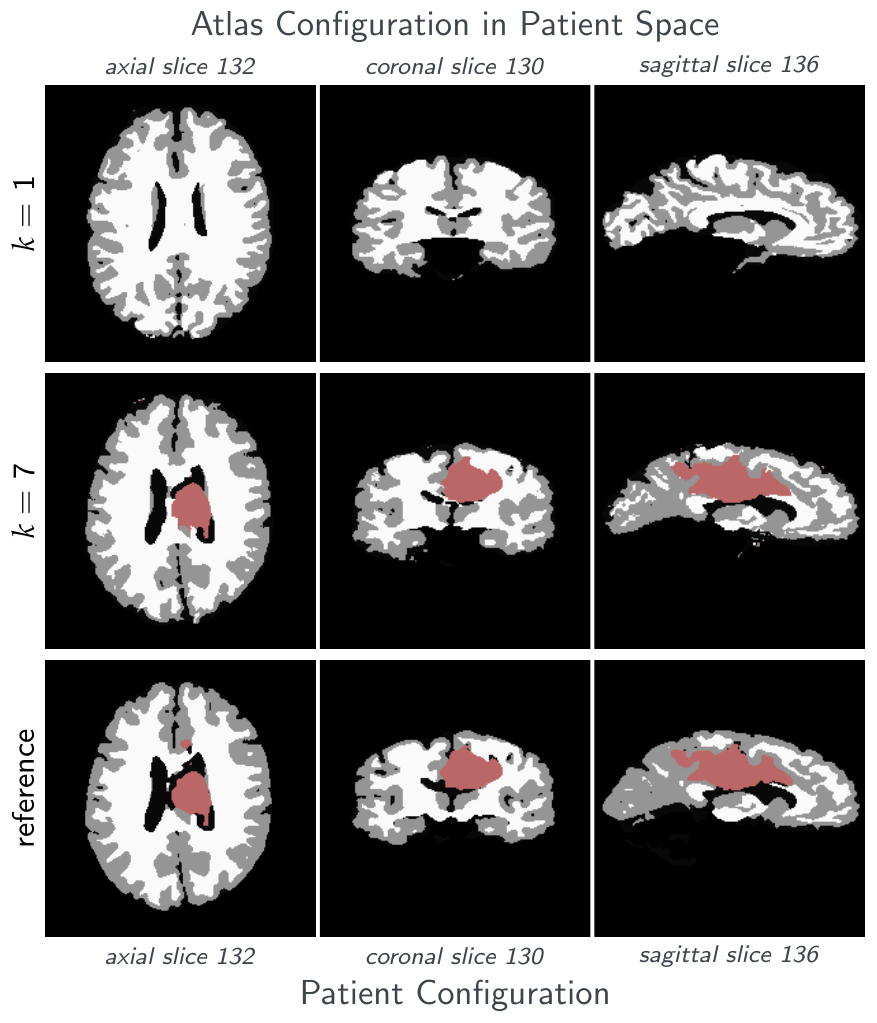}
\end{minipage}
\begin{minipage}{0.49\textwidth}
\textit{AAMP}\\
\includegraphics[scale=.85]{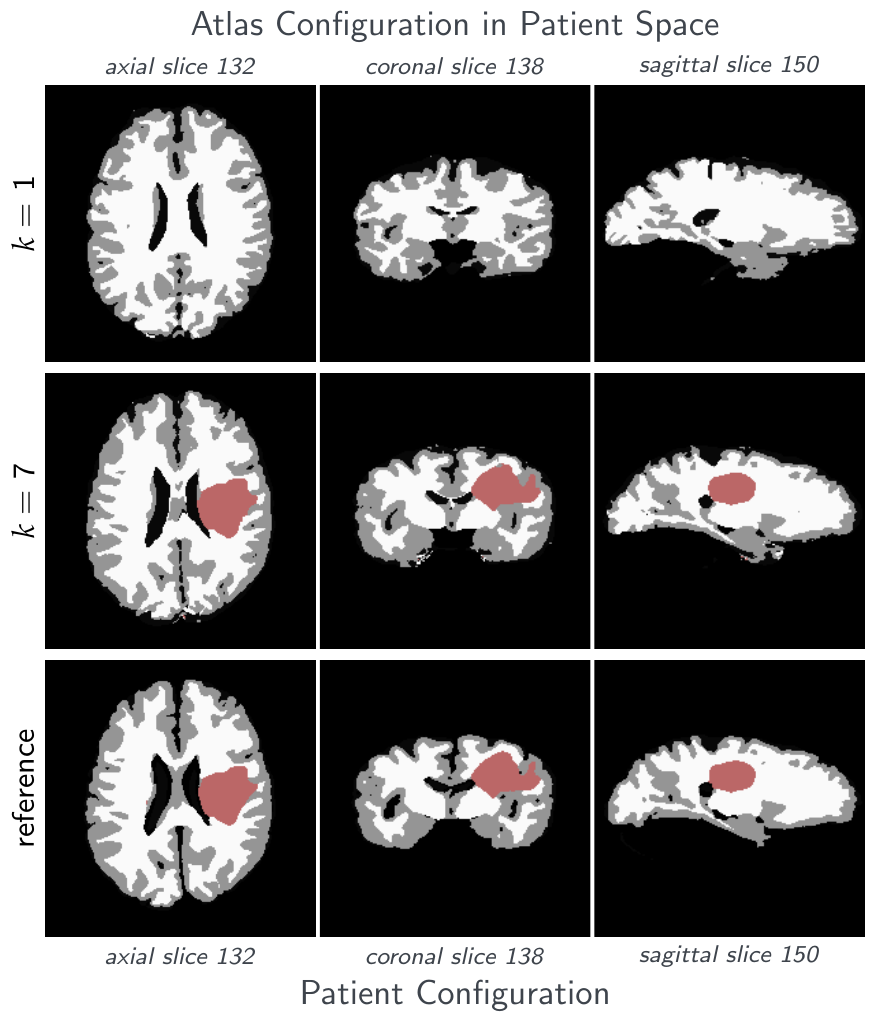}
\end{minipage}\\
\caption{Tumor and brain geometry for the \textbf{real tumor with real velocity (RTRV)} test case;  ground truth ($\rho$ N/A, $\mat{k}$ N/A, $\vect{p}$ N/A, $\vect{v}$ N/A); patients {\bf AAAC, AAMP}.
        We set the parameters for the tumor solver to $\rho_f=15$, $k_f=0$ (reaction-only). We use $n_{\vect{p}}=343$ Gaussians for the inversion. The top row shows the original atlas image. The bottom
        row shows the patient image. The row in the middle shows the solution for our coupled scheme.}
\label{fig:RTRV-6Patients_ax_cor_sag_its_2}
\end{figure*}

\begin{figure*}
\centering
\begin{minipage}{0.49\textwidth}
\textit{AAQD}\\
\includegraphics[scale=.85]{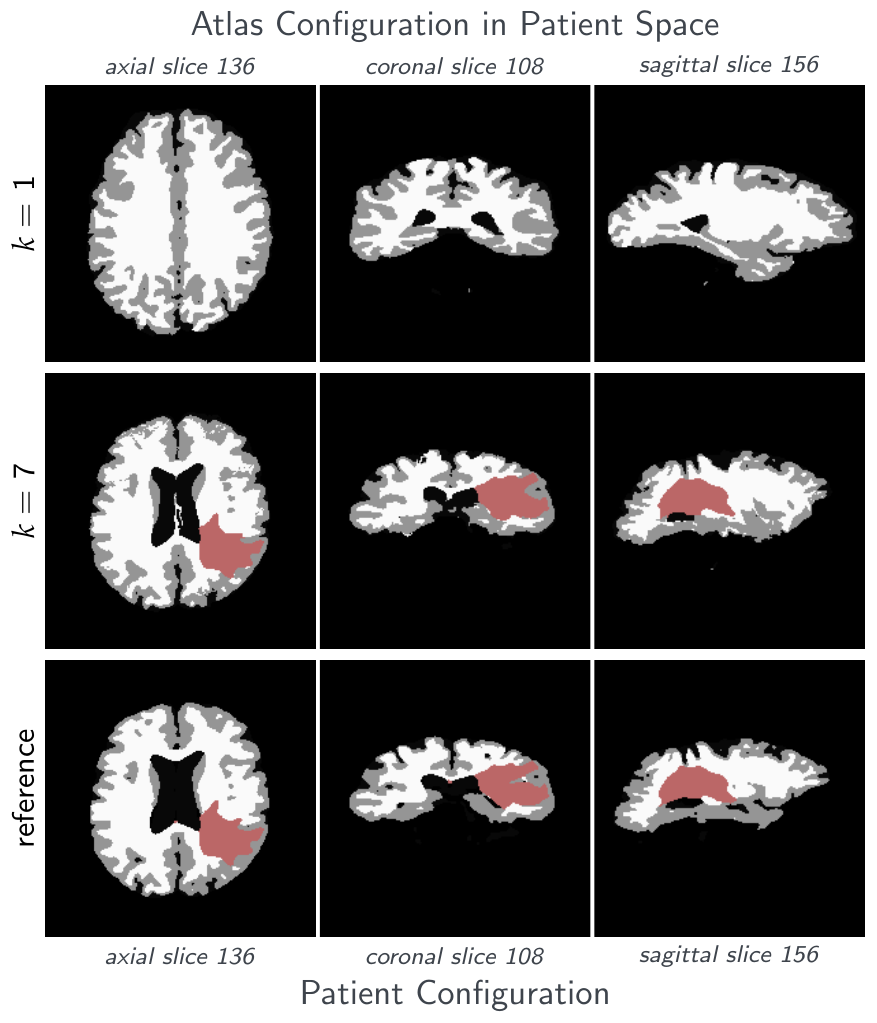}
\end{minipage}
\begin{minipage}{0.49\textwidth}
\textit{AAWI}\\
\includegraphics[scale=.85]{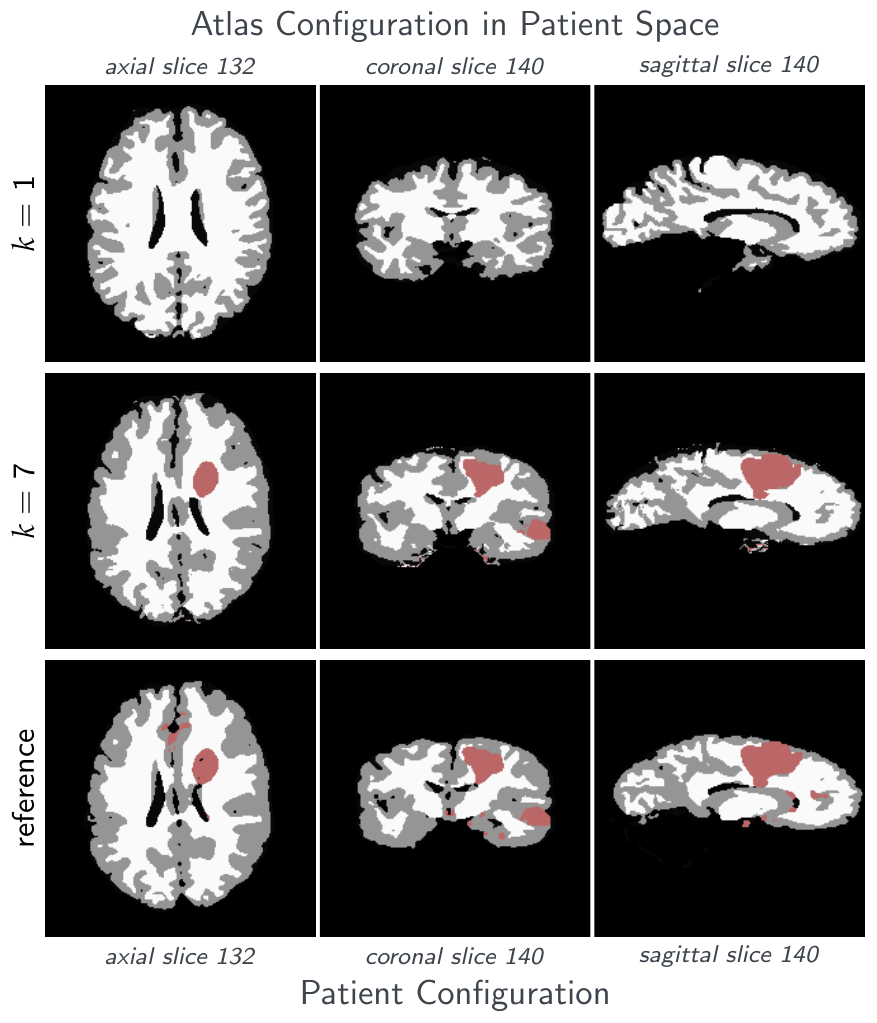}
\end{minipage}\\
\caption{Tumor and brain geometry for the \textbf{real tumor with real velocity (RTRV)} test case;  ground truth ($\rho$ N/A, $\mat{k}$ N/A, $\vect{p}$ N/A, $\vect{v}$ N/A); patients {\bf AAQD, AAWI}.
         We set the parameters for the tumor solver to $\rho_f=15$, $k_f=0$ (reaction-only). We use $n_{\vect{p}}=343$ Gaussians for the inversion. The top row shows the original atlas image. The bottom
         row shows the patient image. The row in the middle shows the solution for our coupled scheme.}
\label{fig:RTRV-6Patients_ax_cor_sag_its_3}
\end{figure*}

\begin{figure*}[ht!]
\centering 
\includegraphics[scale=.85]{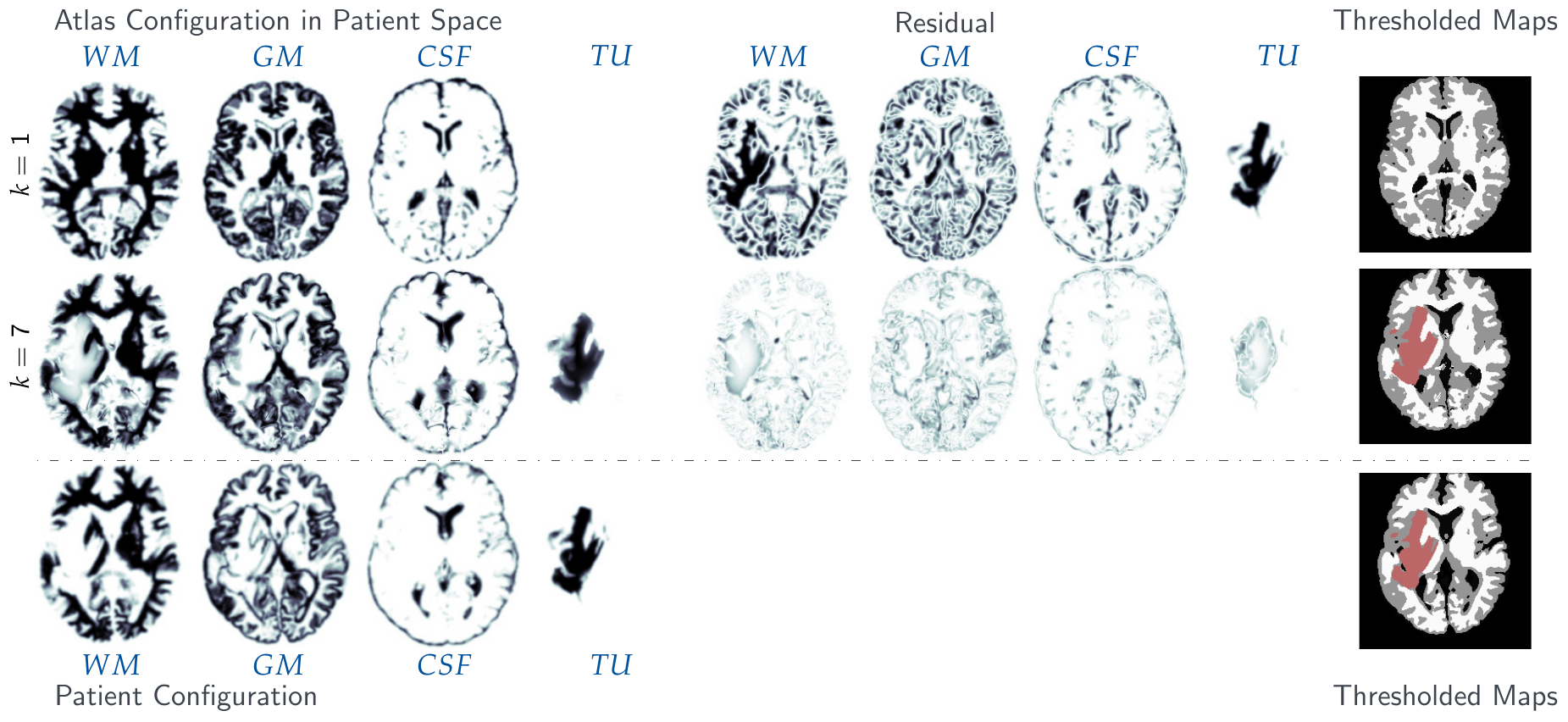}\\

\centering
\begin{small}
\begin{tabular}{rrLlLlllll}
\toprule
iterations & $\beta_{\vect{v}}$ & $\elltwoB$ & $\diceB$ & $\elltwoT$ & $\diceT$ & $\relG$  & $\rtIT$\,[s] & $\rtTUM$\,[\%] & $\rtREG$\,[\%] \\
\midrule
initial &  --  & \num{1.000000e+00} & \num{5.752000e-01} & \num{1.000000e+00} & \num{0.000000e+00} & \num{1.000000e+00} & -- & -- & -- \\
$  1$ &  \num{1e+00}  & \num{9.585000e-01}  & \num{5.945000e-01}  & \num{1.000000e+00}  & \num{0.000000e+00}  & \num{9.987000e-01}  & \num{9.800000e+01}  & \pernum{74.642857}  & \pernum{14.969388}  \\
$  2$ &  \num{1e-01}  & \num{7.424000e-01}  & \num{6.681000e-01}  & \num{3.896000e-01}  & \num{8.431000e-01}  & \num{9.055000e-02}  & \num{4.547000e+01}  & \pernum{51.154607}  & \pernum{47.745766}  \\
$  3$ &  \num{1e-02}  & \num{5.050000e-01}  & \num{7.739000e-01}  & \num{3.132000e-01}  & \num{9.034000e-01}  & \num{4.608000e-02}  & \num{9.586000e+01}  & \pernum{25.756311}  & \pernum{73.722095}  \\
$  4$ &  \num{1e-03}  & \num{3.909000e-01}  & \num{8.227000e-01}  & \num{2.516000e-01}  & \num{9.389000e-01}  & \num{3.685000e-02}  & \num{1.178000e+02}  & \pernum{21.477080}  & \pernum{78.039049}  \\
$  5$ &  \num{1e-04}  & \num{3.599000e-01}  & \num{8.360000e-01}  & \num{2.204000e-01}  & \num{9.513000e-01}  & \num{3.617000e-02}  & \num{1.149000e+02}  & \pernum{21.209748}  & \pernum{78.285466}  \\
$  6$ &  \num{1e-04}  & \num{3.511000e-01}  & \num{8.399000e-01}  & \num{2.051000e-01}  & \num{9.552000e-01}  & \num{3.840000e-02}  & \num{8.994000e+01}  & \pernum{26.995775}  & \pernum{72.437180}  \\
$  7$ &  \num{1e-04}  & \num{3.448000e-01}  & \num{8.425000e-01}  & \num{1.954000e-01}  & \num{9.568000e-01}  & \num{3.921000e-02}  & \num{6.628000e+01}  & --                  & \pernum{99.305975}  \\
\midrule
final &  \num{1e-04}  & \num{3.448000e-01}  & \num{8.425000e-01}  & \num{1.954000e-01}  & \num{9.568000e-01}  & \num{3.921000e-02}  & \num{6.282500e+02}  & \num{1.950500e+02}  & \num{4.345700e+02} \\
\bottomrule
\end{tabular}\\[2mm]
\end{small}
\caption{Results for the \textbf{real tumor/ real velocity (RTRV)} test case, ground truth ($\rho$ N/A, $\mat{k}$ N/A, $\vect{p}$ N/A, $\vect{v}$ N/A) and the \textbf{AAMH} patient.
          The figure shows probability maps for the labels of the healthy atlas brain ($k=1$; top row) and the AAMH patient (target) brain probability maps with tumor (bottom row), along with the reconstructed probability maps, i.e., the final result of our inversion algorithm ($k=7$; middle row) (axial-slice $120$).
          In the table, we report the (summed) mismatch for the brain tissue probability maps ($\elltwoB$) and tumor probability map ($\elltwoT$) in patient space, the mean Dice coefficient for the hard segmentation corresponding to the brain tissue ($\diceB$) and the tumor $\diceT$, respectively, as well as the relative norm of the gradient for the coupled problem \eqref{eq:global_opt} ($\relG$). We also report the run time per iteration in seconds ($\rtIT$), and the percentages ($\rtTUM$) and ($\rtREG$) of this runtime spent in the tumor solver and the image registration solver, respectively. Note that the latter sums up to less than $100\%$ as we do not explicitly measure time spent in additional coupling functionality and in the forward solvers. The last row shows the final state and summed absolute timings for the respective solvers in seconds.}
\label{tab:mismatch_glistr-realdata-nx-256-H1SN-jacbound1E-3_betaIt-1}
\end{figure*}

\ipoint{Results.} \figref{fig:RTRV-6Patients_ax_cor_sag_its_1} -- 
\figref{fig:RTRV-6Patients_ax_cor_sag_its_3} show the healthy atlas ($k=1$) in the top row and the corresponding patient image with tumor in the bottom row for axial, saggital, and coronal orientations. The hard segmentations for the results computed with the proposed approach are shown in the middle row using the same orientations. In~\tabref{tab:mismatch_glistr-realdata-all-patients-nx-256-H1SN-jacbound1E-3_betaIt-1}, we summarize the results for all patient datasets. We report the initial and final values for the mismatch and Dice coefficients associated with the probability maps for the tumor and the brain anatomy, as well as the relative norm of the gradient of the coupled problem in~\eqref{eq:global_opt}. We also report timings for the entire inversion. \figref{tab:mismatch_glistr-realdata-nx-256-H1SN-jacbound1E-3_betaIt-1} shows more detailed images of the probability maps for the patient AAMH (complex and large tumor).
More detailed visual results with information on more iterations of our inversion algorithm for all patients are given in \secref{s:appendix} 
(\figref{fig:appendix:AAAC_axial-over-iterations} through \figref{fig:appendix:AAWI_axial-over-iterations}).
Results for varying reaction and diffusion for AAMH and AAAN are listed in \tabref{tab:mismatch_glistr-realdata-AAMH-AAAN-nx-256-H1SN-jacbound1E-3_betaIt-1_compare_rho}.
Note that all parameter choices refer to the model used for tumor reconstruction in the Picard scheme. The true growth parameters of the tumors are unknown.\\

\begin{table*}
\centering
\caption{Summary of results for the \textbf{real tumor with real velocity (RTRV)} test case,  ground truth
        ($\rho$ N/A, $\mat{k}$ N/A, $\vect{p}$ N/A, $\vect{v}$ N/A); based on real clinical data (taken from \citep{Gooya:2012a}). We set the tumor parameters to $\rho_f=15$, $k_f=0$ (reaction-only). We use a parametrization of tumor initial conditions with $n_{\vect{p}}=343$ Gaussians. We report the (summed) mismatch for the brain tissue probability maps ($\elltwoB$) and tumor probability map ($\elltwoT$) in patient space, the mean Dice coefficient for the hard segmentation corresponding to the brain tissue ($\diceB$) and the tumor ($\diceT$), respectively, as well as the relative norm of the gradient for the coupled problem \eqref{eq:global_opt} ($\relG$). We also report the total run time in seconds ($\rtTOT$), and the run time of the individual components of our Picard scheme, respectively (also in seconds; tumor solver:  $\rtTUM$; image registration: $\rtREG$).  Note that the latter sums up to less than the reported total run time as we do not explicitly measure time spent in additional coupling functionality and forward solvers. We execute our code in parallel on on $11$ nodes using $256$ MPI tasks of HazelHen.}
\label{tab:mismatch_glistr-realdata-all-patients-nx-256-H1SN-jacbound1E-3_betaIt-1}
\begin{small}
\begin{tabular}{rrrLlLlllll}
\toprule
\multicolumn{2}{l}{\textbf{Patient}}  & $\beta_{\vect{v}}$ & $\elltwoB$ & $\diceB$ & $\elltwoT$ & $\diceT$ & $\relG$  & $\rtTOT$\,[s] & $\rtTUM$\,[s] & $\rtREG$\,[s] \\
\midrule
\multirow{ 2}{*}{\textit{AAMH}} &
  initial &  \num{1e+00}  & \num{1.000000e+00} & \num{5.752000e-01} & \num{1.000000e+00} & \num{0.000000e+00} & \num{1.000000e+00} & -- & -- & -- \\
& final   &  \num{1e-04}  & \num{3.448000e-01} & \num{8.425000e-01} & \num{1.954000e-01} & \num{9.568000e-01} & \num{3.921000e-02} & \num{6.282500e+02}  & \num{1.950500e+02}  & \num{4.345700e+02} \\
\midrule
\multirow{ 2}{*}{\textit{AAAN}} &
  initial &  \num{1e+00}  & \num{1.000000e+00}  & \num{5.760000e-01} & \num{1.000000e+00} & \num{0.000000e+00} & \num{1.000000e+00} & -- & -- & -- \\
& final   &  \num{1e-04}  & \num{3.537000e-01}  & \num{8.742000e-01} & \num{3.774000e-01} & \num{8.912000e-01} & \num{1.060000e-01} & \num{6.336300e+02}  & \num{2.154400e+02}  & \num{4.165700e+02} \\
\midrule
\multirow{ 2}{*}{\textit{AAAC}} &
  initial &  \num{1e+00}  & \num{1.000000e+00} & \num{6.094000e-01} & \num{1.000000e+00} & \num{0.000000e+00} & \num{1.000000e+00} & -- & -- & -- \\
& final   &  \num{1e-04}  & \num{3.357000e-01} & \num{8.810000e-01} & \num{2.451000e-01} & \num{9.548000e-01} & \num{4.377000e-02} & \num{4.916900e+02}  & \num{1.591000e+02}  & \num{3.316940e+02} \\
\midrule
\multirow{ 2}{*}{\textit{AAMP}} &
  initial &  \num{1e+00}  & \num{1.000000e+00} & \num{6.037000e-01} & \num{1.000000e+00} & \num{0.000000e+00} & \num{1.000000e+00} & -- & -- & -- \\
& final   &  \num{1e-04}  & \num{3.319000e-01} & \num{8.524000e-01} & \num{1.338000e-01} & \num{9.753000e-01} & \num{3.313000e-02} & \num{6.479100e+02}  & \num{2.359100e+02}  & \num{4.132900e+02} \\
\midrule
\multirow{ 2}{*}{\textit{AAQD}} &
  initial &  \num{1.0e+00} & \num{1.000000e+00} & \num{4.652000e-01} & \num{1.000000e+00} & \num{0.000000e+00} & \num{1.000000e+00} & -- & -- & -- \\
& final   &  \num{1e-04}   & \num{3.241000e-01} & \num{8.506000e-01} & \num{2.549000e-01} & \num{9.306000e-01} & \num{9.884000e-02} & \num{1.100260e+03}  & \num{2.188800e+02}  & \num{8.811100e+02} \\
\midrule
\multirow{ 2}{*}{\textit{AAWI}} &
  initial &  \num{1e+00}  & \num{1.000000e+00} & \num{5.878000e-01} & \num{1.000000e+00} & \num{0.000000e+00} & \num{1.000000e+00} & -- & -- & -- \\
& final   &  \num{1e-04}  & \num{3.453000e-01} & \num{8.384000e-01} & \num{3.916000e-01} & \num{8.737000e-01} & \num{5.262000e-02} & \num{6.838300e+02}  & \num{2.924400e+02}  & \num{3.896800e+02} \\
\bottomrule
\end{tabular}
\end{small}
\end{table*}


\begin{table*}[ht!]\centering
  \caption{Results for the \textbf{AAMH} and the \textbf{AAAN} patient \textbf{real tumor/ real velocity (RTRV)} test case, ground truth
          ($\rho$ N/A, $\mat{k}$ N/A, $\vect{p}$ N/A, $\vect{v}$ N/A) using real clinical input data.
          The table shows the (summed) mismatch for the brain tissue probability maps ($\elltwoB$) and tumor probability map ($\elltwoT$) in patient space, the mean Dice coefficient for the hard segmentation corresponding to the brain tissue ($\diceB$) and the tumor ($\diceT$), respectively for different values for reaction scaling parameter $\rho_f$ and of the diffusion coefficient scaling parameter $k_f$., as well as the relative norm of the gradient for the coupled problem \eqref{eq:global_opt} ($\relG$). We also report the total run time in seconds ($\rtTOT$), and the run time of the individual components of our Picard scheme, respectively (also in seconds; tumor solver:  $\rtTUM$; image registration: $\rtREG$).  Note that the latter sums up to less than the reported total run time as we do not explicitly measure time spent in additional coupling functionality and forward solvers. We execute our code in parallel on on $11$ nodes using $256$ MPI tasks of HazelHen.
          }
\label{tab:mismatch_glistr-realdata-AAMH-AAAN-nx-256-H1SN-jacbound1E-3_betaIt-1_compare_rho}
\begin{footnotesize}
\begin{tabular}{lllrrLlLlllll}
\toprule \multicolumn{11}{l}{\textit{AAMH}}\\
\midrule
$\rho_f$ & $k_f$ & $n_{\vect{p}}$ &  It & $\beta_{\vect{v}}$ & $\elltwoB$ & $\diceB$ & $\elltwoT$ & $\diceT$ & $\relG$  & $\rtTOT$\,[s] & $\rtTUM$\,[s] & $\rtREG$\,[s] \\
\midrule
\multirow{ 2}{*}{$0$} & \multirow{ 2}{*}{$0$} & \multirow{ 2}{*}{$343$} &
    initial &  \num{1e+00}  & \num{1.000000e+00}  & \num{5.752000e-01}  & \num{1.000000e+00}  & \num{0.000000e+00}  & \num{1.000000e+00} & -- & -- & -- \\
& & & final &  \num{1e-04}  & \num{3.396000e-01}  & \num{8.433000e-01}  & \num{1.467000e-01}  & \num{9.688000e-01}  & \num{1.187000e-01}  & \num{6.089700e+02}  & \num{2.129200e+02}  & \num{3.633000e+02} \\
\midrule
\multirow{ 2}{*}{$0$} & \multirow{ 2}{*}{$0$} & \multirow{ 2}{*}{$125$} &
    initial &  \num{1e+00}  & \num{1.000000e+00}  & \num{5.752000e-01}  & \num{1.000000e+00} & \num{0.000000e+00}   & \num{1.000000e+00}  & -- & -- & -- \\
& & & final &  \num{1e-04}  & \num{3.381000e-01}  & \num{8.452000e-01}  & \num{1.734000e-01}  & \num{9.624000e-01}  & \num{1.306000e-01}  & \num{5.091200e+02}  & \num{7.856200e+01}  & \num{4.039080e+02} \\
\midrule
\multirow{ 2}{*}{$15$} & \multirow{ 2}{*}{$0$} & \multirow{ 2}{*}{$343$} &
    initial &  \num{1e+00}  & \num{1.000000e+00}  & \num{5.752000e-01}  & \num{1.000000e+00}  & \num{0.000000e+00}  & \num{1.000000e+00}  & -- & -- & -- \\
& & & final &  \num{1e-04}  & \num{3.448000e-01}  & \num{8.425000e-01}  & \num{1.954000e-01}  & \num{9.568000e-01}  & \num{3.921000e-02}  & \num{6.282500e+02}  & \num{1.950500e+02}  & \num{4.345700e+02} \\
\midrule
\multirow{ 2}{*}{$15$} & \multirow{ 2}{*}{$0$} & \multirow{ 2}{*}{$125$} &
    initial &  \num{1e+00}  & \num{1.000000e+00}  & \num{5.752000e-01}  & \num{1.000000e+00}  & \num{0.000000e+00}  & \num{1.000000e+00}  & -- & -- & -- \\
& & & final &  \num{1e-04}  & \num{3.503000e-01}  & \num{8.429000e-01}  & \num{2.364000e-01}  & \num{9.527000e-01}  & \num{4.828000e-02}  & \num{4.469100e+02}  & \num{7.198300e+01}  & \num{3.479780e+02} \\
\midrule
\multirow{ 2}{*}{$5$} & \multirow{ 2}{*}{$\num{1E-2}$} & \multirow{ 2}{*}{$343$} &
    initial &  \num{1e+00}  & \num{1.000000e+00}  & \num{5.752000e-01}  & \num{1.000000e+00}  & \num{0.000000e+00}  & \num{1.000000e+00}  & -- & -- & -- \\
& & & final &  \num{1e-04}  & \num{3.435000e-01}  & \num{8.432000e-01}  & \num{1.882000e-01}  & \num{9.638000e-01}  & \num{7.679000e-02}  & \num{1.229994e+04}  & \num{1.180600e+04}  & \num{3.406200e+02} \\
\midrule
\multirow{ 2}{*}{$10$} & \multirow{ 2}{*}{$\num{1E-2}$} & \multirow{ 2}{*}{$343$} &
    initial &  \num{1e+00}  & \num{1.000000e+00}  & \num{5.752000e-01}  & \num{1.000000e+00}  & \num{0.000000e+00}  & \num{1.000000e+00}  & -- & -- & -- \\
& & & final &  \num{1e-04}  & \num{3.464000e-01}  & \num{8.433000e-01}  & \num{2.065000e-01}  & \num{9.624000e-01}  & \num{5.158000e-02}  & \num{1.327015e+04}  & \num{1.277270e+04}  & \num{3.438080e+02} \\
\midrule
\multirow{ 2}{*}{$15$} & \multirow{ 2}{*}{$\num{1E-2}$} & \multirow{ 2}{*}{$343$} &
    initial &  \num{1e+00}  & \num{1.000000e+00}  & \num{5.752000e-01}  & \num{1.000000e+00}  & \num{0.000000e+00}  & \num{1.000000e+00}  & -- & -- & -- \\
& & & final &  \num{1e-04}  & \num{3.479000e-01}  & \num{8.442000e-01}  & \num{2.346000e-01}  & \num{9.593000e-01}  & \num{3.466000e-02}  & \num{1.180995e+04}  & \num{1.129390e+04}  & \num{3.632520e+02} \\
\toprule \multicolumn{11}{l}{\textit{AAAN}}\\
\midrule
$\rho_f$ & $k_f$ & $n_{\vect{p}}$ &   It & $\beta_{\vect{v}}$ & $\elltwoB$ & $\diceB$ & $\elltwoT$ & $\diceT$ & $\relG$  & $\rtTOT$\,[s] & $\rtTUM$\,[s] & $\rtREG$\,[s] \\
\midrule
\multirow{ 2}{*}{$0$} & \multirow{ 2}{*}{$0$}  & \multirow{ 2}{*}{$343$} &
    initial &  \num{1e+00}  & \num{1.000000e+00} & \num{5.760000e-01}  & \num{1.000000e+00}  & \num{0.000000e+00}  & \num{1.000000e+00}  & -- & -- & -- \\
& & & final &  \num{1e-04}  & \num{3.419000e-01} & \num{8.785000e-01}  & \num{2.559000e-01}  & \num{9.355000e-01}  & \num{2.194000e-01}  & \num{6.576600e+02}  & \num{2.017600e+02}  & \num{4.226100e+02} \\
\midrule
\multirow{ 2}{*}{$0$} & \multirow{ 2}{*}{$0$}  & \multirow{ 2}{*}{$125$} &
    initial &  \num{1e+00}  & \num{1.000000e+00} & \num{5.760000e-01 } & \num{1.000000e+00}  & \num{0.000000e+00}  & \num{1.000000e+00}  & -- & -- & -- \\
& & & final &  \num{1e-04}  & \num{3.558000e-01} & \num{8.741000e-01}  & \num{4.396000e-01}  & \num{6.678000e-01}  & \num{2.364000e-01}  & \num{4.538200e+02}  & \num{8.890000e+01}  & \num{3.380220e+02} \\
\midrule
\multirow{ 2}{*}{$15$} & \multirow{ 2}{*}{$0$} & \multirow{ 2}{*}{$343$} &
    initial &  \num{1e+00}  & \num{1.000000e+00} & \num{5.760000e-01}  & \num{1.000000e+00}  & \num{0.000000e+00}  & \num{1.000000e+00}  & -- & -- & -- \\
& & & final &  \num{1e-04}  & \num{3.537000e-01} & \num{8.742000e-01}  & \num{3.774000e-01}  & \num{8.912000e-01}  & \num{1.060000e-01}  & \num{6.336300e+02}  & \num{2.154400e+02}  & \num{4.165700e+02} \\
\midrule
\multirow{ 2}{*}{$15$} & \multirow{ 2}{*}{$0$} & \multirow{ 2}{*}{$125$} &
   initial  &  \num{1e+00}  & \num{1.000000e+00} & \num{5.760000e-01}  & \num{1.000000e+00}  & \num{0.000000e+00}  & \num{1.000000e+00}  & -- & -- & -- \\
& & & final &  \num{1e-04}  & \num{3.672000e-01} & \num{8.719000e-01}  & \num{5.711000e-01}  & \num{3.440000e-01}  & \num{1.277000e-01}  & \num{4.673000e+02}  & \num{8.375700e+01}  & \num{3.570700e+02} \\
\midrule
\multirow{ 2}{*}{$5$} & \multirow{ 2}{*}{$\num{1E-2}$} & \multirow{ 2}{*}{$343$} &
    initial &  \num{1e+00}  & \num{1.000000e+00}  & \num{5.760000e-01}  & \num{1.000000e+00}  & \num{0.000000e+00}  & \num{1.000000e+00}  & -- & -- & -- \\
& & & final &  \num{1e-04}  & \num{3.499000e-01}  & \num{8.759000e-01}  & \num{3.170000e-01}  & \num{9.062000e-01}  & \num{1.786000e-01}  & \num{1.816822e+04}  & \num{1.768400e+04}  & \num{3.327960e+02} \\
\midrule
\multirow{ 2}{*}{$10$} & \multirow{ 2}{*}{$\num{1E-2}$} & \multirow{ 2}{*}{$343$} &
    initial &  \num{1e+00}  & \num{1.000000e+00}  & \num{5.760000e-01}  & \num{1.000000e+00}  & \num{0.000000e+00}  & \num{1.000000e+00}  & -- & -- & -- \\
& & & final &  \num{1e-04}  & \num{3.521000e-01}  & \num{8.758000e-01}  & \num{3.596000e-01}  & \num{8.846000e-01}  & \num{1.411000e-01}  & \num{1.875256e+04}  & \num{1.824900e+04}  & \num{3.405200e+02} \\
\midrule
\multirow{ 2}{*}{$15$} & \multirow{ 2}{*}{$\num{1E-2}$} & \multirow{ 2}{*}{$343$} &
    initial &  \num{1e+00}  & \num{1.000000e+00}  & \num{5.760000e-01}  & \num{1.000000e+00}  & \num{0.000000e+00}  & \num{1.000000e+00}  & -- & -- & -- \\
& & & final &  \num{1e-04}  & \num{3.552000e-01}  & \num{8.751000e-01}  & \num{3.975000e-01}  & \num{8.615000e-01}  & \num{1.125000e-01}  & \num{1.849745e+04}  & \num{1.802100e+04}  & \num{3.278940e+02} \\
\bottomrule
\end{tabular}
\end{footnotesize}
\end{table*}

\ipoint{Observations.} The most important observation is that we obtain very good registration results---qualitatively and quantitatively---in what is an extremely challenging registration problem. From visual inspection of the data alone (\figref{fig:RTRV-6Patients_ax_cor_sag_its_1} through \figref{fig:RTRV-6Patients_ax_cor_sag_its_3}) we can immediately see that there are significant anatomical differences between the atlas image and the patient images, and accross patients. The tumors vary significantly in shape and size. Overall, this poses considerable challenges to our framework. The results reported in \figref{fig:RTRV-6Patients_ax_cor_sag_its_1} through \figref{fig:RTRV-6Patients_ax_cor_sag_its_3} and in \figref{tab:mismatch_glistr-realdata-nx-256-H1SN-jacbound1E-3_betaIt-1} clearly demonstrate that the deformed atlases are in very good agreement with the patient data for all six subjects. We reach Dice coefficients between \num{8.384000e-01} to \num{8.810000e-01}  and \num{8.737000e-01} to \num{9.753000e-01} for the probability maps associated with the anatomy and the tumor. These results are slightly worse than those obtained for the artificially grown tumors in the former sections, but still competitive. We also note that the initial Dice coefficients for the anatomy range between \num{4.652000e-01} and \num{6.037000e-01} for these data, which is drastically worse than what we have seen in our synthetic test cases. The runtimes are compareable to our former experiment. We again achieve a reduction of the relative norm of the gradient for the coupled problem in \eqref{eq:global_opt} of about two orders of magnitude (slightly less than what we saw before). The results of our study with varying tumor
model parameters presented in \tabref{tab:mismatch_glistr-realdata-AAMH-AAAN-nx-256-H1SN-jacbound1E-3_betaIt-1_compare_rho}
show that there are slight variances in the results depending on the parameter choices,
but the Picard iteration scheme is successful in all cases. Further enhancements of our environment are required to enable the
identification of the 'correct' tumor growth parametrization.\\

\ipoint{Conclusions:} We have tested our formulation on real data that pose significant challenges due to large inter-subject anatomical variability and a strong variation in the appearance of the tumor, in shape, size, location and growth behaviour. We use a very simple model that only accounts for logistic growth. This, in combination with a flexible, high-dimensional parameterization of the initial condition allows us to overcome these challenges. We achieve extremely promising registration accuracies with a Dice score of up to \num{8.810000e-01} and \num{9.753000e-01} for the label maps associated with the probability maps of the brain anatomy and the tumor in what is an extremely challenging problem. Runs with variation of the growth parameter $\rho_f$ show, on the one hand, that it is important to identify the correct parameters to achieve optimal quality of tumor reconstruction. We, therefore, believe that it will be possible to identify physical tumor growth parameters from our coupled solver if the tumor
model is enhanced (anisotropic diffusion, mass effect, \ldots) and we consider 'correct' time horizons and/or restrict the
initial conditions of tumor to points seeds (see first steps following this idea in ATAV-LD). On the other hand, the results for varying
model parameters show the robustness of our Picard scheme with respect to the model and parameter choice.
Furthermore, the reconstruction results for the AAAN patient data show that we can reconstruct multifocal tumors with comparable quality and computational costs.

%% file: 06-conclusion.tex
We have presented a new method for the registration of images of patients diagnosed with mono- or multi-focal brain tumors to a common reference atlas. Application scenarios are  biophysical model calibration and  image registration.
Our method combines stand-alone forward and inverse tumor~\citep{Gholami:2016a} and diffeomorphic registration~\citep{Mang:2015a,Mang:2016a,Mang:2016c,Mang:2017b} by means of an efficient coupling scheme based on a Picard iteration. It allows us to exploit available, tailored implementations for the solution of the subproblems~\citep{Mang:2016c,Gholami:2017a}, while monitoring convergence using the coupled gradient information. 
We invert simultaneously for the control variables of both problems---a parameterization of the initial condition for the tumor model, and a smooth velocity field to capture the inter-subject variability of brain anatomy.
Here is what we have learned from our experiments on synthetic and real data:
\begin{enumerate}[(i)]
\item Despite the fact that our scheme neglects coupling terms that appear in our coupled optimization problem, we could experimentally show that it efficiently reduces the coupled gradient. A convergence proof of the Picard scheme to a local minimum is beyond the scope of this paper and remains subject to future work.
\item We could demonstrate that our parameterization of the initial condition allows us to generate high-fidelity registrations irrespective of the complexity of the data or the model used for the tumor simulations. If we reduce the number of basis functions used for initial condition parametrization, this is no longer true. 
\item In studies with various models from a pure interpolation with the basis functions that parameterize the initial condition (zero reaction and diffusion coefficient) over a reaction-only, to a full reaction-diffusion model, we could show in our synthetic cases, that
we get the highest accuracy in tumor reconstruction, if we use the correct model. This implies that our framework could eventually serve as a powerful tool for model selection. A rigorous verification of this claim requires significantly more work and remains for the future.
\item Overall, our numerical study, which includes real brain images with real tumors, shows that we can achieve high-fidelity results with an overall low mismatch and high Dice score in particular for the simulated and observed tumor, with Dice coefficients ranging from 93\% for real tumors above a critical size, and up to 97\% for artificially grown tumors in a real brain geometry.
\end{enumerate}
Let us emphasize that our tumor model is currently not sophisticated enough to allow proper parameter identification or tumor growth prediction, but the tumor-registration coupling approach that we present in this work and for which we can show good computational performance and high accuracy for various real brain data test cases lays the basis for further developments with improved tumor solvers and, finally, a parameter identification and growth
prediction tool. The next steps for SIBIA are to improve the biophysical  tumor-growth model by first adding mass-effect and second higher-fidelity tumor growth models.

%% file: 07-appendix.tex
\subsection{Supplementary Material for the ATAV Testcase}
We present additional qualitative data for the ATAV-DIF testcase and the ATAV-LD testcase.
\begin{figure}[htb]
\centering
\includegraphics[width=0.48\textwidth]{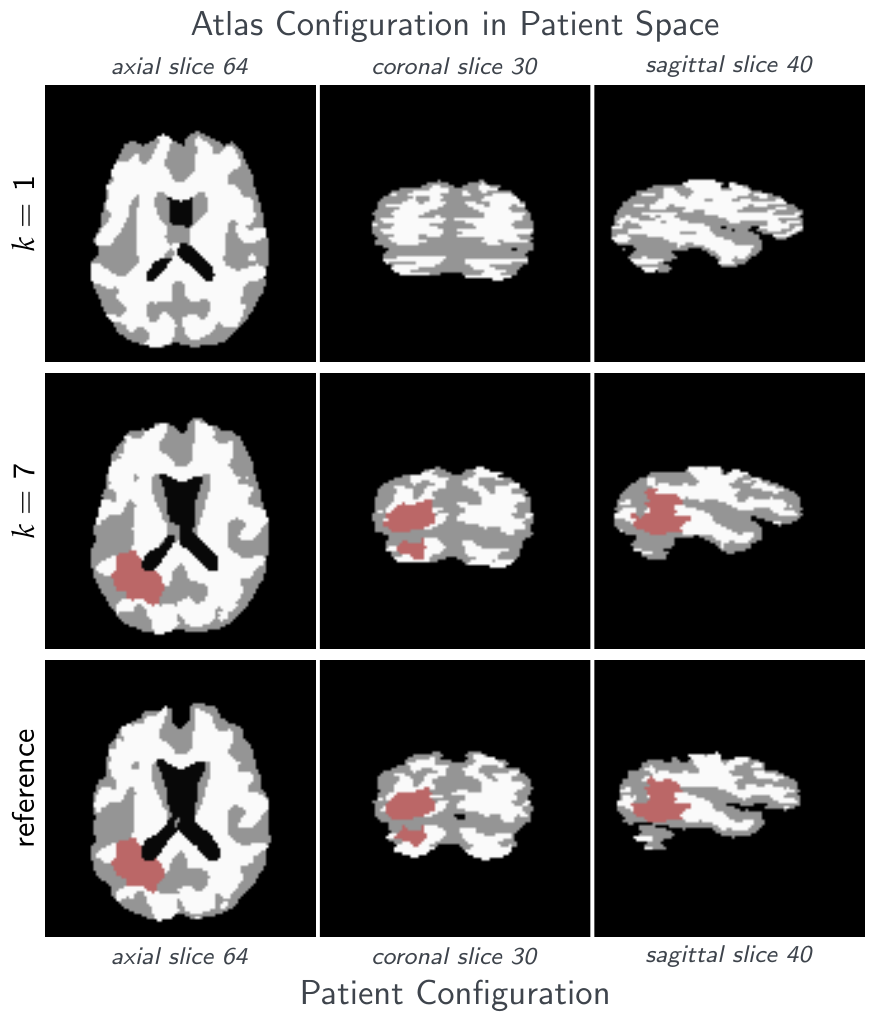}
\caption{Tumor and brain labels (obtained by thresholding probability maps) for the \textbf{analytic tumor with analythic velocity with non-zero diffusion (ATAV-DIF)} test case; ground truth: ($\rho_f=10$, $\rho_w=1$, $\rho_g=0$, $k_f=\num{1.0E-2}$, $k_w=1$, $k_g=0$, $\vect{p}=\vect{p}^\star$, $\vect{v}=-\vect{v}^\star$). We show results for the inversion (velocity and initial condition) if we use the true (`'correct`') tumor parameters used to generate the test case (i.e., $\rho_f=10$ and $k_f=\num{1.0E-2}$). The top row shows the initial label maps for the atlas image. The middle row shows the label maps for the atlas image after registration (transported to the patient space) and the bottom row shows the label maps for the patient data. We can see that the results are qualitatively in excellent agreement. This is confirmed by the values for the mismatch and Dice coefficients for the labels and probability maps for the individual tissue classes reported in \tabref{tab:ATAV-nonzero-diffusion}.}
\label{fig:ATAV-nonzero-diffusion}
\end{figure}

\begin{figure}[htb]
\centering
\includegraphics[width=0.48\textwidth]{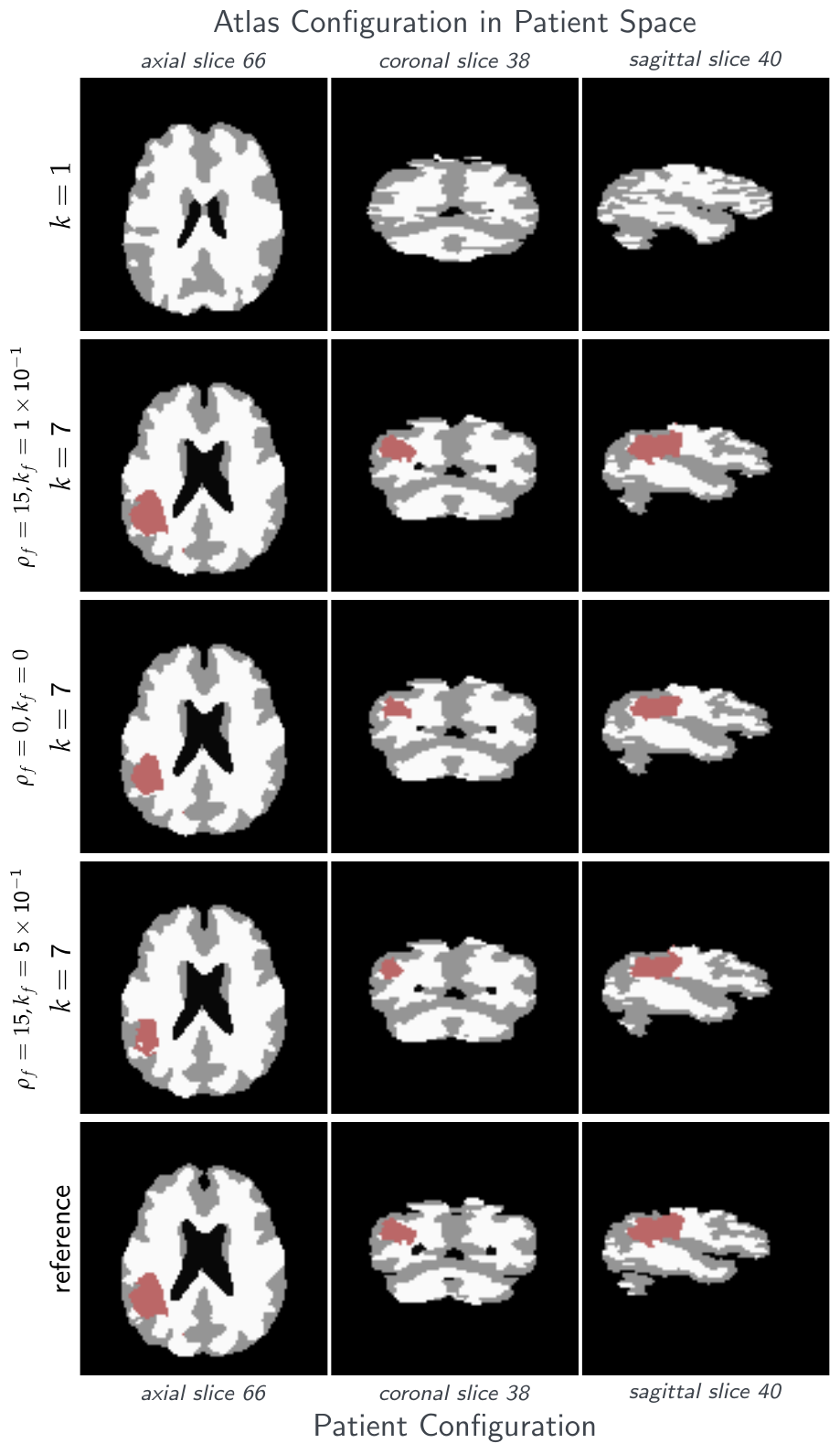}
\caption{Tumor and brain labels for the \textbf{analythic tumor with analythic velocity with non-zero diffusion and low-dimensional initial condition (ATAV-LD)}
test case; ground truth: ($\rho_f=15$, $\rho_w=1$, $\rho_g=0$, $k_f=\num{1.0E-1}$, $k_w=1$, $k_g=0$, $\vect{p}=\vect{p}^\star$ (in patient domain), $\vect{v}$ N/A); for the 'correct' tumor parameters $\rho_f=15$, $k_f=\num{1.0E-1}$, and the two different settings $\rho_f=15$, $k_f=\num{5.0E-1}$ and $\rho_f=k_f=0$.}
\label{fig:ATAV-nonzero-diffusion-sparse}
\end{figure}
\FloatBarrier

\subsection{Supplementary Material for the RTRV Testcase}
\begin{figure*}[ht!]
\includegraphics[scale=0.85]{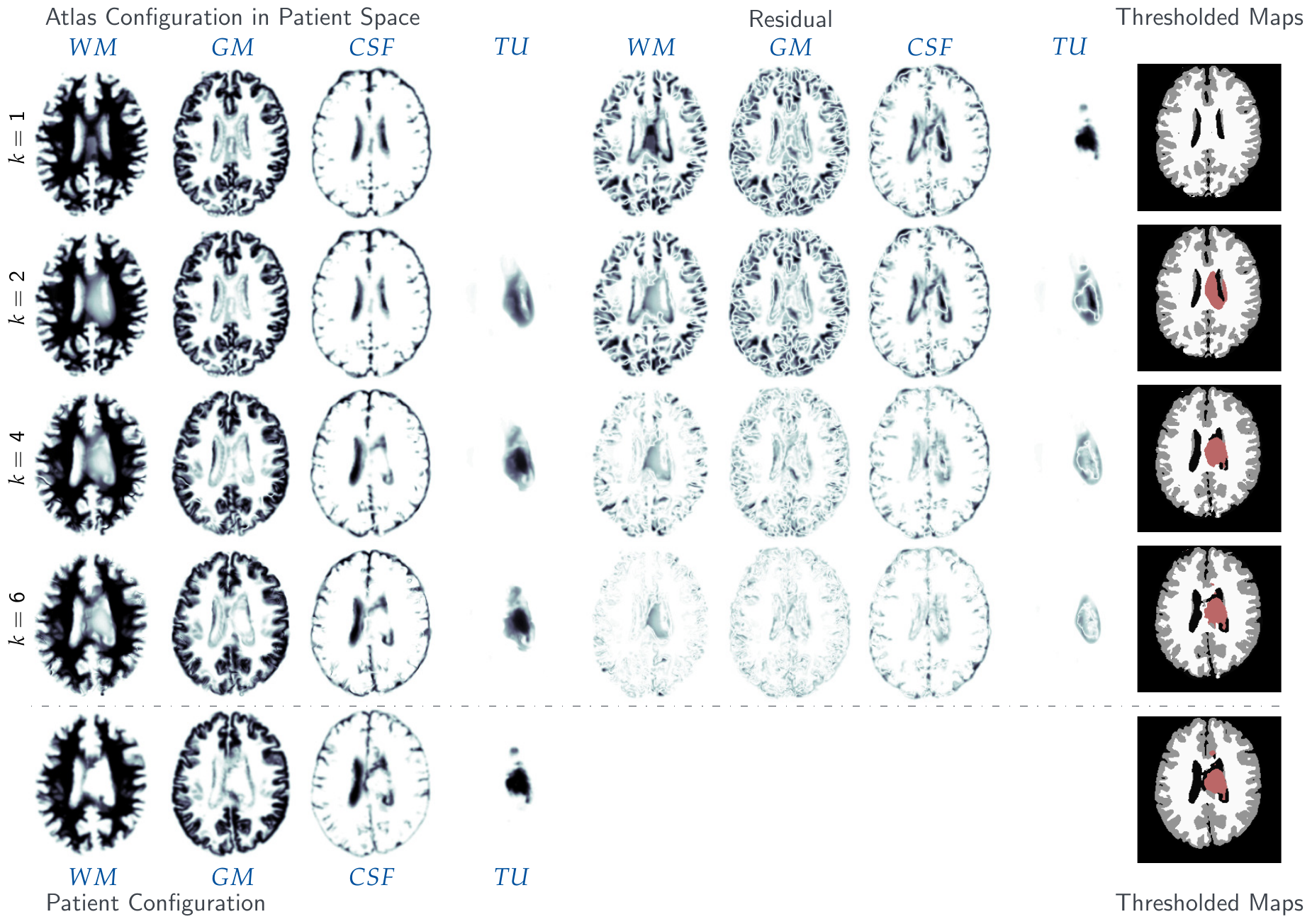}
\caption{Results for the \textbf{real tumor/ real velocity (RTRV)} test case, ground truth ($\rho$ N/A, $\mat{k}$ N/A, $\vect{p}$ N/A, $\vect{v}$ N/A) and the \textbf{AAAC} patient.
         The figure shows probability maps for the labels of the healthy atlas brain ($k=1$; top row) and the AAAN patient (target) brain probability maps with tumor (bottom row),
         along with the reconstructed probability maps throughout the Picard iterations ($k=2,4,6$) (axial-slice $132$).}
         \label{fig:appendix:AAAC_axial-over-iterations}
\end{figure*}

\begin{figure*}[ht!]
\includegraphics[scale=0.85]{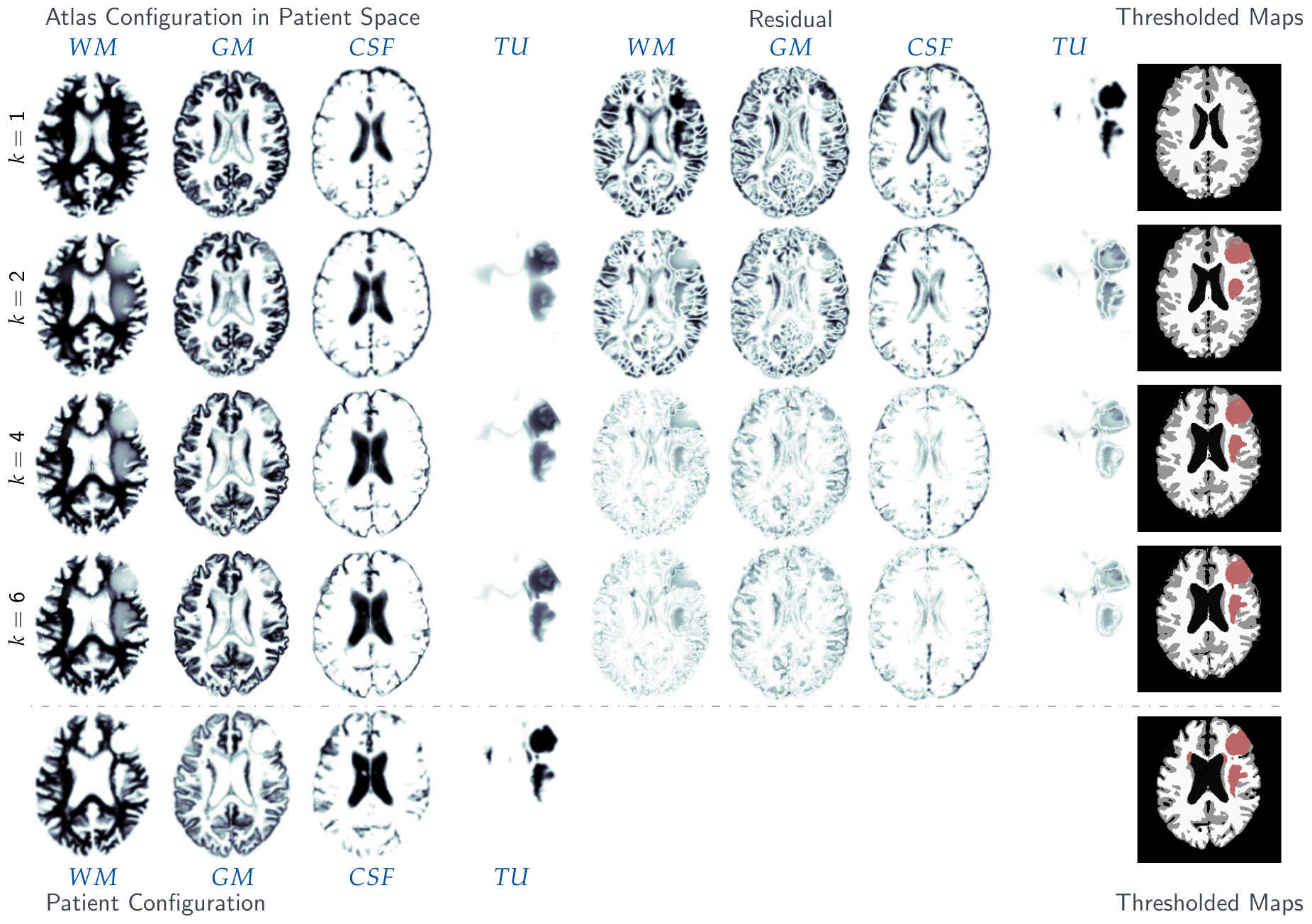}
\caption{Results for the \textbf{real tumor/ real velocity (RTRV)} test case, ground truth ($\rho$ N/A, $\mat{k}$ N/A, $\vect{p}$ N/A, $\vect{v}$ N/A) and the \textbf{AAAN} patient.
         The figure shows probability maps for the labels of the healthy atlas brain ($k=1$; top row) and the AAAN patient (target) brain probability maps with tumor (bottom row),
         along with the reconstructed probability maps throughout the Picard iterations ($k=2,4,6$) (axial-slice $132$).}
         \label{fig:appendix:AAAN_axial-over-iterations}
\end{figure*}

\begin{figure*}[ht!]
\includegraphics[scale=0.85]{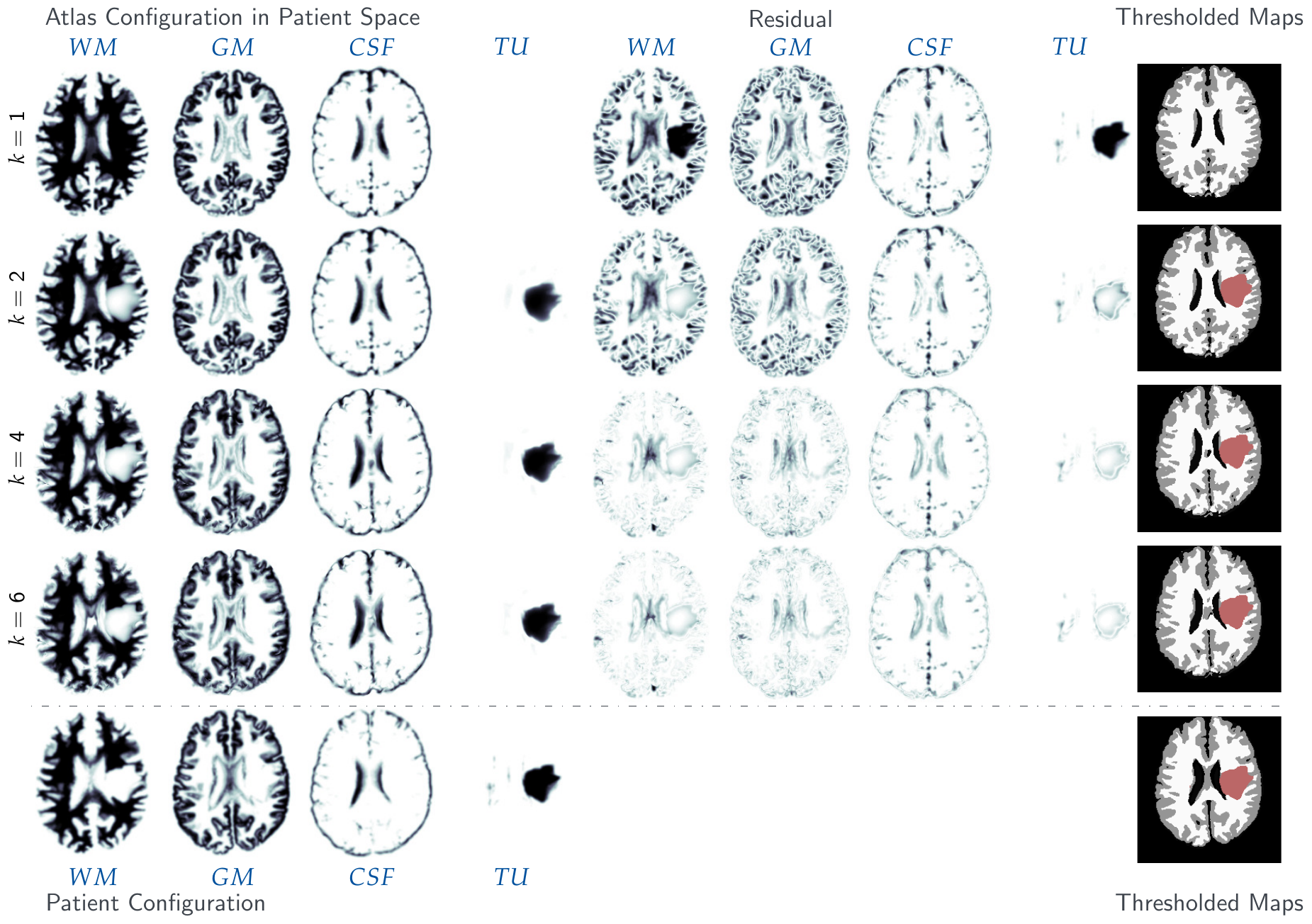}
\caption{Results for the \textbf{real tumor/ real velocity (RTRV)} test case, ground truth ($\rho$ N/A, $\mat{k}$ N/A, $\vect{p}$ N/A, $\vect{v}$ N/A) and the \textbf{AAMP} patient.
         The figure shows probability maps for the labels of the healthy atlas brain ($k=1$; top row) and the AAAN patient (target) brain probability maps with tumor (bottom row),
         along with the reconstructed probability maps throughout the Picard iterations ($k=2,4,6$) (axial-slice $132$).}
         \label{fig:appendix:AAMP_axial-over-iterations}
\end{figure*}

\begin{figure*}[ht!]
\includegraphics[scale=0.85]{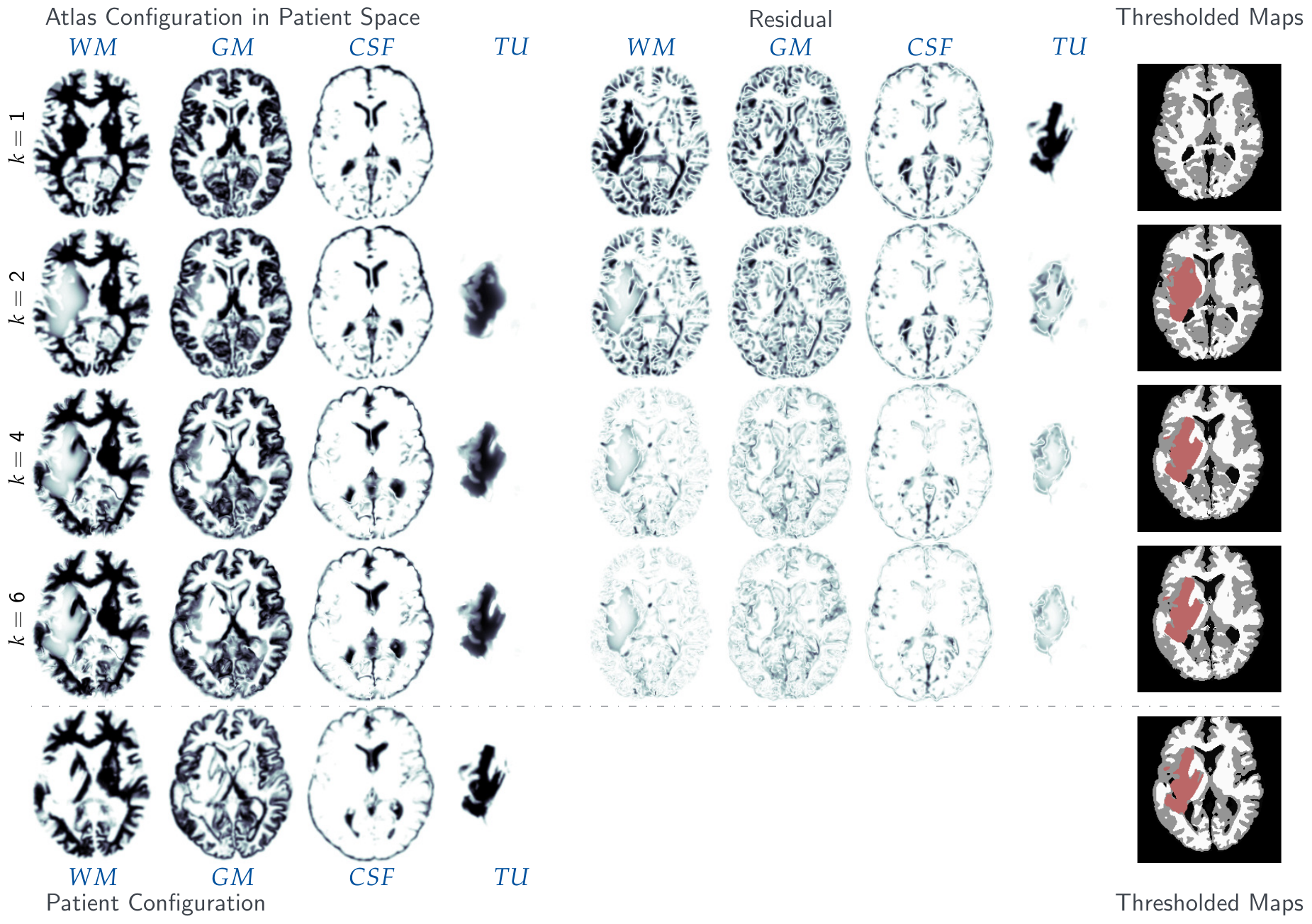}
\caption{Results for the \textbf{real tumor/ real velocity (RTRV)} test case, ground truth ($\rho$ N/A, $\mat{k}$ N/A, $\vect{p}$ N/A, $\vect{v}$ N/A) and the \textbf{AAMH} patient.
         The figure shows probability maps for the labels of the healthy atlas brain ($k=1$; top row) and the AAAN patient (target) brain probability maps with tumor (bottom row),
         along with the reconstructed probability maps throughout the Picard iterations ($k=2,4,6$) (axial-slice $120$).}
         \label{fig:appendix:AAMH_axial-over-iterations}
\end{figure*}

\begin{figure*}[ht!]
\includegraphics[scale=0.85]{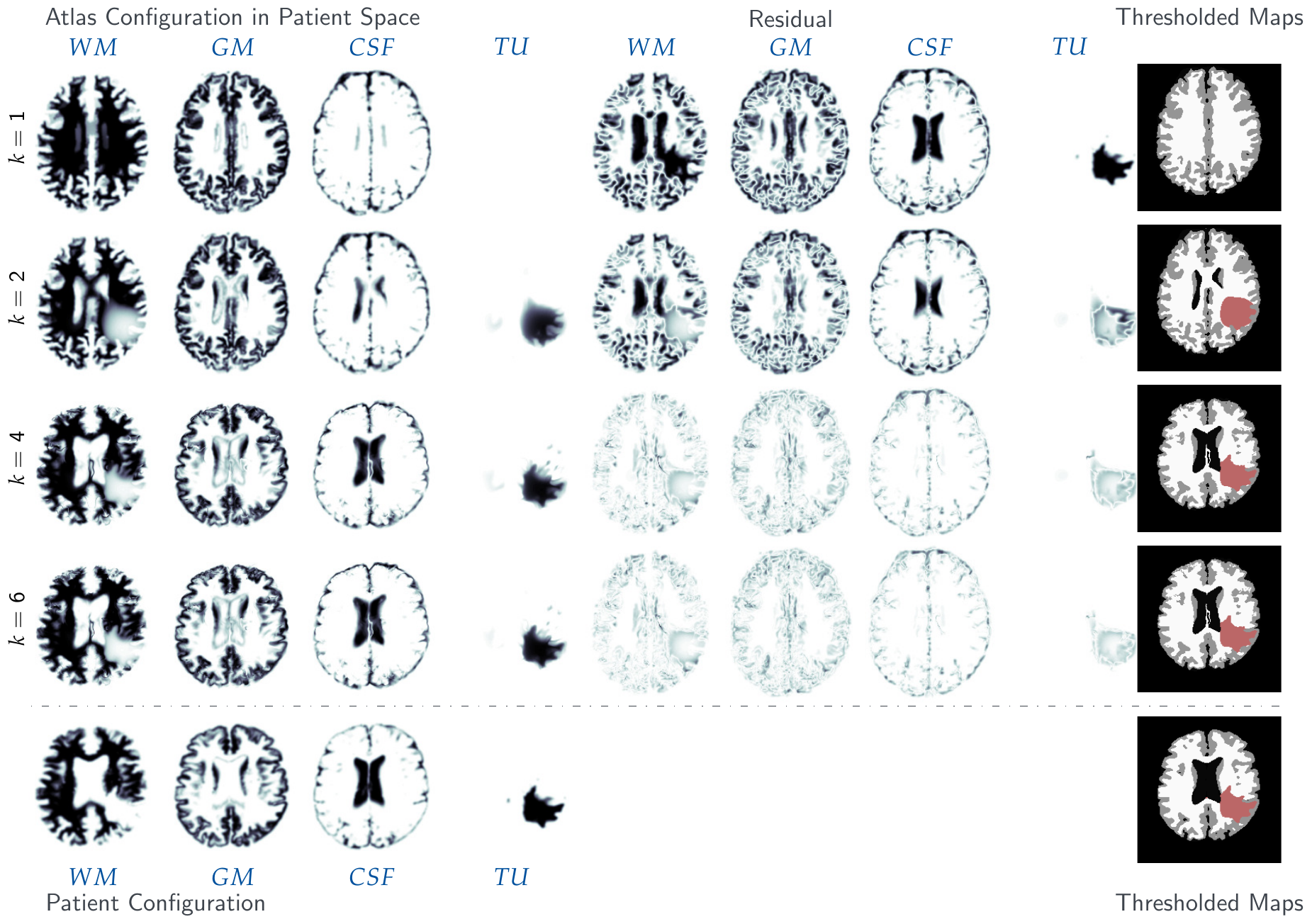}
\caption{Results for the \textbf{real tumor/ real velocity (RTRV)} test case, ground truth ($\rho$ N/A, $\mat{k}$ N/A, $\vect{p}$ N/A, $\vect{v}$ N/A) and the \textbf{AAQD} patient.
         The figure shows probability maps for the labels of the healthy atlas brain ($k=1$; top row) and the AAAN patient (target) brain probability maps with tumor (bottom row),
         along with the reconstructed probability maps throughout the Picard iterations ($k=2,4,6$) (axial-slice $136$).}
         \label{fig:appendix:AAQD_axial-over-iterations}
\end{figure*}

\begin{figure*}[ht!]
\includegraphics[scale=0.85]{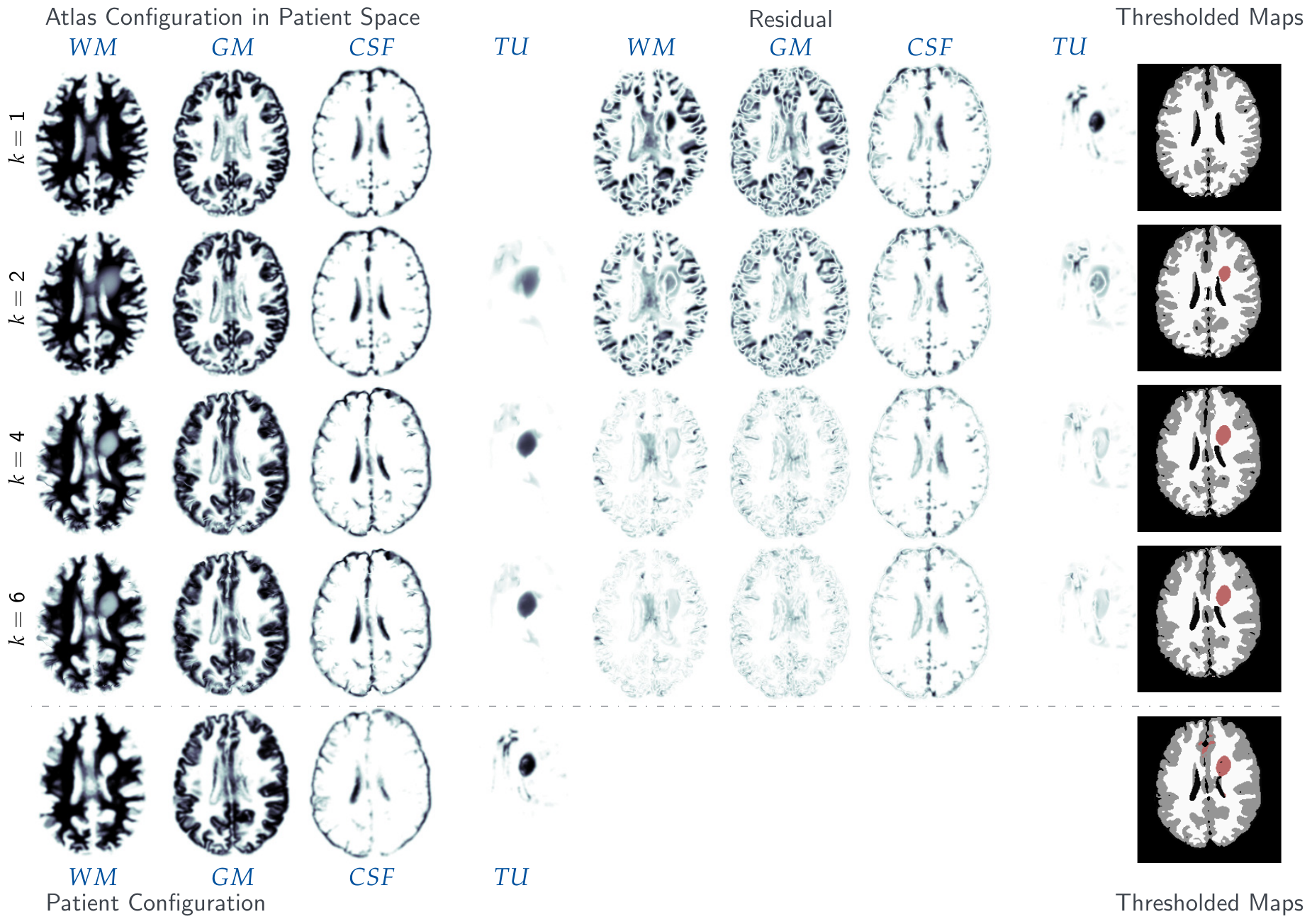}
\caption{Results for the \textbf{real tumor/ real velocity (RTRV)} test case, ground truth ($\rho$ N/A, $\mat{k}$ N/A, $\vect{p}$ N/A, $\vect{v}$ N/A) and the \textbf{AAWI} patient.
         The figure shows probability maps for the labels of the healthy atlas brain ($k=1$; top row) and the AAAN patient (target) brain probability maps with tumor (bottom row),
         along with the reconstructed probability maps throughout the Picard iterations ($k=2,4,6$) (axial-slice $132$).}
         \label{fig:appendix:AAWI_axial-over-iterations}
\end{figure*}

We show detailed results for the RTRV real data test cases for six patient datasets (AAAC, AAAN, AAMP, AAMH, AAQD, AAWI; first proposal for a patient segmentation produced in the first iteration of GLISTR~\citep{Gooya:2012a}).
\figref{fig:appendix:AAAC_axial-over-iterations} through \figref{fig:appendix:AAWI_axial-over-iterations} show axial slices of the evolution of the probability maps of the atlas brain tissue labels (in patient space) and the
reconstructed tumor (in patient space) throughout our inversion algorithm. Brain tissue probability maps, tumor probability map, mismatch for all labels as well as the hard segmentation image are given for Picard iterations
$k=\{1, 2, 4, 6\}$.

